%% file: main.tex
\documentclass[runningheads,envcountsame,envcountsect]{llncs}

\input{preamble}
\arxivtrue

\title{Complete Test Suites for Automata \texorpdfstring{\\}{} in Monoidal Closed Categories\thanks{This research is supported by the NWO grant No.~VI.Vidi.223.096.}}
\author{Bálint Kocsis\ifarxiv\else{*}\fi\orcidID{0000-0003-0570-3031} \and
    Jurriaan Rot\orcidID{0000-0002-1404-6232}}
\institute{Department of Software Science, Radboud University, Nijmegen \\
    \email{\{balint.kocsis,jurriaan.rot\}@ru.nl}}
\authorrunning{Bálint Kocsis and Jurriaan Rot}

\usepackage{enumitem}

\newlist{enum}{enumerate}{3}
\setlist[enum]{label = (\roman*), noitemsep, listparindent = \parindent}
\newlist{items}{itemize}{3}
\setlist[items]{label = $\bullet$, noitemsep, listparindent = \parindent}

\begin{document}
\def\sectionautorefname{Section}
\def\subsectionautorefname{Section}

\maketitle

\begin{abstract}
Conformance testing of automata is about checking the equivalence of a known specification and a black-box implementation. An important notion in conformance testing is that of a complete test suite, which guarantees that if an implementation satisfying certain conditions passes all tests, then it is equivalent to the specification.

We introduce a framework for proving completeness of test suites at the general level of automata in monoidal closed categories. Moreover, we provide a generalization of a classical conformance testing technique, the W-method. We demonstrate the applicability of our results by recovering the W-method for deterministic finite automata, Moore machines, and Mealy machines, and by deriving new instances of complete test suites for weighted automata and deterministic nominal automata.

\end{abstract}

\section{Introduction}
\input{intro}

\section{Overview} \label{sec:overview}
\input{overview}

\section{Words, Languages, and Automata} \label{sec:wla}
\input{wla}

\section{Test Suites and the Generalized W-method} \label{sec:test-suites}
\input{test-suites}

\section{Applications} \label{sec:applications}
\input{applications}

\section{Conclusion and Future Work} \label{sec:conclusion}
\input{conclusion}

\bibliographystyle{splncs04}
\bibliography{refs}

\ifarxiv
\clearpage
\appendix
\input{appendix}
\fi

\end{document}

%% file: preamble.tex
\usepackage[utf8]{inputenc}
\usepackage[T1]{fontenc}
\usepackage[USenglish]{babel}
\usepackage{comment}
\usepackage[hidelinks]{hyperref}

\usepackage{amsmath}
\usepackage{amssymb}
\usepackage{quiver}
\usepackage{tikz}
\usepackage{tikz-cd}
\usepackage{wrapfig}
\usepackage{xparse}

\usetikzlibrary{automata,positioning}

\newif\ifarxiv
\NewDocumentCommand{\appendixref}{m}
  {\ifarxiv Appendix~\ref{#1}%
  \fi}

\NewDocumentCommand{\xto}{O{} m}{\xrightarrow[#1]{#2}}
\renewcommand{\epsilon}{\varepsilon}
\renewcommand{\rho}{\varrho}

\NewDocumentCommand{\name}{m}{\mathrm{#1}}

\NewDocumentCommand{\N}{}{\mathbb{N}}
\NewDocumentCommand{\R}{}{\mathbb{R}}
\NewDocumentCommand{\1}{}{1}
\NewDocumentCommand{\2}{}{2}
\NewDocumentCommand{\setof}{m m}{\left\{ #1\ |\ #2 \right\}}
\NewDocumentCommand{\card}{m}{|#1|}

\NewDocumentCommand{\funcset}{m m}{#2^{#1}}
\NewDocumentCommand{\im}{m o}{\IfValueTF{#2}{#1\hspace{-2pt}\left[#2\right]}{\name{im}(#1)}}

\NewDocumentCommand{\cat}{m}{\mathcal{#1}}
\NewDocumentCommand{\Hom}{O{\mathrm{Hom}} m m}{#1(#2, #3)}
\NewDocumentCommand{\id}{s m}{1\IfBooleanF{#1}{_{#2}}}
\NewDocumentCommand{\inv}{s m}{\IfBooleanTF{#1}{(#2)^{-1}}{#2^{-1}}}

\NewDocumentCommand{\mprod}{m m}{\langle #1, #2 \rangle}
\NewDocumentCommand{\minj}{m}{\kappa_{#1}}
\NewDocumentCommand{\mcoprod}{m m}{[#1, #2]}
\NewDocumentCommand{\mfcoprod}{m m}{[#2]_{#1}}

\NewDocumentCommand{\Alg}{m}{\name{Alg}(#1)}
\NewDocumentCommand{\Coalg}{m}{\name{Coalg}(#1)}

\NewDocumentCommand{\Set}{}{\mathbf{Set}}
\NewDocumentCommand{\Pos}{}{\mathbf{Pos}}
\NewDocumentCommand{\Vect}{m}{\mathbf{Vect}_{#1}}
\NewDocumentCommand{\Nom}{}{\mathbf{Nom}}

\NewDocumentCommand{\op}{m}{#1^\mathrm{op}}

\NewDocumentCommand{\tprod}{m m}{#1 \otimes #2}
\NewDocumentCommand{\tunit}{}{I}
\NewDocumentCommand{\inthom}{m m}{[#1, #2]}
\NewDocumentCommand{\assoc}{s m m m}{\alpha\IfBooleanF{#1}{_{#2, #3, #4}}}
\NewDocumentCommand{\lunit}{s m}{\lambda\IfBooleanF{#1}{_{#2}}}
\NewDocumentCommand{\runit}{s m}{\rho\IfBooleanF{#1}{_{#2}}}
\NewDocumentCommand{\ev}{s m m}{\epsilon\IfBooleanF{#1}{_{#2, #3}}}

\NewDocumentCommand{\absepis}{}{\mathcal{E}}
\NewDocumentCommand{\absmonos}{}{\mathcal{M}}

\NewDocumentCommand{\aut}{m}{\mathcal{#1}}
\NewDocumentCommand{\acclang}{m o}{L_{#1}\IfValueT{#2}{(#2)}}
\NewDocumentCommand{\stequiv}{o m m}{#2 \sim\IfValueT{#1}{_{#1}} #3}
\NewDocumentCommand{\autequiv}{m m}{#1 \sim #2}
\NewDocumentCommand{\fdom}{m}{\mathcal{#1}}
\NewDocumentCommand{\agreets}{m m m}{#2 \sim_{#1} #3}
\NewDocumentCommand{\W}{m m m}{T^{#1}_{#2, #3}}
\NewDocumentCommand{\fdommcomp}{m}{\fdom{U}_{#1}}

\NewDocumentCommand{\adj}{s m m}{\tau\IfBooleanT{#1}{^{-1}}_{#2}(#3)}

\NewDocumentCommand{\abcobj}{}{\Sigma}
\NewDocumentCommand{\outobj}{}{O}

\NewDocumentCommand{\words}{o O{\abcobj}}{#2^\IfValueTF{#1}{{#1}}{*}}
\NewDocumentCommand{\langs}{}{\inthom{\words}{\outobj}}
\NewDocumentCommand{\wincl}{m}{j_{#1}}

\NewDocumentCommand{\cons}{o}{\name{cons}\IfValueT{#1}{_{#1}}}
\NewDocumentCommand{\snoc}{}{\name{snoc}}
\NewDocumentCommand{\append}{s o}{\mu\IfBooleanT{#1}{^\le}\IfValueT{#2}{_{#2}}}

\NewDocumentCommand{\evemp}{}{\mathrm{ev}_0}
\NewDocumentCommand{\langderiv}{}{d}

\NewDocumentCommand{\algfunc}{}{F}
\NewDocumentCommand{\alg}{o}{\alpha\IfValueT{#1}{_{#1}}}
\NewDocumentCommand{\malg}{m}{r_{#1}}
\NewDocumentCommand{\coalgfunc}{}{G}
\NewDocumentCommand{\coalg}{o}{\gamma\IfValueT{#1}{_{#1}}}
\NewDocumentCommand{\mcoalg}{m}{l_{#1}}

\NewDocumentCommand{\initmor}{m}{i_{#1}}
\NewDocumentCommand{\transmor}{s m}{\delta\IfBooleanT{#1}{^*}_{#2}}
\NewDocumentCommand{\finmor}{m}{f_{#1}}
\NewDocumentCommand{\reclang}{m o}{L_{#1}\IfValueT{#2}{(#2)}}

\NewDocumentCommand{\specname}{}{S}
\NewDocumentCommand{\implname}{}{M}
\NewDocumentCommand{\mW}{m m m}{t^{#1}_{#2, #3}}
\NewDocumentCommand{\fdomgcomp}{o m}{\fdom{U}\IfValueT{#1}{^{#1}}_{#2}}

\NewDocumentCommand{\fld}{}{\mathbb{K}}
\NewDocumentCommand{\wamatrix}{m m}{M^{#1}_{#2}}
\NewDocumentCommand{\unitmatrix}{}{I}
\NewDocumentCommand{\transp}{m}{#1^t}
\NewDocumentCommand{\vspan}{m}{\name{span}(#1)}
\NewDocumentCommand{\fdomwa}{m m}{\fdom{U}^{#1}_{#2}}

\NewDocumentCommand{\atoms}{}{\mathbb{A}}
\NewDocumentCommand{\Perm}{m}{\name{Perm}(#1)}

%% file: intro.tex
In \emph{conformance testing} of deterministic automata, the problem is to check the equivalence of a known specification and a black-box implementation~\cite{DBLP:conf/dagstuhl/testing05}. This form of testing is widely used, for instance, in practical implementations of \emph{automata learning}, where one aims to infer an automaton model by systematically interacting with a black-box system~\cite{DBLP:conf/dagstuhl/HowarS16,DBLP:phd/dnb/Isberner15,DBLP:journals/cacm/Vaandrager17}.
In particular, conformance testing is a standard technique to evaluate whether a learned model is correct. In Angluin's~\cite{DBLP:journals/iandc/Angluin87} minimally adequate teacher (MAT) framework for active automata learning, this is referred to as an \emph{equivalence query} --- which, when the system under learning really is a black-box, can only be implemented through testing.

In a black-box setting, testing can never be exhaustive~\cite{moore1956gedanken}: for any (finite) collection of tests, we can construct a faulty implementation that passes all the tests. However, under certain assumptions on the implementation, it is possible to construct test suites which guarantee equivalence. These test suites are called \emph{complete}. In particular, an \emph{$m$-complete} test suite is complete with respect to all implementations with at most $m$ states. The W-method~\cite{DBLP:journals/tse/Chow78,vasilevskii1973failure} is a classical construction of an $m$-complete test suite, which has been extended and improved in numerous ways (see~\cite{DBLP:journals/infsof/DorofeevaEMCY10,DBLP:journals/pieee/LeeY96,moerman2019nominal} for an overview). State-of-the-art libraries for automata learning~\cite{IsbernerHS15,MuskardinAPPT21} implement various (randomised) versions of these test suites to evaluate the equivalence query.

Active automata learning has been extended from deterministic finite automata~\cite{DBLP:journals/iandc/Angluin87} to a wide range of system types, such as Mealy machines~\cite{MargariaNRS04}, 
non-deterministic automata~\cite{BolligHKL09},
register automata~\cite{CEGAR12,BolligHLM13,CasselHJS16},
nominal automata~\cite{MoermanS0KS17},
weighted automata~\cite{BalleM15,Bergadano,DBLP:conf/fossacs/HeerdtKR020},
automata over infinite words~\cite{AngluinAF19,MalerP95}, and
probabilistic models~\cite{TapplerA0EL21}. 
Moreover, automata learning algorithms have been generalized to the abstract level of category theory~\cite{DBLP:conf/fossacs/BarloccoKR19,DBLP:conf/csl/ColcombetPS21,DBLP:conf/cmcs/HeerdtSS20,DBLP:conf/lics/UrbatS20}. However, most variants of complete test suites are only developed for Mealy machines and DFAs. While the abstract frameworks for automata learning rely on equivalence queries, a general theory of evaluating these in practice through conformance testing is missing. Developing such a theory is relevant, as evaluating the equivalence query is often the bottleneck in automata learning~\cite{AslamCSB20}.

In this paper, we study conformance testing at the generality of \emph{machines in a category}, more specifically in monoidal closed categories. This setting goes back to classical work by Arbib and Manes~\cite{arbib1975adjoint}, Goguen~\cite{DBLP:journals/jcss/Goguen75}, and Adámek and Trnková~\cite{adamek1990automata}. 
This setting allows us to conveniently define reachability and language equivalence, by viewing automata both as algebras and as coalgebras.

Our main contribution is to define the W-method and to provide a framework for proving $(m)$-completeness of test suites at the general level of automata in monoidal closed categories.
To this end, we generalize part of the bisimulation-based proof presented in~\cite{DBLP:conf/tacas/KrugerJR24} for Mealy machines, which ultimately goes back to~\cite{DBLP:journals/tse/Chow78,vasilevskii1973failure}. In particular, our main result allows us to prove completeness under the assumption that the we can reach all states in the implementation, a property guaranteed by the W-method in concrete cases.
We instantiate our approach to recover known complete test suites for deterministic finite automata, Mealy machines, and Moore machines. We go on to instantiate it to weighted automata~\cite{DBLP:books/ems/21/DrosteK21} and deterministic nominal automata~\cite{DBLP:journals/corr/BojanczykKL14}, providing, to the best of our knowledge, the first account of complete test suites for these systems.

\paragraph{Related work.}
Our approach is inspired by the categorical automata learning framework of Urbat and Schr\"{o}der~\cite{DBLP:conf/lics/UrbatS20}, which similarly relies on the possibility of viewing automata as both algebras and coalgebras in the more general setting of adjoint automata. We address conformance testing instead of automata learning.

There is extensive literature on advanced complete test suites for concrete models, in particular for DFAs and Mealy machines (but also for e.g. NFAs~\cite{DBLP:conf/hase/PetrenkoY14} and other conformance relations such as ioco~\cite{BosJM19}). In this paper, we focus on the W-method, which conceptually underlies many of these techniques, as a first step towards a categorical theory of conformance testing.

In~\cite{DBLP:conf/ictac/KansoABT10}, a coalgebraic conformance testing theory is developed for Mealy machines with monadic effects. The focus in \emph{op. cit.} is on modelling test purposes and execution at that level, but not on $m$-completeness.

\paragraph{Outline.}
In \autoref{sec:overview}, we provide background on conformance testing and an overview of our approach. Then, in \autoref{sec:wla}, we set the scene for the generalized test suite construction by recalling automata in monoidal closed categories and related concepts such as words and languages. We introduce our framework for proving completeness of test suites in \autoref{sec:test-suites}. In \autoref{sec:applications}, we use our results to obtain complete test suites for all the abovementioned classes of automata. Finally, \autoref{sec:conclusion} summarizes our results and discusses further work.

We assume familiarity with basic category theory~\cite{mac2013categories}. Some notions we rely on are recalled in \appendixref{sec:cat-prelims}. Most proofs are also deferred to the appendix.

\paragraph{Notation.}
We write $gf$ or $g \circ f$ for the composition of morphisms $g$ and $f$. We denote the product of $X$ and $Y$ by $X \times Y$. For $f \colon Z \to X$ and $g \colon Z \to Y$, we let $\mprod{f}{g} \colon Z \to X \times Y$ denote the tupling of $f$ and $g$. The coproduct of $X$ and $Y$ is written as $X + Y$. For $f \colon X \to Z$ and $g \colon Y \to Z$, we denote by $\mcoprod{f}{g} \colon X + Y \to Z$ the cotupling of $f$ and $g$. The coproduct of a family of objects $(X_i)_{i \in I}$ is written as $\sum_{i \in I}{X_i}$. For a family of morphisms $f_i \colon X_i \to Y$ indexed by $i \in I$, we write $\mfcoprod{i \in I}{f_i} \colon \sum_{i \in I}{X_i} \to Y$ for the cotupling of the family. Furthermore, we write $\minj{i} \colon X_i \to \sum_{i \in I}{X_i}$ for the coproduct injections.

%% file: overview.tex
In this section, we review some background on conformance testing and the W-method in particular, and we explain our approach to generalizing the test suite construction and the proof of its completeness.

\paragraph{Complete Test Suites.}
Let us introduce some terminology for conformance testing in the case of deterministic finite automata. For this section, we fix a set $\abcobj$ of input symbols.
A \emph{deterministic finite automaton} (DFA)~\cite{DBLP:journals/ibmrd/RabinS59} is a tuple $(Q, q_0, \delta, F)$, where $Q$ is a finite set of \emph{states}, $q_0 \in Q$ is an \emph{initial state}, $\delta \colon Q \times \abcobj \to Q$ is a \emph{transition function}, and $F \subseteq Q$ is a set of \emph{final states}.
The \emph{size} of a DFA $\aut{A} = (Q, q_0, \delta, F)$ is $\card{\aut{A}} = \card{Q}$. The transition function $\delta$ of $\aut{A}$ extends to words over $\abcobj$ as a function $\delta^* \colon Q \times \words \to Q$ as usual. We write $q \in \aut{A}$ to mean $q \in Q$.

\begin{definition}
Let $\aut{A} = (Q, q_0, \delta, F)$ and $\aut{B} = (Q', q_0', \delta', F')$ be two DFAs.
\begin{enum}
\item The \emph{accepted language} of $q \in Q$ is 
$\acclang{\aut{A}}[q] = \setof{w \in \words}{\delta^*(q, w) \in F}$. The \emph{accepted language} of $\aut{A}$ is $\acclang{\aut{A}} = \acclang{\aut{A}}[q_0]$.

\item Given a set $L \subseteq \words$ of words, we say that two states $p \in Q$ and $q \in Q'$ are \emph{$L$-equivalent}, denoted by $\stequiv[L]{p}{q}$, if $\acclang{\aut{A}}[p] \cap L = \acclang{\aut{B}}[q] \cap L$. In the case $L = \words$, we say that $p$ and $q$ are \emph{equivalent}, denoted by $\stequiv{p}{q}$. We write $\agreets{L}{\aut{A}}{\aut{B}}$ for $\stequiv[L]{q_0}{q_0'}$. We say that $\aut{A}$ and $\aut{B}$ are \emph{equivalent}, denoted by $\autequiv{\aut{A}}{\aut{B}}$, if $\stequiv{q_0}{q_0'}$.

\item We say that $\aut{A}$ is \emph{minimal} if for all $p, q \in Q$, $\stequiv{p}{q}$ implies $p = q$.
\end{enum}
\end{definition}

We now define the vocabulary for conformance testing. A \emph{test suite} is a finite set $T \subseteq \words$ of words. In the following, we use the symbol $\aut{\specname}$ for the specification and the symbol $\aut{\implname}$ for the implementation. We say that $\aut{\specname}$ and $\aut{\implname}$ \emph{agree on} a test suite $T$ if $\agreets{T}{\aut{\specname}}{\aut{\implname}}$.

Since for any test suite, there exists a faulty implementation that agrees on all the tests, we need to restrict the space of possible faulty implementations to achieve completeness. This motivates the definition of a fault domain.
\begin{definition}
\begin{enum}

\item A \emph{fault domain} is a collection of DFAs.

\item Let $\aut{\specname}$ be an automaton. A test suite $T \subseteq \words$ is \emph{complete} for $\aut{\specname}$ with respect to a fault domain $\fdom{U}$ if for all DFAs $\aut{\implname} \in \fdom{U}$, $\agreets{T}{\aut{\specname}}{\aut{\implname}}$ implies $\autequiv{\aut{\specname}}{\aut{\implname}}$.
\end{enum}
\end{definition}
Intuitively, a test suite is complete if it contains a failing test case for every inequivalent DFA in the given fault domain.

A fault domain that has often been considered in the literature~\cite{DBLP:journals/tse/Chow78,DBLP:journals/infsof/DorofeevaEMCY10,moerman2019nominal,vasilevskii1973failure}
is the collection $\fdommcomp{m}$ of all DFAs that have at most $m$ states, for some $m \in \N$. Test suites that are complete with respect to this fault domain are called \emph{$m$-complete}.

\paragraph{The W-method.}
The method of Vasilevskii and Chow, termed the \emph{W-method}, constructs a test suite based on the specification using sets of words with special properties, called a \emph{state cover} and a \emph{characterization set}. A state cover $P$ is a set of words containing sequences with which we can cover the whole state space, and a characterization set contains words that distinguish inequivalent states.
\begin{definition}
\begin{enum}
\item A \emph{state cover} for a DFA $\aut{\specname} = (Q, q_0, \delta, F)$ is a finite set $P \subseteq \words$ of words that contains the empty word and contains, for each state $q \in Q$, an \emph{access sequence} for $q$, i.e., a word $w \in \words$ such that $\delta^*(q_0, w) = q$.

\item A \emph{characterization set} for a DFA $\aut{\specname}$ is a finite set $W \subseteq \words$ of words such that $\epsilon \in W$ and for all states $p, q \in \aut{\specname}$, $\stequiv[W]{p}{q}$ implies $\stequiv{p}{q}$.
\end{enum}
\end{definition}
Intuitively, a characterization set contains a so-called \emph{distinguishing sequence} for any pair of inequivalent states $p$ and $q$, i.e. a word $w \in \words$ such that $w \in \acclang{\aut{\specname}}[p]$ and $w \notin \acclang{\aut{\specname}}[q]$ or vice-versa. Note that, by definition, such a distinguishing sequence always exists for inequivalent states.

We are now in the position to define the W-method. In the following definition, the symbol $\cdot$ denotes concatenation of languages.
\begin{definition} \label{def:W-method}
Let $\aut{\specname}$ be a DFA. Let $P$ be a state cover for $\aut{\specname}$, let $W$ be a characterization set for $\aut{\specname}$, and let $k$ be a natural number. Then the \emph{W test suite} of order $k$ associated to $P$ and $W$ is defined as $\W{k}{P}{W} = P \cdot \words[\le k+1] \cdot W$.
\end{definition}
The state cover $P$ makes sure that we reach all states in the specification. The role of the infixes $\words[\le k+1]$ is to reach states in the implementation. Finally, the characterization set $W$ is used to distinguish the states reached in the specification and the implementation after reading a word from $P \cdot \words[\le k+1]$.

It is a classical result~\cite{DBLP:journals/tse/Chow78,vasilevskii1973failure} that the W-method of order $k$ is $n + k$-complete, where $n$ is the number of states of the specification. This can be proven in two steps; this proof strategy is also used in~\cite{moerman2019nominal} and~\cite{DBLP:conf/tacas/KrugerJR24}. The first step is to construct a state cover for the implementation from a state cover of the specification.
\begin{lemma} \label{lem:reachability-impl-dfa}
Let $\aut{\specname}$ be a DFA with $n = \card{\aut{\specname}}$, and let $\aut{\implname} \in \fdommcomp{n + k}$ for some $k \in \N$. Suppose that $P \subseteq \words$ is a state cover for $\aut{\specname}$ and $W \subseteq \words$ is a characterization set for $\aut{\specname}$. Suppose furthermore that $\aut{\specname}$ is minimal and $\agreets{P \cdot W}{\aut{\specname}}{\aut{\implname}}$. Then $P \cdot \words[\le k]$ is a state cover for $\aut{\implname}$.
\end{lemma}

Second, we prove equivalence of two DFAs using a state cover for the implementation and the assumption that the two machines agree on suitable tests.
\begin{lemma} \label{lem:agreets-equiv-dfa}
Let $\aut{\specname}$ and $\aut{\implname}$ be two DFAs, and suppose $C \subseteq \words$ is a state cover for $\aut{\implname}$ and $W \subseteq \words$ is a characterization set for $\aut{\specname}$. Assume that $\aut{\specname}$ is minimal, and let $T = C \cdot \words[\le 1] \cdot W$. Then $\agreets{T}{\aut{\specname}}{\aut{\implname}}$ implies $\autequiv{\aut{\specname}}{\aut{\implname}}$.
\end{lemma}
For a proof of the above two lemmas in the case of Mealy machines, we refer the reader to~\cite{DBLP:conf/tacas/KrugerJR24}. In both lemmas, we assumed minimality of the specification: this is fine, as we can apply any minimization algorithm to a potential non-minimal specification. Furthermore, the hypotheses produced in active learning algorithms are generally minimal.

Altogether, we obtain $n+k$-completeness of the W-method.
\begin{corollary} \label{cor:W-comp}
Let $\aut{\specname}$ be a minimal DFA with $n = \card{\aut{\specname}}$. Then for all state covers $P \subseteq \words$ for $\aut{\specname}$, for all characterization sets $W \subseteq \words$ for $\aut{\specname}$, and for all $k \in \N$, the test suite $\W{k}{P}{W}$ is $n+k$-complete for $\aut{\specname}$.
\end{corollary}
\begin{proof}
Let $\aut{\implname} \in \fdommcomp{n+k}$, and suppose $\agreets{\W{k}{P}{W}}{\aut{\specname}}{\aut{\implname}}$. Since $P \cdot W \subseteq \W{k}{P}{W}$, we also have $\agreets{P \cdot W}{\aut{\specname}}{\aut{\implname}}$. Hence, by \autoref{lem:reachability-impl-dfa}, $P \cdot \words[\le k]$ is a state cover for $\aut{\implname}$. Then, since $\W{k}{P}{W} = (P \cdot \words[\le k]) \cdot \words[\le 1] \cdot W$, $\autequiv{\aut{\specname}}{\aut{\implname}}$ follows by \autoref{lem:agreets-equiv-dfa}. \qed
\end{proof}

\begin{wrapfigure}{r}[0pt]{0.35\textwidth}
\centering
\vspace{-7mm}
\resizebox{120pt}{!}{
\begin{tikzpicture}[auto,initial text=,bend angle=15]
    \node[state,initial,accepting] (0)              {$q_0$};
    \node[state,accepting]         (1) [right=of 0] {$q_1$};
    \node[state,accepting]         (2) [right=of 1] {$q_2$};
    \node[state]                   (3) [below=of 1] {$q_3$};
    \path[->] (0) edge                 node [swap] {$c,e$}   (3)
                  edge [bend right]    node [swap] {$1$}     (1)
              (1) edge [bend right]    node [swap] {$c$}     (0)
                  edge                 node        {$e$}     (3)
                  edge [bend right]    node [swap] {$1$}     (2)
              (2) edge [bend right]    node [swap] {$c$}     (1)
                  edge [bend right=50] node [swap] {$e$}     (0)
                  edge                 node        {$1$}     (3)
              (3) edge [loop below]    node        {$c,e,1$} ();
\end{tikzpicture}
}
\caption{A DFA $\aut{\specname}$ for a coffee machine}
\label{fig:ex-dfa}
\vspace{-2mm}
\end{wrapfigure}
Let us illustrate the W-method on a toy example. (This example is based on a Mealy machine in~\cite{DBLP:conf/tacas/KrugerJR24}.) Suppose we have a coffee machine that can dispense coffee or espresso. Coffee costs 1 coin and espresso costs 2 coins. Suppose that the machine breaks whenever we try to order something without enough money, or when we insert more than 2 coins. Then the DFA $\aut{\specname}$ over the alphabet $\abcobj = \{c, e, 1\}$ depicted in \autoref{fig:ex-dfa} accepts precisely those interaction sequences that do not break the coffee machine.

By taking the shortest access sequences for all the states, we see that a state cover for $\aut{\specname}$ is given by $P = \{\epsilon, c, 1, 11\}$. The input sequence $c$ distinguishes $q_0$ and $q_1$, as well as $q_2$ and $q_3$, and the input sequence $1$ distinguishes $q_0$ and $q_2$, as well as $q_1$ and $q_2$, $q_0$ and $q_3$, and $q_1$ and $q_3$. Hence, a characterization set for this DFA is $W = \{\epsilon, c, 1\}$. Thus, setting $k = 0$, the W-method gives the test suite $\W{0}{P}{W} = P \cdot \words[\le 1] \cdot W$.

By \autoref{cor:W-comp}, $\W{0}{P}{W}$ is $4$-complete for $\aut{\specname}$. For instance, consider the implementation $\aut{\implname}_1$ that is the same as $\aut{\specname}$ except that the $c$-transition of $q_2$ goes to $q_2$. Then the test case $11c1$ (and only this) detects this fault. Similarly, the faulty implementation $\aut{\implname}_2$ that is the same as $\aut{\specname}$ except that the $e$-transition of $q_2$ goes to $q_1$ is rejected by the test case $11ec$ (and only this).

\paragraph{Generalizing the W-method.}
We wish to generalize the W-method to a categorical setting. The first hurdle towards achieving this goal is giving definitions of a state cover and of a characterization set. This is challenging because the nature of the two sets is fundamentally different: the state cover concerns \emph{reachability} of states, whereas the characterization set is about \emph{equivalence} of states. These two notions are \emph{dual} to each other~\cite{arbib1975adjoint}: the former is described using free algebras, while the latter by cofree algebras.

We observe that the setting of automata in monoidal closed categories (called \emph{machines} by Goguen~\cite{DBLP:journals/jcss/Goguen75} and \emph{sequential $\Sigma$-automata} by Adámek and Trnková~\cite{adamek1990automata} in the Cartesian monoidal case) provides an adequate framework for our goals. This setting defines an abstract notion of automaton which can be viewed both as an algebra and as a coalgebra for suitable endofunctors, giving a straightforward way to define reachability and equivalence of states. It covers many kinds of automata, including DFAs, Moore machines, Mealy machines, linear weighted automata, and deterministic nominal automata.

This setting is particularly nice since it allows us to work with concrete descriptions of the initial algebra and of the final coalgebra for the functors describing the algebraic and coalgebraic presentations of automata, respectively. Additionally, these descriptions are direct generalizations of the corresponding and well-known~\cite{jacobs1997tutorial,DBLP:journals/tcs/Rutten00} concepts in the case of DFAs: the initial algebra arises as the object of words, and the final coalgebra as the object of languages. Crucially, this provides a link between these two objects, making it more or less straightforward to define generalizations of a state cover and of a characterization set.

In this categorical setting, we provide a generalization of \autoref{lem:agreets-equiv-dfa}. We do not generalize \autoref{lem:reachability-impl-dfa}, as it relies on notions of size that are better handled on a case-by-case basis. As applications, we will be able to derive complete test suites for all the abovementioned classes of of automata.

%% file: wla.tex
In this section, we define the ingredients necessary to carry out our generalized test suite construction. In particular, we define words, languages, and automata in monoidal closed categories, and operations on these objects. We recall that words form an initial algebra~\cite{DBLP:journals/jcss/Goguen75} and that languages form a final coalgebra~\cite{DBLP:conf/lics/UrbatS20} of certain endofunctors. We show how to give definitions in such a way that the resulting algebraic and coalgebraic language semantics for automata coincide.

For the remainder of this paper, we fix a monoidal closed category $\cat{C}$ with tensor product $\tprod{-}{-}$, tensor unit $\tunit$, associator $\assoc*{}{}{}$, left unitor $\lunit*{}$, and right unitor $\runit*{}$. The internal hom of $\cat{C}$ is denoted by $\inthom{-}{-}$. We denote by $\adj{X}{-}$ the natural bijection $\Hom[\cat{C}]{\tprod{Z}{X}}{Y} \xto{\cong} \Hom[\cat{C}]{Z}{\inthom{X}{Y}}$ arising from the adjunction $\tprod{-}{X} \dashv \inthom{X}{-}$. We denote the counit of same adjunction by $\ev{X}{Y} \colon \tprod{\inthom{X}{Y}}{X} \to Y$. Furthermore, we fix two objects $\abcobj, \outobj \in \cat{C}$. We think of $\abcobj$ as the \emph{alphabet object} and of $\outobj$ as the \emph{output object}. The alphabet object generalizes the set of input symbols for DFAs. In the case of DFAs, the output of a state is simply a Boolean value stating whether the state is accepting or not. The output object generalizes this concept, so that the output of a state can be an arbitrary value. We assume that $\cat{C}$ has binary products and countable coproducts. Finally, we assume that for any $X \in \cat{C}$, the functor $\tprod{X}{-}$ preserves countable coproducts. This assumption is satisfied, for instance, if the functor $\tprod{X}{-}$ also has a right adjoint, as is the case in any braided monoidal closed category.

We note that, intuitively, the internal hom $\inthom{-}{-}$ internalizes the homsets of the category $\cat{C}$ as objects of $\cat{C}$, in the sense that for any pair of objects $X$ and $Y$, there is a natural bijection $\Hom[\cat{C}]{X}{Y} \cong \Hom[\cat{C}]{\tunit}{\inthom{X}{Y}}$.

\subsection{Words}

We begin by defining words and operations on words.
\begin{definition} \label{def:words}
\begin{enum}
\item For an object $X \in \cat{C}$, define its $n$-fold tensor product by recursion on $n$: $X^0 = \tunit$ and $X^{n+1} = \tprod{X^n}{X}$.

\item For a natural number $k \in \N$, let $X^{\le k} = \sum_{n \le k}{X^n}$.

\item Finally, let $X^* = \sum_{n \in N}{X^n}$, and define $\wincl{k} \colon X^{\le k} \to X^*$ as $\mfcoprod{n \le k}{\minj{n}}$.
\end{enum}
\end{definition}
We think of $\words$ as the \emph{object of words} over $\abcobj$.

At this point, we note that both functors $\tprod{-}{X}$ and $\tprod{X}{-}$ preserve countable coproducts (the former by having a right adjoint $\inthom{X}{-}$, and the latter by assumption). In particular, for any object $X$, we have isomorphisms
\[ \mfcoprod{n \in \N}{\tprod{\minj{n}}{\id{X}}} \colon \sum_{n \in N}{\tprod{\words[n]}{X}} \xto{\cong} \tprod{\words}{X}, \quad \mfcoprod{n \in \N}{\tprod{\id{X}}{\minj{n}}} \colon \sum_{n \in N}{\tprod{X}{\words[n]}} \xto{\cong} \tprod{X}{\words}. \]
Combining these yields the isomorphism
\[ \mfcoprod{m,n \in \N}{\tprod{\minj{m}}{\minj{n}}} \colon \sum_{m,n \in \N}{\tprod{\words[m]}{\words[n]}} \xto{\cong} \tprod{\words}{\words}. \]

We define two operations on words, $\cons$ and $\snoc$. Intuitively, $\cons$ prepends a letter to the beginning of a word, and $\snoc$ appends a letter to the end.
\begin{definition}
\begin{enum}
\item We define a family of morphisms $\cons[n] \colon \tprod{\abcobj}{\words[n]} \to \words[n+1]$ by recursion on $n$:
$\cons[0] = \inv{\lunit{\abcobj}} \circ \runit{\abcobj}$ and $\cons[n+1] = (\tprod{\cons[n]}{\id{\abcobj}}) \circ \inv{\assoc{\abcobj}{\words[n]}{\abcobj}}$.

\item Let $\cons \colon \tprod{\abcobj}{\words} \to \words$ be defined as the composite
\[ \tprod{\abcobj}{\words} \xto{\cong} \sum_{n \in \N}{\tprod{\abcobj}{\words[n]}} \xto{\sum_{n \in \N}{\cons[n]}} \sum_{n \in \N}{\words[n+1]} \xto{\mfcoprod{n \in \N}{\minj{n+1}}} \words. \]

\item The map $\snoc \colon \tprod{\words}{\abcobj} \to \words$ is defined as the composition of the isomorphism $\tprod{\words}{\abcobj} \xto{\cong} \sum_{n \in \N}{\tprod{\words[n]}{\abcobj}}$ and $\mfcoprod{n \in \N}{\minj{n+1}} \colon \sum_{n \in \N}{\words[n+1]} \to \words$.
\end{enum}
\end{definition}

We can also define concatenation of words.
\begin{definition}
\begin{enum}
\item We define a family of maps $\append[m,n] \colon \tprod{\words[m]}{\words[n]} \to \words[m+n]$ by recursion on n:
$\append[m,0] = \runit{\words[m]}$ and $\append[m,n+1] = (\tprod{\append[m,n]}{\id{\abcobj}}) \circ \inv{\assoc{\words[m]}{\words[n]}{\abcobj}}$.

\item We define the map $\append \colon \tprod{\words}{\words} \to \words$ as the composite
\[ \tprod{\words}{\words} \xto{\cong} \sum_{m,n \in \N}{\tprod{\words[m]}{\words[n]}} \xto{\sum_{m,n \in \N}{\append[m,n]}} \sum_{m,n \in \N}{\words[m+n]} \xto{\mfcoprod{m,n \in \N}{\minj{m+n}}} \words. \]
\end{enum}
\end{definition}

In the next section, we will use the following concatenation operation on morphisms into $\words$, which generalizes concatenation of languages.
\begin{definition}
Let $a \colon X \to \words$, $b \colon Y \to \words$ be two morphisms. The \emph{concatenation} of $a$ and $b$, denoted by $a \cdot b$, is defined as the composite
\[ \tprod{X}{Y} \xto{\tprod{a}{b}} \tprod{\words}{\words} \xto{\append} \words. \]
\end{definition}
By convention, $\cdot$ associates to the left.

We can characterize $\words$ as the initial algebra of a certain endofunctor on $\cat{C}$. From now on, we fix the functor $\algfunc \colon \cat{C} \to \cat{C}$ given on objects by $\algfunc X = \tunit + \tprod{X}{\abcobj}$.
\begin{proposition} \label{prop:words-init-alg}
Let $\alg = \mcoprod{\minj{0}}{\snoc} \colon \tunit + \tprod{\words}{\abcobj} \to \words$. Then the algebra $(\words, \alg)$ is an initial $\algfunc$-algebra.
\end{proposition}
For an $\algfunc$-algebra $(A, \alg[A])$, we denote the unique $\algfunc$-algebra homomorphism from $(\words, \alg)$ to $(A, \alg[A])$ by $\malg{\alg[A]} \colon \words \to A$.

\subsection{Languages}

We now turn to languages. In the set-theoretical case, languages can be defined as functions $\words \to \2$. Replacing the output set $\2$ by an arbitrary object yields the following definition.
\begin{definition}
A \emph{language} is a morphism $\words \to \outobj$.
\end{definition}
Internalizing the above notion, we may think of $\langs$ as the \emph{object of languages}.

We define two operations on languages. The first one computes the output for the empty word. The second one corresponds to the derivative of languages~\cite{DBLP:journals/jacm/Brzozowski64}.
\begin{definition} \label{def:evemp-langderiv}
\begin{enum}
\item
Define the morphism $\evemp \colon \langs \to \outobj$ as the composite
$\evemp = \ev{\words}{\outobj} \circ (\tprod{\id{\langs}}{\minj{0}}) \circ \inv{\runit{\langs}}$.

\item Define the morphism $\langderiv \colon \langs \to \inthom{\abcobj}{\langs}$ as $\adj{\abcobj}{\adj{\words}{d'}}$, where $d'$ is the composite $d' = \ev{\words}{\outobj} \circ (\tprod{\id{\langs}}{\cons}) \circ \assoc{\langs}{\abcobj}{\words}$.
\end{enum}
\end{definition}

We can characterize $\langs$ as the final coalgebra for a certain endofunctor. From now on, we fix the functor $\coalgfunc \colon \cat{C} \to \cat{C}$ given on objects by $\coalgfunc X = \outobj \times \inthom{\abcobj}{X}$.
\begin{proposition} \label{prop:lang-fin-coalg}
Let $\coalg = \mprod{\evemp}{\langderiv} \colon \langs \to \outobj \times \inthom{\abcobj}{\langs}$. Then the coalgebra $(\langs, \coalg)$ is a final $\coalgfunc$-coalgebra.
\end{proposition}
For a $\coalgfunc$-coalgebra $(C, \coalg[C])$, we denote the unique $\coalgfunc$-coalgebra homomorphism from $(C, \coalg[C])$ to $(\langs, \coalg)$ by $\mcoalg{\coalg[C]} \colon C \to \langs$.

\subsection{Automata}

We conclude this section with a discussion of automata in monoidal closed categories. These are very similar to classical DFAs. The difference is that we replace Cartesian product with tensor product, and we generalize the initial state to an initial state morphism and the set of final states to an output morphism.
\begin{definition}
An \emph{automaton} over $\abcobj$ in $\cat{C}$ is a tuple $(Q, \initmor{Q}, \transmor{Q}, \finmor{Q})$, where $Q \in \cat{C}$ and $\initmor{Q} \colon \tunit \to Q$, $\transmor{Q} \colon \tprod{Q}{\abcobj} \to Q$, $\finmor{Q} \colon Q \to \outobj$ are morphisms in $\cat{C}$.
\end{definition}
We think of $Q$ as an object of \emph{states}, of $\initmor{Q}$ as selecting an \emph{initial state}, of $\transmor{Q}$ as a \emph{transition morphism}, and of $\finmor{Q}$ as an \emph{output morphism}.

We can view any automaton $\aut{A} = (Q, \initmor{Q}, \transmor{Q}, \finmor{Q})$ as an $\algfunc$-algebra $(Q, \mcoprod{\initmor{Q}}{\transmor{Q}})$ equipped with an output morphism $\finmor{Q}$. Hence, by the initiality of $(\words, \alg)$, we get the unique $\algfunc$-algebra homomorphism $\malg{\mcoprod{\initmor{Q}}{\transmor{Q}}} \colon \words \to Q$. We denote this morphism by $\malg{\aut{A}}$. Intuitively, the morphism $\malg{\aut{A}}$ maps a word $w$ to the state of $\aut{A}$ reached from the initial state upon reading the input $w$.

\begin{definition}
\begin{enum}
\item Given an automaton $\aut{A} = (Q, \initmor{Q}, \transmor{Q}, \finmor{Q})$, the \emph{recognized language} of $\aut{A}$ is defined as $\reclang{\aut{A}} = \words \xto{\malg{\aut{A}}} Q \xto{\finmor{Q}} \outobj$.

\item We say that two automata $\aut{A}$ and $\aut{B}$ are \emph{equivalent}, denoted by $\autequiv{\aut{A}}{\aut{B}}$, if they recognize the same language, i.e. $\reclang{\aut{A}} = \reclang{\aut{B}}$.
\end{enum}
\end{definition}
The recognized language of an automaton generalizes the usual concept of accepted language of DFAs: the language assigns to a given word the output of the state reached from the initial state upon reading that word.

We can also view any automaton $\aut{A} = (Q, \initmor{Q}, \transmor{Q}, \finmor{Q})$ as a $\coalgfunc$-coalgebra $(Q, \mprod{\finmor{Q}}{\adj{\abcobj}{\transmor{Q}}})$ equipped with an initial state $\initmor{Q}$. By the finality of $(\langs, \coalg)$, we get the unique $\coalgfunc$-coalgebra homomorphism $\mcoalg{\mprod{\finmor{Q}}{\adj{\abcobj}{\transmor{Q}}}} \colon Q \to \langs$. We denote this morphism by $\mcoalg{\aut{A}}$. Intuitively, the morphism $\mcoalg{\aut{A}}$ maps a state of $\aut{A}$ to the language accepted from that state.

Note that the definition of recognized language relies on the initial algebra morphism $\malg{\aut{A}}$. Alternatively, it can be defined coalgebraically as the morphism
\[ \reclang{\aut{A}}' = \tunit \xto{\initmor{Q}} Q \xto{\mcoalg{\aut{A}}} \langs. \]
The following proposition asserts that these two presentations are equivalent.
\begin{proposition} \label{prop:rec-lang-equiv}
The bijection $\Hom[\cat{C}]{\words}{\outobj} \cong \Hom[\cat{C}]{\tunit}{\inthom{\words}{\outobj}}$ sends $\reclang{\aut{A}}$ to $\reclang{\aut{A}}'$.
\end{proposition}

Note that, by definition, a DFA $\aut{A}$ is minimal if and only if the function $q \mapsto \acclang{\aut{A}}[q]$ is injective. As a direct generalization, we define an automaton $\aut{A}$ to be \emph{minimal} if $\mcoalg{\aut{A}}$ is a monomorphism. (Goguen~\cite{DBLP:journals/jcss/Goguen75} calls such an automaton \emph{reduced}, while Arbib and Manes~\cite{arbib1975adjoint} call such an automaton \emph{observable}.)

%% file: test-suites.tex
We now generalize the test suite construction to our categorical setting. We continue to use the symbol $\aut{\specname}$ for the specification and $\aut{\implname}$ for the implementation.

Since in the case of DFAs, test suites, state covers, and characterization sets are sets of words, it would make sense to define the generalizations of these as subobjects of $\words$. However, it turns out that the theory becomes simpler if we consider arbitrary morphisms instead of monos. In particular, this allows us to bypass the need for factorization systems (cf. \autoref{rem:factorizations}).

We define a \emph{test suite} to be a morphism $t \colon T \to \words$. We say that two automata $\aut{\specname}$ and $\aut{\implname}$ \emph{agree on} $t$, denoted by $\agreets{t}{\aut{\specname}}{\aut{\implname}}$, if $\reclang{\aut{\specname}} \circ t = \reclang{\aut{\implname}} \circ t$.
Note that we removed the condition of finiteness from our generalized test suite definition. Similarly, we will drop the assumption of finiteness in the definitions of the generalizations of state cover and characterization set. We choose to do this in order to make the theory simpler. In practice, all these components will still be finitary in some sense: for instance, a finite-dimensional vector space or an orbit-finite nominal set.

\begin{definition}
\begin{enum}
\item A \emph{fault domain} is a collection of automata (in $\cat{C}$).

\item A test suite $t \colon T \to \words$ is \emph{complete} for an automaton $\aut{\specname}$ with respect to a fault domain $\fdom{U}$ if for all automata $\aut{\implname} \in \fdom{U}$, $\agreets{t}{\aut{\specname}}{\aut{\implname}}$ implies $\autequiv{\aut{\specname}}{\aut{\implname}}$.
\end{enum}
\end{definition}

We now define the generalization of state covers. Before we do so, we introduce an auxiliary notion.
\begin{definition}
We say that a morphism $a \colon A \to \words$ \emph{contains the empty word} if the morphism $\minj{0} \colon \tunit \to \words$ factors through $a$, i.e. there exists a morphism $i_A \colon \tunit \to A$ such that $\minj{0} = a i_A$.
\end{definition}
\begin{definition}
Let $\aut{\specname} = (\specname, \initmor{\specname}, \transmor{\specname}, \finmor{\specname})$ be an automaton. A \emph{state cover} for $\aut{\specname}$ is a morphism $p \colon P \to \words$ containing the empty word such that the composite $P \xto{p} \words \xto{\malg{\aut{\specname}}} \specname$ is a split epi.
\end{definition}
Intuitively, a state cover for $\aut{\specname}$ is a collection of words such that we can effectively assign an access sequence to each state in $\aut{\specname}$. 

Unfortunately, being a state cover is a relatively strong condition which cannot be satisfied in some of our examples of interest, such as ordered automata (cf. \autoref{rem:state-cov-ord-aut}). It turns out that in order to prove the generalization of \autoref{lem:agreets-equiv-dfa}, a weaker albeit more complicated condition suffices.
\begin{definition}
Let $\aut{\specname} = (\specname, \initmor{\specname}, \transmor{\specname}, \finmor{\specname})$ be an automaton. A \emph{weak state cover} for $\aut{\specname}$ is a morphism $p \colon P \to \words$ containing the empty word together with a map $\delta_P \colon \tprod{P}{\abcobj} \to P$ such that the following diagram commutes.
\[\begin{tikzcd}
	{\tprod{P}{\abcobj}} & {\tprod{\words}{\abcobj}} & \words \\
	P & \words & \specname
	\arrow["{\tprod{p}{\id{\abcobj}}}", from=1-1, to=1-2]
	\arrow["{\delta_P}"', from=1-1, to=2-1]
	\arrow["\snoc", from=1-2, to=1-3]
	\arrow["{\malg{\aut{\specname}}}", from=1-3, to=2-3]
	\arrow["p"', from=2-1, to=2-2]
	\arrow["{\malg{\aut{\specname}}}"', from=2-2, to=2-3]
\end{tikzcd}\]
\end{definition}
Intuitively, a weak state cover assigns to each word $w$ in $P$ and symbol $a$ in $\abcobj$ a word $v = \delta_P(w, a)$ in $P$ such that we reach the same state upon reading $wa$ and $v$. Thus, it provides a transition structure on a restricted collection of words $p \colon P \to \words$ that follows the transitions of the automaton $\aut{\specname}$.

\begin{remark} \label{rem:state-cov-ord-aut}
For motivation of why we need the weaker definition of state covers, consider the example of ordered automata~\cite{DBLP:conf/lata/KlimaP19}. These are automata in the category $\Pos$ of posets and monotone maps equipped with the Cartesian monoidal structure. In this example, we choose the alphabet $\abcobj$ to be a discrete finite poset. Hence, the object of words $\words$ is also a discrete poset. This implies that, apart from degenerate cases, a state cover $p \colon P \to \words$ cannot exist. For, suppose we have an ordered automaton $\aut{\specname}$ with state space $S$ and a morphism $p \colon P \to \words$. For $p$ to be a state cover, we need to find a right inverse $s \colon S \to P$ of $\malg{\aut{\specname}} \circ p$. Such a morphism must map related elements of $S$ to related elements of $P$. But if two elements of $P$ are related, then they are necessarily sent by $p$ to the same word due to the discrete order on $\words$. Hence, the composite $\malg{\aut{\specname}} \circ p \circ s$ sends every connected subset of $S$ (viewed as an undirected graph) to the same state. Thus, $s$ is a right inverse only if each connected component of $S$ has size $1$, i.e. if $S$ is a discrete poset.
\end{remark}

\begin{proposition} \label{prop:sc-is-weak-sc}
Let $\aut{\specname}$ be an automaton. Then every state cover for $\aut{\specname}$ is a weak state cover for $\aut{\specname}$.
\end{proposition}
\begin{proof}
Let $p \colon P \to \words$ be a state cover for $\aut{\specname} = (\specname, \initmor{\specname}, \transmor{\specname}, \finmor{\specname})$. Then $p$ contains the empty word. Since $\malg{\aut{\specname}} \circ p$ is a split epi, it has a right inverse $s \colon \specname \to P$. Defining $\delta_P$ as $s \circ \malg{\specname} \circ \snoc \circ (\tprod{p}{\id{\abcobj}})$ makes the required diagram commute. \qed
\end{proof}

Characterization sets are generalized as follows.
\begin{definition}
Let $\aut{\specname} = (\specname, \initmor{\specname}, \transmor{\specname}, \finmor{\specname})$ be an automaton. A \emph{characterization morphism} for $\aut{\specname}$ is a morphism $w \colon W \to \words$ containing the empty word such that for all $f, g \colon X \to \specname$, if $\inthom{w}{\id{\outobj}} \circ \mcoalg{\aut{\specname}} \circ f = \inthom{w}{\id{\outobj}} \circ \mcoalg{\aut{\specname}} \circ g \colon X \to \inthom{W}{\outobj}$, then $\mcoalg{\aut{\specname}} \circ f = \mcoalg{\aut{\specname}} \circ g \colon X \to \langs$.
\end{definition}
\begin{remark} \label{rem:char-mor-def}
For motivation of the definition of characterization morphisms, note that, given an automaton $\aut{\specname} = (\specname, \initmor{\specname}, \transmor{\specname}, \finmor{\specname})$, the categorical counterpart of the relation $\stequiv{}{}$ for $\aut{\specname}$ from \autoref{sec:overview} is the kernel pair $(e_1, e_2)$ of $\mcoalg{\aut{\specname}} \colon S \to \langs$, and that the counterpart of the relation $\stequiv[W]{}{}$ for a morphism $w \colon W \to \words$ is the kernel pair $(e_1^w, e_2^w)$ of $\inthom{w}{\id{\outobj}} \circ \mcoalg{\aut{\specname}} \colon S \to \inthom{W}{\outobj}$. Thus, the condition $\stequiv[W]{p}{q} \implies \stequiv{p}{q}$ can be expressed by saying that $\mprod{e_1^w}{e_2^w}$ factors through $\mprod{e_1}{e_2}$. This is equivalent to the definition given above.
\end{remark}

We now state our main theorem, which is a generalization of \autoref{lem:agreets-equiv-dfa}.
\begin{theorem} \label{thm:main}
Let $\aut{\specname}$ and $\aut{\implname}$ be two automata. Suppose $(c, \delta_C)$ is a weak state cover for $\aut{\implname}$ and $w$ is a characterization morphism for $\aut{\specname}$. Assume that $\aut{\specname}$ is minimal and let $t = c \cdot \wincl{1} \cdot w$. Then $\agreets{t}{\aut{\specname}}{\aut{\implname}}$ implies $\autequiv{\aut{\specname}}{\aut{\implname}}$.
\end{theorem}

To state completeness of the generalized test suite, we introduce an abstract fault domain, which is in a sense the most general fault domain with respect to which the test suite is complete.
\begin{definition} \label{def:gen-fdom}
For a morphism $c \colon C \to \words$, define the fault domain $\fdomgcomp{c}$ as
\[ \fdomgcomp{c} = \setof{\aut{\implname}}{\exists \delta_C.\ (c, \delta_C) \text{ is a weak state cover for $\aut{\implname}$}}. \]
\end{definition}
\begin{corollary}
Let $\aut{\specname}$ be a minimal automaton. Then for all $c \colon C \to \words$ and for all characterization morphisms $w \colon W \to \words$ for $\aut{\specname}$, the test suite $t = c \cdot \wincl{1} \cdot w$ is complete for $\aut{\specname}$ with respect to $\fdomgcomp{c}$.
\end{corollary}
\begin{proof}
Let $\aut{\implname} \in \fdomgcomp{c}$, and assume $\agreets{t}{\aut{\specname}}{\aut{\implname}}$. Then there exists a $\delta_C$ such that $(c, \delta_C)$ is a weak state cover for $\aut{\implname}$. Thus, $\autequiv{\aut{\specname}}{\aut{\implname}}$ follows by \autoref{thm:main}. \qed
\end{proof}

We now turn to the generalized W-method.
\begin{definition} \label{def:gen-W-ts}
Let $p \colon P \to \words$ and $w \colon W \to \words$ be two morphisms, and let $k \in \N$. Then the \emph{W test suite} of order $k$ associated to $p$ and $w$ is defined as the morphism $\mW{k}{p}{w} = p \cdot \wincl{k+1} \cdot w$.
\end{definition}

We state completeness of the generalized W-method with respect to a fault domain which is a special case of the one given in \autoref{def:gen-fdom}, namely, $\fdomgcomp[k]{p} = \fdomgcomp{p \cdot \wincl{k}}$. Completeness with respect to this fault domain in the case of Mealy machines has already been considered by Maarse~\cite{maarse2020}. The related notion of $k$-$A$-completeness for Mealy machines has been introduced by Vaandrager, Fiter\v{a}u-Bro\c{s}tean, and Melse~\cite{vaandrager2024}.
\begin{corollary} \label{cor:gen-W-comp}
Let $\aut{\specname}$ be a minimal automaton. Then for all $p \colon P \to \words$, for all characterization morphisms $w \colon W \to \words$ for $\aut{\specname}$, and for all $k \in \N$, the test suite $\mW{k}{p}{w}$ is complete for $\aut{\specname}$ with respect to $\fdomgcomp[k]{p}$.
\end{corollary}

\begin{remark} \label{rem:factorizations}
In order to simplify the theory, we have opted for defining test suites (as well as state covers and characterization morphisms) as arbitrary morphisms instead of subobjects. However, we note that our theory would also work out if we defined test suites to be $\absmonos$-morphisms for some factorization system $(\absepis, \absmonos)$ such that all morphisms in $\absepis$ are epimorphisms. To obtain the adjusted test suite, we simply need to take the $(\absepis, \absmonos)$-factorization of the original test suite. Whether we restrict test suites to $\absmonos$-morphisms or not does not influence the completeness of test suites due to the following proposition.
\begin{proposition}
Let $(\absepis, \absmonos)$ be a factorization system on $\cat{C}$ such that all morphisms in $\absepis$ are epimorphisms. Let $t \colon T \to \words$ be a test suite and let $T \xto{e} T' \xto{t'} \words$ be its $(\absepis, \absmonos)$-factorization. Then for all automata $\aut{\specname}$ and $\aut{\implname}$, $\agreets{t}{\aut{\specname}}{\aut{\implname}}$ if and only if $\agreets{t'}{\aut{\specname}}{\aut{\implname}}$.
\end{proposition}
\begin{proof}
Follows immediately from the definition of $\agreets{t}{\aut{\specname}}{\aut{\implname}}$ and the fact that $e \in \absepis$ is an epimorphism. \qed
\end{proof}
\begin{corollary} \label{cor:fact-comp}
Let $t$ and $t'$ be defined as in the previous proposition. Then for all automata $\aut{\specname}$ and for all fault domains $\fdom{U}$, $t$ is complete for $\aut{\specname}$ with respect to $\fdom{U}$ if and only if $t'$ is complete for $\aut{\specname}$ with respect to $\fdom{U}$.
\end{corollary}

In view of \autoref{cor:fact-comp}, we may replace the original test suite $t$ with the factorized test suite $t'$ when dealing with completeness. We will use this fact when considering instances of the general framework in the next section.
\end{remark}

%% file: applications.tex
In this section, we use our general framework to derive complete test suites for various kinds of automata, including weighted automata and deterministic nominal automata. Details for the examples can be found in \appendixref{sec:applications-details}.

\subsection{DFAs, Moore Machines, and Mealy Machines}

DFAs are covered by our framework: the categorical definitions of \autoref{sec:wla} and \autoref{sec:test-suites} specialize to the familiar ones in \autoref{sec:overview} if we take as $\cat{C}$ the category $\Set$ of sets and functions with the cartesian product as monoidal structure and if we set $\outobj = \2$. A difference is that, as mentioned in \autoref{sec:test-suites}, the specialized notions of test suite, state cover, and characterization morphism are functions into $\words$ rather than mere subsets. We recover the original definitions by taking the images of the functions (cf. \autoref{rem:factorizations}). We also recover the W-method and \autoref{lem:agreets-equiv-dfa}.

Moore machines are essentially the same as DFAs, except that the set of final states $F \subseteq Q$ is generalized to an output function $f \colon Q \to \outobj$ for some output set $\outobj$. In the abstract framework, this amounts to changing the output object from $\2$ to $\outobj$. Also in this case, we recover the W-method and an analogue of \autoref{lem:agreets-equiv-dfa}. As an example, take the Moore machine $\aut{\specname}$ with states and transitions identical to those of the DFA in \autoref{fig:ex-dfa}, and with the output function $f \colon Q \to \N \cup \{-1\}$ that assigns to each state the amount of coins stored in the coffee machine in that state: $f(q_0) = 0$, $f(q_1) = 1$, $f(q_2) = 2$, and $f(q_3) = -1$, where $-1$ is used as a value indicating an error. Then $P = \{\epsilon, c, 1, 11\}$ is still a state cover for $\aut{\specname}$, $W = \{\epsilon, c, 1\}$ is still a characterization set for $\aut{\specname}$, and the test suite $\W{0}{P}{W}$ is still $4$-complete for $\aut{\specname}$. A difference in this example is that the input sequence $1$ also distinguishes $q_0$ and $q_1$. Hence, we could also use the alternative and smaller characterization set $W' = \{\epsilon, 1\}$. This results in the smaller $4$-complete test suite $\W{0}{P}{W'} = P \cdot \words[\le 1] \cdot W'$.

Mealy machines are similar to Moore machines, but they assign an output to each transition instead of each state. Formally, this amounts to replacing the output morphism $f \colon Q \to \outobj$ by $\lambda \colon Q \times \abcobj \to \outobj$. Mealy machines are covered by our framework by considering the category $\Set$ of sets and functions with the cartesian monoidal structure and by taking the output object to be the set $\funcset{\abcobj}{\outobj}$. In particular, we recover~\cite[Lemma~3.8]{DBLP:conf/tacas/KrugerJR24}.

We mention a subtle difference between our presentation of the W-method and the usual one found in the literature. Traditionally, the output function of a Mealy machine is extended to words as a function $\lambda^* \colon Q \times \words \to \outobj^*$ that records all the outputs encountered during a run of the machine. In contrast, the function $\reclang{\aut{A}} \colon \words \to \funcset{\abcobj}{\outobj}$ for a Mealy machine $\aut{A}$ maps a word $w$ and an input symbol $a$ to the output of the \emph{last} transition during the run of the input word $wa$. Hence, the assumption $\agreets{T}{\aut{\specname}}{\aut{\implname}}$ only guarantees that the last transitions of $\aut{\specname}$ and $\aut{\implname}$ match on input words from $T$. We can recover the stronger guarantee that $\lambda^*(q_0, w) = \lambda^*(q_0', w)$ for all $w \in T$ by making sure that the test suite is prefix-closed (i.e. any prefix of any word in $T$ is also in $T$).

\subsection{Weighted Automata} \label{sec:applications-wa}

Weighted automata~\cite{DBLP:books/ems/21/DrosteK21} are generalizations of DFAs where a weight is assigned to each transition, expressing e.g. the cost or reliability of its execution. In this subsection, we fix a field $\fld$ and a finite set $\abcobj$ of input symbols.
\begin{definition}
A \emph{weighted automaton} (WA) is a tuple $(Q, s_0, \delta, f)$, where $Q$ is a finite set of \emph{states}, $s_0 \colon Q \to \fld$ is an \emph{input weight function}, $\delta \colon Q \times \abcobj \times Q \to \fld$ is a \emph{transition weight function}, and $f \colon Q \to \fld$ is an \emph{output weight function}.
\end{definition}

The transition function $\delta$ of a WA $(Q, s_0, \delta, f)$ can be conveniently represented as a $\abcobj$-indexed family of $\fld$-valued matrices $\wamatrix{\delta}{a} \in \funcset{Q \times Q}{\fld}\ (a \in \abcobj)$ defined as $\wamatrix{\delta}{a}(p, q) = \delta(q, a, p)$. Note that the column index specifies the source state and the row index the target state. Similarly, the input and output weight functions $s_0$ and $f$ can be thought of as $\fld$-valued column vectors of cardinality $Q$.

WAs are covered by our framework: letting $\abcobj'$ be the free vector space over $\abcobj$, WAs correspond to automata over $\abcobj'$ in the category $\Vect{\fld}$ of vector spaces over $\fld$ and linear maps equipped with the tensor product of vector spaces as monoidal structure, taking the output object $\outobj = \fld$. The corresponding automaton in $\Vect{\fld}$ has state space $\funcset{Q}{\fld}$. The correspondence is spelled out in \appendixref{sec:wa-equiv-char}. In the remainder of the section, we identify WAs and automata in $\Vect{\fld}$ via this correspondence.

We recall that WAs recognize \emph{weighted languages}, i.e. functions $\words \to \fld$. For a WA $\aut{A} = (Q, s_0, \delta, f)$ and a word $w \in \words$, define the matrix $\wamatrix{\delta}{w}$ by recursion on $w$: $\wamatrix{\delta}{\epsilon} = \unitmatrix$ (where $\unitmatrix$ denotes the unit matrix) and $\wamatrix{\delta}{wa} = \wamatrix{\delta}{a} \wamatrix{\delta}{w}$. The \emph{recognized language} $\reclang{\aut{A}}[s]$ of an input weight vector $s \in \funcset{Q}{\fld}$ is then defined as $\reclang{\aut{A}}[s](w) = \transp{f} \wamatrix{\delta}{w} s$, where $\transp{f}$ denotes the transpose of $f$. The \emph{recognized language} of $\aut{A}$ is $\reclang{\aut{A}} = \reclang{\aut{A}}[s_0]$. We note furthermore that the function $s \mapsto \reclang{\aut{A}}[s]$ is a linear map $\funcset{Q}{\fld} \to \funcset{\words}{\fld}$, which follows directly from the definition of $\reclang{\aut{A}}[s]$.

\begin{wrapfigure}{r}[0pt]{0.4\textwidth}
\vspace{-7mm}
\centering
\resizebox{120pt}{!}{
\begin{tikzpicture}[auto,accepting/.style=accepting by arrow]
    \node[state,initial,initial text={$1$}]  (0)              {$q_0$};
    \node[state,accepting,accepting text={$1$}] (1) [right=of 0] {$q_1$};
    \path[->] (0) edge [loop above] node {$a/1, b/1$} ()
                  edge              node {$b/1$}      (1)
              (1) edge [loop above] node {$a/2, b/2$} ();
\end{tikzpicture}
}
\caption{A WA $\aut{\specname}$ for computing the decimal value of a binary number}
\label{fig:ex-wa}
\vspace{-7mm}
\end{wrapfigure}
As an example, consider the WA $\aut{\specname}$ over the alphabet $\{a,b\}$ and field $\R$ depicted in \autoref{fig:ex-wa}. (This example is taken from~\cite{DBLP:books/ems/21/DrosteK21}.) The input, output, and transition weights not shown are equal to 0. This automaton recognizes the weighted language $L \colon \words \to \R$ that maps a word $w$ to the decimal value of $w$ if understood as a binary number with $a$ standing for the digit 0 and $b$ for the digit 1.

Let us instantiate the abstract framework to the case of WAs. We start by noting that if the vector spaces $V$ and $W$ have generating sets $B$ and $C$, respectively, then $\tprod{V}{W}$ is generated by $\setof{\tprod{a}{b}}{a \in A, b \in B} \cong B \times C$. Furthermore, if $V_i$ is a vector space with generating set $B_i$ for all $i \in I$, then the direct sum $\bigoplus_{i \in I}{V_i}$ is generated by $\bigcup_{i \in I}{\setof{\minj{i}(b)}{b \in B_i}} \cong \coprod_{i \in I}{B_i}$ (where $\coprod$ denotes the disjoint union of sets). Hence, the vector space $(\abcobj')^n$ (where $\abcobj'$ is the free vector space over $\abcobj$) is generated by the set $\words[n]$, $(\abcobj')^{\le k}$ by $\words[\le k]$, and $(\abcobj')^*$ by $\words$. Furthermore, the map $\wincl{k} \colon (\abcobj')^{\le k} \to (\abcobj')^*$ is the inclusion of the subspace $(\abcobj')^{\le k}$. Since a linear map $(\abcobj')^* \to \fld$ is completely determined by its values on the basis vectors, the categorical notion of language is in bijective correspondence with weighted languages. The following proposition states how various notions from \autoref{sec:wla} specialize to the case of WAs.
\begin{proposition} \label{prop:wa-defs-wla}
Let $\aut{A} = (Q, s_0, \delta, f)$ be a WA.
\begin{enum}
\item The morphism $\append \colon \tprod{(\abcobj')^*}{(\abcobj')^*} \to (\abcobj')^*$ corresponds to the bilinear map $(\abcobj')^* \times (\abcobj')^* \to (\abcobj')^*$ defined on basis elements by $(u, v) \mapsto u v$.

\item The morphism $\malg{\aut{A}} \colon (\abcobj')^* \to \funcset{Q}{\fld}$ is the linear extension of the map $\words \to \funcset{Q}{\fld}$ sending $w$ to $\wamatrix{\delta}{w} s_0$.

\item The recognized language of $\aut{A}$ in the categorical sense is the linear extension of the recognized language of $\aut{A}$.

\item The morphism $\mcoalg{\aut{A}} \colon \funcset{Q}{\fld} \to \Hom[\Vect{\fld}]{(\abcobj')^*}{\fld}$ sends an input weight vector $s \in \funcset{Q}{\fld}$ to the linear extension of the weighted language $\reclang{\aut{A}}[s]$.

\item The automaton $\aut{A}$ is minimal iff for all $s \in \funcset{Q}{\fld}$, $\reclang{\aut{A}}[s] = 0$ implies $s = 0$.
\end{enum}
\end{proposition}

We now turn to the discussion of test suites. For the remainder of this section, we fix a specification WA $\aut{\specname} = (Q, s_0, \delta, f)$. Test suites, state covers, and characterization morphisms specialize to linear maps with codomain $(\abcobj')^*$. We focus on the case where these morphisms are actually subspace inclusions. Hence, we say that a subspace \emph{is} a test suite, and we may refer to a certain subspace as a state cover or as a characterization \emph{space}, the meaning being that the corresponding subspace inclusion is a state cover or a characterization morphism. We concentrate on those subspaces that are generated by subsets of $\words$.

\begin{proposition} \label{prop:wa-agree-ts}
Let $\aut{\implname}$ be a WA, let $T \subseteq \words$, and let $t \colon \vspan{T} \to (\abcobj')^*$ be the subspace inclusion. Then $\agreets{t}{\aut{\specname}}{\aut{\implname}}$ if and only if $\reclang{\aut{\specname}}|_T = \reclang{\aut{\implname}}|_T$.
\end{proposition}
Thus, we obtain a notion of completeness of test suites for WAs: a test suite $T \subseteq \words$ is complete for $\aut{\specname}$ with respect to a fault domain $\fdom{U}$ if and only if $\reclang{\aut{\specname}}|_T = \reclang{\aut{\implname}}|_T$ implies $\reclang{\aut{\specname}} = \reclang{\aut{\implname}}$ for all $\aut{\implname} \in \fdom{U}$. This notion is completely analogous to that for DFAs.

The following proposition characterizes the key components of the generalized W-method in the case of WAs in terms of subsets of $\words$.
\begin{proposition} \label{prop:wa-sc-char}
\begin{enum}
\item Let $P \subseteq \words$. Then $\vspan{P}$ is a state cover for $\aut{\specname}$ if and only if $\epsilon \in P$ and $\funcset{Q}{\fld}$ is generated by $\setof{\wamatrix{\delta}{w} s_0}{w \in P}$.

\item Let $W \subseteq \words$. Then $\vspan{W}$ is a characterization space for $\aut{\specname}$ if and only if $\epsilon \in W$ and for all $s \in \funcset{Q}{\fld}$, $\reclang{\aut{\specname}}[s]|_W = 0$ implies $\reclang{\aut{\specname}}[s] = 0$.
\end{enum}
\end{proposition}
Motivated by the previous proposition, if $P \subseteq \words$ is such that $\vspan{P}$ is a state cover for $\aut{\specname}$, then we also call $P$ itself a state cover. Similarly, if $\vspan{W}$ is a characterization space, we call $W$ a characterization \emph{set}. Notice the similarity of the characterizations with the corresponding definitions in \autoref{sec:overview} for DFAs.

We are now ready to derive a specialized W-method for WAs. It turns out that, under the identification of subspaces of $(\abcobj')^*$ and subsets of $\words$, it coincides with the classical W-method (\autoref{def:W-method}). Furthermore, it is characterized as the image of the generalized W test suite from \autoref{def:gen-W-ts}.
\begin{lemma} \label{lem:wa-ts-image}
Let $P, W \subseteq \words$, let $k \in \N$, and let $p \colon \vspan{P} \to (\abcobj')^*$ and $w \colon \vspan{W} \to (\abcobj')^*$ denote the inclusions. Then $\vspan{\W{k}{P}{W}} = \im{\mW{k}{p}{w}}$.
\end{lemma}

To state completeness of the specialized W-method, we define an appropriate fault domain, which is a restriction of the fault domain $\fdomgcomp[k]{p}$.
\begin{definition}
For a set $P \subseteq \words$ and natural number $k$, define
\begin{align*}
\fdomwa{k}{P} = \setof{\aut{\implname}}{P \cdot \words[\le k] \text{ is a state cover for $\aut{\implname}$}}.
\end{align*}
\end{definition}
The following theorem asserts completeness of the W-method for WAs.
\begin{theorem} \label{thm:w-method-lwa}
Suppose $\aut{\specname}$ is minimal. Then for all $P \subseteq \words$, for all characterization sets $W \subseteq \words$ for $\aut{\specname}$, and for all $k \in \N$, the test suite $\W{k}{P}{W}$ (\autoref{def:W-method}) is complete for $\aut{\specname}$ with respect to $\fdomwa{k}{P}$.
\end{theorem}

Continuing our previous example, we derive a complete test suite for the WA $\aut{\specname}$ of \autoref{fig:ex-wa}. We have $\wamatrix{\delta}{\epsilon} s_0 = \wamatrix{\delta}{\epsilon} q_0 = q_0$ and $\wamatrix{\delta}{b} s_0 - \wamatrix{\delta}{\epsilon} s_0 = \wamatrix{\delta}{b} q_0 - \wamatrix{\delta}{\epsilon} q_0 = (q_0 + q_1) - q_0 = q_1$. Since $\{ q_0, q_1 \}$ is a basis for $\funcset{Q}{\fld}$, by \autoref{prop:wa-sc-char} (i), $P = \{\epsilon, b\}$ is a state cover for $\aut{\specname}$.

Next, we show that $W = \{\epsilon, b\}$ is a characterization set for $\aut{\specname}$. For this, we recall that $\reclang{\aut{\specname}}[q_0] = \reclang{\aut{\specname}}$ maps a word $w$ to the decimal value of $w$ understood as a binary number with $a$ standing for $0$ and $b$ for $1$. Furthermore, $\reclang{\aut{\specname}}[q_1]$ maps a word $w$ to $2^{|w|}$. Thus, if $s = k_0q_0 + k_1q_1$ such that $\reclang{\aut{\specname}}[s]|_W = 0$, then $\reclang{\aut{\specname}}[s](\epsilon) = k_0 \reclang{\aut{\specname}}[q_0](\epsilon) + k_1 \reclang{\aut{\specname}}[q_1](\epsilon) = k_1 = 0$. Moreover, we have $\reclang{\aut{\specname}}[s](b) = k_0 \reclang{\aut{\specname}}[q_0](b) + k_1 \reclang{\aut{\specname}}[q_1](b) = k_0 + 2k_1 = k_0 = 0$. Thus, $\reclang{\aut{\specname}}[s]|_W = 0$ implies $s = 0$. Hence, by \autoref{prop:wa-sc-char} (ii), $W = \{ \epsilon, b \}$ is a characterization set for $\aut{\specname}$.

By \autoref{thm:w-method-lwa}, we conclude that $\W{1}{P}{W} = P \cdot \words[\le 2] \cdot W$ is a complete test suite for $\aut{\specname}$ with respect to $\fdomwa{1}{P}$. See \appendixref{sec:ex-wa-impl} for an example of how a possible faulty implementation is rejected.

\subsection{Nominal Automata}

Nominal automata~\cite{DBLP:journals/corr/BojanczykKL14} are a model of computation over potentially infinite alphabets. They are based on the notion of \emph{nominal sets}, which are essentially sets with certain symmetries. The idea is that we demand that the transition function of a nominal automaton respect the symmetries of the state space, so that we can represent it via finite manners.

In this subsection, we fix a countable set $\atoms$ of \emph{atoms}. We denote by $\Perm{\atoms}$ the group of \emph{finite permutations} over $\atoms$ (i.e. bijections $\pi \colon \atoms \to \atoms$ such that $\pi(a) = a$ for all but finitely many $a \in \atoms$). For a $\Perm{\atoms}$-set $X$, we denote the action of $\pi$ on an element $x \in X$ by $\pi \cdot x$. We recall that an \emph{equivariant function} $f \colon X \to Y$ between $\Perm{\atoms}$-sets satisfies $f(\pi \cdot x) = \pi \cdot f(x)$, an \emph{equivariant subset} $A \subseteq X$ is such that $x \in A \implies \pi \cdot x \in A$, and an \emph{equivariant element} $a \in X$ is such that $\{a\}$ is an equivariant subset. For background on nominal sets in general, we refer to~\cite{pitts2013nominal}. Finally, we fix an orbit-finite nominal set $\abcobj$ acting as the alphabet.

\begin{definition}
A \emph{deterministic nominal automaton} (or DNA for short) is a tuple $(Q, q_0, \delta, F)$, where $Q$ is an orbit-finite nominal set of \emph{states}, $q_0 \in Q$ is an equivariant \emph{initial state}, $\delta \colon Q \times \Sigma \to Q$ is an equivariant \emph{transition map}, and $F \subseteq Q$ is an equivariant subset of \emph{final states}.
\end{definition}
For DNA $\aut{A} = (Q, q_0, \delta, F)$, we write $q \in \aut{A}$ for $q \in Q$.

DNAs are covered by our framework, since they correspond precisely to automata over $\abcobj$ in the category $\Nom$ of nominal sets and equivariant maps with the cartesian product as monoidal structure, taking the output object $\outobj = \2$, the discrete nominal set with 2 elements. To see this, note that, given a nominal set $Q$, an equivariant element $q_0 \in Q$ corresponds to an equivariant map $\1 \to Q$ (where \1 is the discrete nominal set with 1 element), and that an equivariant subset $F \subseteq Q$ corresponds to an equivariant map $Q \to \2$. We identify DNAs with automata in $\Nom$ via this correspondence.

DNAs recognize \emph{nominal languages}: these are equivariant subsets of $\words$, the nominal set of words over $\abcobj$ with the pointwise group action. The transition map $\delta$ of a nominal automaton $\aut{A} = (Q, q_0, \delta, F)$ can be extended to an equivariant map $\delta^* \colon Q \times \words \to Q$ just as in the case of DFAs. The \emph{accepted language} $\acclang{\aut{A}}[q]$ of a state $q \in Q$ is also defined analogously to DFAs as the set $\acclang{\aut{A}}[q] = \setof{w \in \words}{\delta^*(q, w) \in F}$, and the accepted language $\acclang{\aut{A}}$ of $\aut{A}$ is $\acclang{\aut{A}}[q_0]$.

\begin{wrapfigure}{r}[0pt]{0.5\textwidth}
\centering
\vspace{-7mm}
\begin{tikzpicture}[auto,initial text=]
    \node[state,initial]   (0)              {$q_0$};
    \node[state]           (1) [right=of 0] {$q_{1,a}$};
    \node[state,accepting] (2) [right=of 1] {$q_2$};
    \node[state]           (3) [below=of 2] {$q_3$};
    \path[->] (0) edge              node        {$a$}                      (1)
              (1) edge              node        {$a$}                      (2)
                  edge              node [swap] {$\atoms \setminus \{a\}$} (3)
              (2) edge              node        {$\atoms$}                 (3)
              (3) edge [loop below] node        {$\atoms$}                 ();
\end{tikzpicture}
\caption{An example DNA $\aut{\specname}$ accepting the language $L = \setof{aa}{a \in \atoms}$}
\label{fig:ex-dna}
\vspace{-5mm}
\end{wrapfigure}
As an example, consider the DNA $\aut{\specname}$ over the alphabet $\atoms$ of atoms (regarded as a nominal set with group action $\pi \cdot a = \pi(a)$) depicted in \autoref{fig:ex-dna}. (This example is taken from~\cite{moerman2019nominal}.) The automaton has state space $Q = \{q_0,q_2,q_3\} \cup \setof{q_{1,a}}{a \in \atoms}$; the infinite amount of states $\setof{q_{1,a}}{a \in \atoms}$ is depicted with one symbolic state $q_{1,a}$. The transitions $q_0 \xto{a} q_{1,a}$ and $q_{1,a} \xto{a} q_2$ are to be understood as one transition for each $a \in \atoms$. Furthermore, transitions of the form $p \xto{A} q$ for $A \subseteq \atoms$ stand for a number of transitions, one for each $a \in A$. The automaton $\aut{\specname}$ accepts the nominal language $L = \setof{aa}{a \in \atoms}$.

We now instantiate the abstract framework to DNAs. We first note that the categorical construction of the nominal set $\words$ coincides with the description given above. Thus, languages $\words \to \2$ in the categorical sense correspond precisely to nominal languages. Furthermore, the map $\wincl{k} \colon \words[\le k] \to \words$ is simply the inclusion of a subset. The map $\mu \colon \words \times \words \to \words$ is given by concatenation of words. For any DNA $\aut{A}$, the morphism $\malg{\aut{A}}$ maps a word $w \in \words$ to $\delta^*(q_0, w)$. Thus, the categorical notion of recognized language corresponds to the accepted language of $\aut{A}$ as defined above. Moreover, the morphism $\mcoalg{\aut{A}}$ maps a state $q \in \aut{A}$ to $\acclang{\aut{A}}[q]$. Finally, an automaton $\aut{A}$ is minimal (in the categorical sense) if for all $p, q \in \aut{A}$, $\acclang{\aut{A}}[p] = \acclang{\aut{A}}[q]$ implies $p = q$.

For the remainder of this section, we fix a specification DNA $\aut{\specname} = (Q, q_0, \delta, F)$. Similarly to the case of WAs, we focus on the case where test suites, state covers, and characterization morphisms are actually subset inclusions. A subset inclusion is equivariant if and only if the subset is equivariant. Thus, we say that an equivariant subset of $\words$ \emph{is} a test suite, state cover, or characterization \emph{set}. If $T \subseteq \words$ is a test suite, and $t \colon T \to \words$ is the corresponding inclusion, then $\agreets{t}{\aut{\specname}}{\aut{\implname}}$ if and only if $\acclang{\aut{\specname}} \cap T = \acclang{\aut{\implname}} \cap T$ for any DNA $\aut{\implname}$. Hence, the notion of completeness of test suites for DNAs coincides with that for DFAs. A set $W \subseteq \words$ is a characterization set for $\aut{\specname}$ if and only if $\epsilon \in W$ and for all $p, q \in Q$, $\acclang{\aut{\specname}}[p] \cap W = \acclang{\aut{\specname}}[q] \cap W$ implies $\acclang{\aut{\specname}}[p] = \acclang{\aut{\specname}}[q]$. Thus, characterization sets are exactly the same as for DFAs.

The specialized W-method for DNAs coincides with the W-method for DFAs. Furthermore, just as for WAs, it is characterized as the image of the generalized W test suite. Namely, if $P, W \subseteq \words$, $p \colon P \to \words$ and $w \colon W \to \words$ are the subset inclusions, and $k \in \N$, then $\W{k}{P}{W} = \im{\mW{k}{p}{w}}$.

The following theorem asserts the completeness of the W-method for DNAs, and it is an immediate consequence of \autoref{cor:gen-W-comp} and \autoref{cor:fact-comp}.
\begin{theorem}
Suppose $\aut{\specname}$ is minimal. Then for all $P \subseteq \words$, for all characterization sets $W \subseteq \words$ for $\aut{\specname}$, and for all $k \in \N$, the test suite $\W{k}{P}{W}$ is complete for $\aut{\specname}$ with respect to $\fdomgcomp[k]{p}$ (where $p \colon P \to \words$ is the inclusion).
\end{theorem}

Continuing the previous example, we derive a complete test suite for the DNA $\aut{\specname}$. We first show that $P = \{\epsilon\} \cup \setof{a}{a \in \atoms} \cup \setof{aa}{a \in \atoms} \cup \setof{aab}{a, b \in \atoms}$ is a weak state cover for $\aut{\specname}$. Define the map $\delta_P \colon P \times \abcobj \to P$ as
\begin{equation*}
\delta_P(w, c) = \begin{cases}
                     c   & \text{ if $w = \epsilon$} \\
                     aa  & \text{ if $w = a$ for $a \in \atoms$ and $a = c$} \\
                     aac & \text{ if $w = a$ for $a \in \atoms$ and $a \ne c$} \\
                     aac & \text{ if $w = aa$ for $a \in \atoms$} \\
                     aab & \text{ if $w = aab$ for $a,b \in \atoms$}
                 \end{cases}.
\end{equation*}
It can be checked that $\delta_P$ is equivariant. Furthermore, we have $\delta^*(q_0, \delta_P(w, c)) = \delta^*(q_0, wc)$, as can readily be verified.

Next, we note that $W = \{\epsilon\} \cup \setof{a}{a \in \atoms} \cup \setof{aa}{a \in \atoms}$ is a characterization set for $\aut{\specname}$. Hence, we obtain the complete test suite $\W{0}{P}{W} = P \cdot \atoms^{\le 1} \cdot W$. We remark that $\W{0}{P}{W}$ is an infinite but orbit-finite subset of $\words$.

%% file: conclusion.tex
We introduced a general framework for studying the W-method and proving completeness of test suites, at the categorical
level of automata in monoidal closed categories. Besides recovering basic instances such as DFAs, Mealy machines, and Moore machines,
our framework instantiates to complete test suites for the new cases of weighted automata and deterministic nominal automata.

There are several open questions and avenues for future work. We have focused on the classical W-method, but it would
be useful to also cover refinements thereof such as Wp~\cite{FujiwaraTesting1991}, where distinguishing sequences
depend on the state reached, or Hybrid ADS~\cite{moerman2019nominal}. The latter requires an understanding
of adaptive distinguishing sequences at an abstract level, which is of interest on its own.
A further direction is to include in our framework a notion of \emph{finiteness} of test suites using, e.g., finitely presentable objects~\cite{AR94}.
Moreover, a key aspect of the W-method is that the size of the state cover and characterization set
are polynomial in the size of the specification, and it would be useful to capture this
at a more general level. 

A possible direction for future work is to move from weighted automata over a field 
to weighted automata over a \emph{semiring}, and complement for instance the learning
algorithms for weighted automata over the integers proposed in~\cite{DBLP:journals/pacmpl/BunaMargineanCSW24,DBLP:conf/fossacs/HeerdtKR020}. Instantiating
to the Boolean semiring could then connect to more concrete work on complete test
suites for non-deterministic automata~\cite{DBLP:conf/hase/PetrenkoY14}.

Our results assume a state cover and a characterization set as input. We have not touched upon the topic on how to acquire these sets from the specification. It would be of practical relevance to develop algorithms for computing state covers and characterization sets for weighted automata and for deterministic nominal automata.

%% file: appendix.tex
\section{Categorical preliminaries} \label{sec:cat-prelims}

In this section, we recall some categorical notions used in this paper: algebras and coalgebras for a functor, and monoidal closed categories.

\begin{definition}
Let $F \colon \cat{C} \to \cat{C}$ be a functor.
\begin{enum}
\item An \emph{$F$-algebra} is a pair $(X, a)$, where $X \in \cat{C}$ and $a \colon FX \to X$. A \emph{homomorphism} of $F$-algebras $(X, a)$ and $(Y, b)$ is a map $h \colon X \to Y$ such that the following diagram commutes.
\[\begin{tikzcd}
	FX & FY \\
	X & Y
	\arrow["Fh", from=1-1, to=1-2]
	\arrow["a"', from=1-1, to=2-1]
	\arrow["b", from=1-2, to=2-2]
	\arrow["h"', from=2-1, to=2-2]
\end{tikzcd}\]

\item Dually, an \emph{$F$-coalgebra} is a pair $(X, c)$, where $X \in \cat{C}$ and $c \colon X \to FX$. A \emph{homomorphism} of $F$-coalgebras $(X, c)$ and $(Y, d)$ is a map $h \colon X \to Y$ such that the following diagram commutes.
\[\begin{tikzcd}
	X & Y \\
	FX & FY
	\arrow["h", from=1-1, to=1-2]
	\arrow["c"', from=1-1, to=2-1]
	\arrow["d", from=1-2, to=2-2]
	\arrow["Fh"', from=2-1, to=2-2]
\end{tikzcd}\]
\end{enum}
\end{definition}

For a functor $F$, the collection of $F$-algebras and their homomorphisms form a category $\Alg{F}$. An initial object in this category is called an \emph{initial algebra}. Similarly, there is a category $\Coalg{F}$ of $F$-coalgebras and their homomorphisms. A final object in $\Coalg{F}$ is called a \emph{final coalgebra}.

\begin{definition}
A \emph{monoidal category} is a category $\cat{C}$ equipped with a bifunctor $\tprod{-}{-} \colon \cat{C} \times \cat{C} \to \cat{C}$ called the \emph{tensor product}, a specified object $\tunit \in \cat{C}$ called the \emph{tensor unit}, a natural isomorphism $\assoc{X}{Y}{Z} \colon \tprod{(\tprod{X}{Y})}{Z} \xto{\cong} \tprod{X}{(\tprod{Y}{Z})}$ called the \emph{associator}, a natural isomorphism $\lunit{X} \colon \tprod{\tunit}{X} \xto{\cong} X$ called the \emph{left unitor}, and a natural isomorphism $\runit{X} \colon \tprod{X}{\tunit} \xto{\cong} X$ called the \emph{right unitor}, such that the following two diagrams (called the \emph{pentagon identity} and the \emph{triangle identity}, respectively) commute.
\[\begin{tikzcd}
	& {\tprod{(\tprod{W}{X})}{(\tprod{Y}{Z})}} \\
	{\tprod{(\tprod{(\tprod{W}{X})}{Y})}{Z}} && {\tprod{W}{(\tprod{X}{(\tprod{Y}{Z})})}} \\
	{\tprod{(\tprod{W}{(\tprod{X}{Y})})}{Z}} && {\tprod{W}{(\tprod{(\tprod{X}{Y})}{Z})}}
	\arrow["{\assoc{W}{X}{\tprod{Y}{Z}}}", from=1-2, to=2-3]
	\arrow["{\assoc{\tprod{W}{X}}{Y}{Z}}", from=2-1, to=1-2]
	\arrow["{\tprod{\assoc{W}{X}{Y}}{\id{Z}}}"', from=2-1, to=3-1]
	\arrow["{\assoc{W}{\tprod{X}{Y}}{Z}}"', from=3-1, to=3-3]
	\arrow["{\tprod{\id{W}}{\assoc{X}{Y}{Z}}}"', from=3-3, to=2-3]
\end{tikzcd}\]
\[\begin{tikzcd}
	{\tprod{(\tprod{X}{\tunit})}{Y}} && {\tprod{X}{(\tprod{\tunit}{Y})}} \\
	& {\tprod{X}{Y}}
	\arrow["{\assoc{X}{\tunit}{Y}}", from=1-1, to=1-3]
	\arrow["{\tprod{\runit{X}}{\id{Y}}}"', from=1-1, to=2-2]
	\arrow["{\tprod{\id{X}}{\lunit{Y}}}", from=1-3, to=2-2]
\end{tikzcd}\]
\end{definition}

\begin{definition}
A monoidal category $(\cat{C}, \tprod{-}{-}, \tunit, \assoc*{}{}{}, \lunit*{}, \runit*{})$ is \emph{closed} if, for every object $X \in \cat{C}$, the functor $\tprod{-}{X} \colon \cat{C} \to \cat{C}$ has a right adjoint $\inthom{X}{-} \colon \cat{C} \to \cat{C}$. This right adjoint is referred to as the \emph{internal hom}.
\end{definition}
Given a monoidal closed category $\cat{C}$, the internal hom extends to a functor $\inthom{-}{-} \colon \op{\cat{C}} \times \cat{C} \to \cat{C}$ defined on morphisms $f \colon X' \to X$ and $g \colon Y \to Y'$ by $\inthom{f}{g} = \adj{X'}{g \circ \ev{X}{Y} \circ (\tprod{\id{\inthom{X}{Y}}}{f})}$.

\section{Proofs for \autoref{sec:wla}}

We first show some lemmas that are used in later proofs.
\begin{lemma} \label{lem:consn-iso}
The morphism $\cons[n] \colon \tprod{\abcobj}{\words[n]} \to \words[n+1]$ is an isomorphism.
\end{lemma}
\begin{proof}
We prove the statement by induction on $n$. We have that $\cons[0] = \inv{\lunit{\abcobj}} \runit{\abcobj}$ is an isomorphism since it is a composite of isomorphisms. Moreover, $\cons[n+1] = (\tprod{\cons[n]}{\id{\abcobj}}) \inv{\assoc{\abcobj}{\words[n]}{\abcobj}}$ is an isomorphism since $\cons[n]$ is an isomorphism by the induction hypothesis, and functors and composition preserve isomorphisms. \qed
\end{proof}

\begin{lemma} \label{lem:snoc-cons-append-comp}
The diagrams
\[\begin{tikzcd}
	{\tprod{\abcobj}{\words[n]}} & {\words[n+1]} & {\tprod{\words[n]}{\abcobj}} & {\words[n+1]} \\
	{\tprod{\abcobj}{\words}} & \words & {\tprod{\words}{\abcobj}} & \words \\
	& {\tprod{\words[m]}{\words[n]}} & {\words[m+n]} \\
	& {\tprod{\words}{\words}} & \words
	\arrow["{\cons[n]}", from=1-1, to=1-2]
	\arrow["{\tprod{\id{\abcobj}}{\minj{n}}}"', from=1-1, to=2-1]
	\arrow["{\minj{n+1}}", from=1-2, to=2-2]
	\arrow[Rightarrow, no head, from=1-3, to=1-4]
	\arrow["{\tprod{\minj{n}}{\id{\abcobj}}}"', from=1-3, to=2-3]
	\arrow["{\minj{n+1}}", from=1-4, to=2-4]
	\arrow["\cons"', from=2-1, to=2-2]
	\arrow["\snoc"', from=2-3, to=2-4]
	\arrow["{\append[m,n]}", from=3-2, to=3-3]
	\arrow["{\tprod{\minj{m}}{\minj{n}}}"', from=3-2, to=4-2]
	\arrow["{\minj{m+n}}", from=3-3, to=4-3]
	\arrow["\append"', from=4-2, to=4-3]
\end{tikzcd}\]
commute.
\end{lemma}
\begin{proof}
The first diagram is the outer square of the following commuting diagram.
\[\begin{tikzcd}
	& {\tprod{\abcobj}{\words[n]}} && {\words[n+1]} \\
	\\
	{\tprod{\abcobj}{\words}} & {\sum_{n \in \N}{\tprod{\abcobj}{\words[n]}}} && {\sum_{n \in \N}{\words[n+1]}} && \words
	\arrow["{\cons[n]}", from=1-2, to=1-4]
	\arrow["{\tprod{\id{\abcobj}}{\minj{n}}}"', from=1-2, to=3-1]
	\arrow["{\minj{n}}"', from=1-2, to=3-2]
	\arrow["{\minj{n}}"', from=1-4, to=3-4]
	\arrow["{\minj{n+1}}", from=1-4, to=3-6]
	\arrow["\cong", from=3-1, to=3-2]
	\arrow["\cons"', curve={height=24pt}, from=3-1, to=3-6]
	\arrow["{\sum_{n \in \N}{\cons[n]}}", from=3-2, to=3-4]
	\arrow["{\mfcoprod{n \in \N}{\minj{n+1}}}", from=3-4, to=3-6]
\end{tikzcd}\]
Similarly, the second diagram is the outer square of the following diagram.
\[\begin{tikzcd}
	& {\tprod{\words[n]}{\abcobj}} & {\words[n+1]} \\
	\\
	{\tprod{\words}{\abcobj}} & {\sum_{n \in \N}{\tprod{\words[n]}{\abcobj}}} & {\sum_{n \in \N}{\words[n+1]}} && \words
	\arrow[Rightarrow, no head, from=1-2, to=1-3]
	\arrow["{\tprod{\minj{n}}{\id{\abcobj}}}"', from=1-2, to=3-1]
	\arrow["{\minj{n}}"', from=1-2, to=3-2]
	\arrow["{\minj{n}}"', from=1-3, to=3-3]
	\arrow["{\minj{n+1}}", from=1-3, to=3-5]
	\arrow["\cong", from=3-1, to=3-2]
	\arrow["\snoc"', curve={height=24pt}, from=3-1, to=3-5]
	\arrow[Rightarrow, no head, from=3-2, to=3-3]
	\arrow["{\mfcoprod{n \in \N}{\minj{n+1}}}", from=3-3, to=3-5]
\end{tikzcd}\]
Finally, the third diagram is the outer square of the following diagram.
\[\begin{tikzcd}[column sep = 2.3em, row sep = 2.7em]
	& {\tprod{\words[m]}{\words[n]}} && {\words[m+n]} \\
	\\
	{\tprod{\words}{\words}} & {\sum_{m,n \in \N}{\tprod{\words[m]}{\words[n]}}} && {\sum_{m,n \in \N}{\words[m+n]}} && \words
	\arrow["{\append[m,n]}", from=1-2, to=1-4]
	\arrow["{\tprod{\minj{m}}{\minj{n}}}"', from=1-2, to=3-1]
	\arrow["{\minj{m,n}}"', from=1-2, to=3-2]
	\arrow["{\minj{m,n}}"', from=1-4, to=3-4]
	\arrow["{\minj{m+n}}", from=1-4, to=3-6]
	\arrow["\cong", from=3-1, to=3-2]
	\arrow["\append"', curve={height=24pt}, from=3-1, to=3-6]
	\arrow["{\sum_{m,n \in \N}{\append[m,n]}}", from=3-2, to=3-4]
	\arrow["{\mfcoprod{m,n \in \N}{\minj{m+n}}}", from=3-4, to=3-6]
\end{tikzcd}\]
\qed
\end{proof}

\begin{lemma} \label{lem:append-snoc}
The following diagram commutes.
\[\begin{tikzcd}
	{\tprod{\words}{\abcobj}} && {\tprod{\words}{\words}} \\
	&& \words
	\arrow["{\tprod{\id{\words}}{\minj{1} \inv{\lunit{\abcobj}}}}", from=1-1, to=1-3]
	\arrow["\snoc"', from=1-1, to=2-3]
	\arrow["\append", from=1-3, to=2-3]
\end{tikzcd}\]
\end{lemma}
\begin{proof}
Since $\mfcoprod{n \in \N}{\tprod{\minj{n}}{\id{\abcobj}}} \colon \sum_{n \in \N}{\tprod{\words[n]}{\abcobj}} \to \tprod{\words}{\abcobj}$ is an isomorphism, it suffices to show
\[ \append \circ (\tprod{\id{\words}}{\minj{1} \inv{\lunit{\abcobj}}}) \circ \mfcoprod{n \in \N}{\tprod{\minj{n}}{\id{\abcobj}}} = \snoc \circ \mfcoprod{n \in \N}{\tprod{\minj{n}}{\id{\abcobj}}}. \]
This holds if and only if
\[ \append \circ (\tprod{\id{\words}}{\minj{1} \inv{\lunit{\abcobj}}}) \circ (\tprod{\minj{n}}{\id{\abcobj}}) = \snoc \circ (\tprod{\minj{n}}{\id{\abcobj}}) \]
for all $n \in \N$. But
\begin{align*}
\append \circ (\tprod{\id{\words}&}{\minj{1} \inv{\lunit{\abcobj}}}) \circ (\tprod{\minj{n}}{\id{\abcobj}})
	 = \append \circ (\tprod{\minj{n}}{\minj{1}}) \circ (\tprod{\id{\words[n]}}{\inv{\lunit{\abcobj}}}) \\
	&= \minj{n+1} \circ \append[n,1] \circ (\tprod{\id{\words[n]}}{\inv{\lunit{\abcobj}}}) \\
	&= \minj{n+1} \circ (\tprod{\runit{\words[n]}}{\id{\abcobj}}) \circ \inv{\assoc{\words[n]}{\tunit}{\abcobj}} \circ (\tprod{\id{\words[n]}}{\inv{\lunit{\abcobj}}}) \\
	&= \minj{n+1} \circ \id{\tprod{\words[n]}{\abcobj}} = \minj{n+1} = \snoc \circ (\tprod{\minj{n}}{\id{\abcobj}})
\end{align*}
by \autoref{lem:snoc-cons-append-comp} and the definition of $\append[n,1]$. \qed
\end{proof}

\begin{lemma} \label{lem:append-zero}
For all $n \in \N$, we have $\append[0,n] = \lunit{\words[n]} \colon \tprod{\tunit}{\words[n]} \to \words[n]$.
\end{lemma}
\begin{proof}
We prove the claim by induction on $n$.
\begin{items}
\item Case $0$: we have
\[ \append[0,0] = \runit{\tunit} = \lunit{\tunit}. \]

\item Case $n+1$: we have
\[ \append[0,n+1] = (\tprod{\append[0,n]}{\id{\abcobj}}) \circ \inv{\assoc{\tunit}{\words[n]}{\abcobj}} = (\tprod{\lunit{\words[n]}}{\id{\abcobj}}) \circ \inv{\assoc{\tunit}{\words[n]}{\abcobj}} = \lunit{\tprod{\words[n]}{\abcobj}} \]
using the induction hypothesis. \qed
\end{items}
\end{proof}

\subsection{Proof of \autoref{prop:words-init-alg}}

Let $(A, \alg[A])$ be an $\algfunc$-algebra. Write $\alg[A] = \mcoprod{i_A}{\delta_A}$ for some $i_A \colon \tunit \to A$ and $\delta_A \colon \tprod{A}{\abcobj} \to A$. We construct an $\algfunc$-algebra homomorphism $\malg{\alg[A]} \colon \words \to A$ and show its uniqueness. We define the morphism $\malg{\alg[A]}$ in stages, by first defining morphisms $\malg{\alg[A],n} \colon \abcobj^n \to Q$ by recursion on $n$.
\begin{align*}
\malg{\alg[A],0} &= \tunit \xto{i_A} A \\
\malg{\alg[A],n+1} &= \tprod{\abcobj^n}{\abcobj} \xto{\tprod{\malg{\alg[A],n}}{\id{\abcobj}}} \tprod{A}{\abcobj} \xto{\delta_A} A
\end{align*}
Then, $\malg{\alg[A]}$ is defined as the cotuple $\mfcoprod{n \in \N}{\malg{\alg[A],n}}$. 

We need to show that $\malg{\alg[A]}$ is an $\algfunc$-algebra homomorphism, i.e. that the following diagram commutes.
\[\begin{tikzcd}
	{\tunit + \tprod{\words}{\abcobj}} && {\tunit + \tprod{A}{\abcobj}} \\
	\words && A
	\arrow["{\id{\tunit} + \tprod{\malg{\alg[A]}}{\id{\abcobj}}}", from=1-1, to=1-3]
	\arrow["{\mcoprod{\minj{0}}{\snoc}}"', from=1-1, to=2-1]
	\arrow["{\mcoprod{i_A}{\delta_A}}", from=1-3, to=2-3]
	\arrow["{\malg{\alg[A]}}"', from=2-1, to=2-3]
\end{tikzcd}\]
The commutativity of this diagram is equivalent to the commutativity of the following two diagrams.
\[\begin{tikzcd}
	\tunit && {\tprod{\words}{\abcobj}} & {\tprod{A}{\abcobj}} \\
	\words & A & \words & A
	\arrow["{\minj{0}}"', from=1-1, to=2-1]
	\arrow["{i_A}", from=1-1, to=2-2]
	\arrow["{\tprod{\malg{\alg[A]}}{\id{\abcobj}}}", from=1-3, to=1-4]
	\arrow["\snoc"', from=1-3, to=2-3]
	\arrow["{\delta_A}", from=1-4, to=2-4]
	\arrow["{\malg{\alg[A]}}"', from=2-1, to=2-2]
	\arrow["{\malg{\alg[A]}}"', from=2-3, to=2-4]
\end{tikzcd}\]
The first diagram commutes by the definition of $\malg{\alg[A]}$ and $\malg{\alg[A],0}$. We precompose the second diagram by the isomorphism $\sum_{n \in \N}{\tprod{\words[n]}{\abcobj}} \xto{\cong} \tprod{\words}{\abcobj}$. The two composites are now equal if and only if the morphisms obtained by precomposing them with $\minj{n}$ are equal for all $n \in \N$. Thus, we need to check commutativity of the following diagram.
\[\begin{tikzcd}[column sep = 3em]
	{\tprod{\words[n]}{\abcobj}} & {\tprod{\words}{\abcobj}} & {\tprod{A}{\abcobj}} \\
	{\tprod{\words}{\abcobj}} & \words & A
	\arrow["{\tprod{\minj{n}}{\id{\abcobj}}}", from=1-1, to=1-2]
	\arrow["{\tprod{\minj{n}}{\id{\abcobj}}}"', from=1-1, to=2-1]
	\arrow["{\tprod{\malg{\alg[A]}}{\id{\abcobj}}}", from=1-2, to=1-3]
	\arrow["{\delta_A}", from=1-3, to=2-3]
	\arrow["\snoc"', from=2-1, to=2-2]
	\arrow["{\malg{\alg[A]}}"', from=2-2, to=2-3]
\end{tikzcd}\]
This follows from the commuting diagram below, applying \autoref{lem:snoc-cons-append-comp} and the definition of $\malg{\alg[A],n+1}$.
\[\begin{tikzcd}
	&& {\tprod{\words}{\abcobj}} \\
	& {\tprod{\words[n]}{\abcobj}} && {\tprod{A}{\abcobj}} \\
	{\tprod{\words}{\abcobj}} & \words && A
	\arrow["{\tprod{\malg{\alg[A]}}{\id{\abcobj}}}", from=1-3, to=2-4]
	\arrow["{\tprod{\minj{n}}{\id{\abcobj}}}", from=2-2, to=1-3]
	\arrow["{\tprod{\malg{\alg[A],n}}{\id{\abcobj}}}", from=2-2, to=2-4]
	\arrow["{\tprod{\minj{n}}{\id{\abcobj}}}"', from=2-2, to=3-1]
	\arrow["{\minj{n+1}}"'{pos=0.7}, from=2-2, to=3-2]
	\arrow["{\malg{\alg[A],n+1}}", from=2-2, to=3-4]
	\arrow["{\delta_A}", from=2-4, to=3-4]
	\arrow["\snoc"', from=3-1, to=3-2]
	\arrow["{\malg{\alg[A]}}"', from=3-2, to=3-4]
\end{tikzcd}\]

Finally, we need to show uniqueness of $\malg{\alg[A]}$. Let $h \colon \words \to A$ be an $\algfunc$-algebra homomorphism from $[\minj{0}, \snoc]$ to $\alg[A]$. Then, by similar reasoning as above (using the universal property of the coproduct and using the isomorphism $\sum_{n \in \N}{\tprod{\words[n]}{\abcobj}} \xto{\cong} \tprod{\words}{\abcobj}$), the condition of being an $\algfunc$-algebra homomorphism is equivalent to the commutativity of the following diagrams.
\[\begin{tikzcd}
	&&& {\tprod{\words}{\abcobj}} \\
	I && {\tprod{\words[n]}{\abcobj}} && {\tprod{A}{\abcobj}} \\
	\words & A & {\tprod{\words}{\abcobj}} & \words & A
	\arrow["{\tprod{h}{\id{\abcobj}}}", from=1-4, to=2-5]
	\arrow["{\minj{0}}"', from=2-1, to=3-1]
	\arrow["{i_A}", from=2-1, to=3-2]
	\arrow["{\tprod{\minj{n}}{\id{\abcobj}}}", from=2-3, to=1-4]
	\arrow["{\tprod{h\minj{n}}{\id{\abcobj}}}", from=2-3, to=2-5]
	\arrow["{\tprod{\minj{n}}{\id{\abcobj}}}"', from=2-3, to=3-3]
	\arrow["{\minj{n+1}}", from=2-3, to=3-4]
	\arrow["{\delta_A}", from=2-5, to=3-5]
	\arrow["h"', from=3-1, to=3-2]
	\arrow["\snoc"', from=3-3, to=3-4]
	\arrow["h"', from=3-4, to=3-5]
\end{tikzcd}\]

To prove $h = \malg{\alg[A]}$, it suffices to prove $h\minj{n} = \malg{\alg[A],n}$ for all $n \in \N$. We do this by induction on $n$. We have
\[ h\minj{0} = i_A = \malg{\alg[A],0} \]
by the first diagram and the definition of $\malg{\alg[A],0}$, and
\[ h\minj{n+1} = \delta_A(\tprod{h\minj{n}}{\id{\abcobj}}) = \delta_A(\tprod{\malg{\alg[A],n}}{\id{\abcobj}}) = \malg{\alg[A],n+1} \]
by the second diagram, the induction hypothesis, and the definition of $\malg{\alg[A],n+1}$. \qed

\subsection{Proof of \autoref{prop:lang-fin-coalg}} \label{sec:proof-lang-fin-coalg}

Let $(C, \coalg[C])$ be a $\coalgfunc$-coalgebra. Write $\coalg[C] = \mprod{f_C}{\overline{\delta_C}}$ for some $f \colon C \to \outobj$ and $\overline{\delta_C} \colon C \to \inthom{\abcobj}{C}$. Furthermore, write $\delta_C = \adj*{\abcobj}{\overline{\delta_C}} \colon \tprod{C}{\abcobj} \to C$. We construct a morphism $\mcoalg{\coalg[C]} \colon C \to \langs$ as the adjunct of $\mcoalg{\coalg[C]}' \colon \tprod{C}{\words} \to \outobj$ defined as the composite
\[ \tprod{C}{\words} \xto{\cong} \sum_{n \in \N}{\tprod{C}{\words[n]}} \xto{\mfcoprod{n \in \N}{\mcoalg{\coalg[C],n}}} \outobj, \]
where $\mcoalg{\coalg[C],n} \colon \tprod{C}{\words[n]} \to \outobj$ is defined by recursion on $n$ using \autoref{lem:consn-iso}.
\begin{align*}
\mcoalg{\coalg[C],0} &= \tprod{C}{\tunit} \xto{\runit{C}} C \xto{f_C} \outobj \\
\mcoalg{\coalg[C],n+1} &= \tprod{C}{\words[n+1]} \xto{\tprod{\id{C}}{\inv{\cons[n]}}} \tprod{C}{(\tprod{\abcobj}{\words[n]})} \to \\
  &\hspace{6em}\xto{\inv{\assoc{C}{\abcobj}{\words[n]}}} \tprod{(\tprod{C}{\abcobj})}{\words[n]} \xto{\tprod{\delta_C}{\id{\words[n]}}} \tprod{C}{\words[n]} \xto{\mcoalg{\coalg[C],n}} \outobj
\end{align*}

We check that $\mcoalg{\coalg[C]}$ is a $\coalgfunc$-coalgebra homomorphism, i.e. that the following diagram commutes.
\[\begin{tikzcd}
	QC && \langs \\
	{\outobj \times \inthom{\abcobj}{C}} && {\outobj \times \inthom{\abcobj}{\langs}}
	\arrow["{\mcoalg{\coalg[C]}}", from=1-1, to=1-3]
	\arrow["{\mprod{f_Q}{\overline{\delta_Q}}}"', from=1-1, to=2-1]
	\arrow["{\mprod{\evemp}{\langderiv}}", from=1-3, to=2-3]
	\arrow["{\id{\outobj} \times \inthom{\id{\abcobj}}{\mcoalg{\coalg[C]}}}"', from=2-1, to=2-3]
\end{tikzcd}\]
The commutativity of the diagram above is equivalent to the commutativity of the following two diagrams.
\[\begin{tikzcd}
	C && \langs & C && \langs \\
	&& \outobj & {\inthom{\abcobj}{C}} && {\inthom{\abcobj}{\langs}}
	\arrow["{\mcoalg{\coalg[C]}}", from=1-1, to=1-3]
	\arrow["{f_C}"', from=1-1, to=2-3]
	\arrow["\evemp", from=1-3, to=2-3]
	\arrow["{\mcoalg{\coalg[C]}}", from=1-4, to=1-6]
	\arrow["{\overline{\delta_C}}"', from=1-4, to=2-4]
	\arrow["\langderiv", from=1-6, to=2-6]
	\arrow["{\inthom{\id{\abcobj}}{\mcoalg{\coalg[C]}}}"', from=2-4, to=2-6]
\end{tikzcd}\]
The first diagram commutes since it is the outline of the following commutative diagram.
\[\begin{tikzcd}
	& C & \langs \\
	& {\tprod{C}{\tunit}} & {\tprod{\langs}{\tunit}} \\
	& {\tprod{C}{\words}} & {\tprod{\langs}{\words}} \\
	{\tprod{C}{\tunit}} & {\sum_{n \in \N}{\tprod{C}{\words[n]}}} & \outobj \\
	& C
	\arrow["{\mcoalg{\coalg[C]}}", from=1-2, to=1-3]
	\arrow["{\inv{\runit{C}}}", from=1-2, to=2-2]
	\arrow["{\inv{\runit{C}}}"', from=1-2, to=4-1]
	\arrow["{\inv{\runit{\langs}}}", from=1-3, to=2-3]
	\arrow["{\tprod{\mcoalg{\coalg[C]}}{\id{\tunit}}}", from=2-2, to=2-3]
	\arrow["{\tprod{\id{C}}{\minj{0}}}"', from=2-2, to=3-2]
	\arrow["{\tprod{\id{\langs}}{\minj{0}}}", from=2-3, to=3-3]
	\arrow["{\tprod{\mcoalg{\coalg[C]}}{\id{\words}}}", from=3-2, to=3-3]
	\arrow["\cong"', from=3-2, to=4-2]
	\arrow["{\ev{\words}{\outobj}}", from=3-3, to=4-3]
	\arrow["{\tprod{\id{C}}{\minj{0}}}"{description}, from=4-1, to=3-2]
	\arrow["{\minj{0}}"', from=4-1, to=4-2]
	\arrow["{\runit{C}}"', from=4-1, to=5-2]
	\arrow["{\mfcoprod{n \in \N}{\mcoalg{\coalg[C],n}}}", from=4-2, to=4-3]
	\arrow["{f_C}"', from=5-2, to=4-3]
\end{tikzcd}\]
Taking the adjunct of the second diagram twice, we need to show commutativity of the diagram
\[\begin{tikzcd}
	{\tprod{(\tprod{C}{\abcobj})}{\words}} &&& {\tprod{(\tprod{\langs}{\abcobj})}{\words}} \\
	{\tprod{C}{\words}} &&& \outobj
	\arrow["{\tprod{(\tprod{\mcoalg{\coalg[C]}}{\id{\abcobj}})}{\id{\words}}}", from=1-1, to=1-4]
	\arrow["{\tprod{\delta_C}{\id{\words}}}"', from=1-1, to=2-1]
	\arrow["{d'}", from=1-4, to=2-4]
	\arrow["{\mcoalg{\coalg[C]}'}"', from=2-1, to=2-4]
\end{tikzcd}\]
where $d'$ is defined in \autoref{def:evemp-langderiv}. This diagram is the outline of the following diagram.
\[\begin{tikzcd}
	{\tprod{(\tprod{C}{\abcobj})}{\words}} &&& {\tprod{(\tprod{\langs}{\abcobj})}{\words}} \\
	& {\tprod{C}{(\tprod{\abcobj}{\words})}} && {\tprod{ \langs}{(\tprod{\abcobj}{\words})}} \\
	& {\tprod{C}{\words}} && {\tprod{\langs}{\words}} \\
	{\tprod{C}{\words}} &&& \outobj
	\arrow["{\tprod{(\tprod{\mcoalg{\coalg[C]}}{\id{\abcobj}})}{\id{\words}}}", from=1-1, to=1-4]
	\arrow["{\assoc{C}{\abcobj}{\words}}"', from=1-1, to=2-2]
	\arrow["{\tprod{\delta_C}{\id{\words}}}"', from=1-1, to=4-1]
	\arrow["{\assoc{\langs}{\abcobj}{\words}}", from=1-4, to=2-4]
	\arrow["{\tprod{\mcoalg{\coalg[C]}}{\id{\tprod{\abcobj}{\words}}}}", from=2-2, to=2-4]
	\arrow["{\tprod{\id{C}}{\cons}}"', from=2-2, to=3-2]
	\arrow["{\tprod{\id{\langs}}{\cons}}", from=2-4, to=3-4]
	\arrow["{\tprod{\mcoalg{\coalg[C]}}{\id{\words}}}", from=3-2, to=3-4]
	\arrow["{\mcoalg{\coalg[C]}'}"'{pos=0.3}, from=3-2, to=4-4]
	\arrow["{\ev{\words}{\outobj}}", from=3-4, to=4-4]
	\arrow["{\mcoalg{\coalg[C]}'}"', from=4-1, to=4-4]
\end{tikzcd}\]
We still need to show the commutativity of the leftmost pentagon. Since we have an isomorphism $\mfcoprod{n \in \N}{\tprod{\id{\tprod{C}{\abcobj}}}{\minj{n}}} \colon \sum_{n \in \N}{\tprod{(\tprod{C}{\abcobj})}{\words[n]}} \xto{\cong} \tprod{(\tprod{C}{\abcobj})}{\words}$, it suffices to show that the two sides of the pentagon become equal after precomposing with this isomorphism. This condition is equivalent to the commutativity of the following diagram for all $n \in \N$.
\[\hspace{-0.1cm}\begin{tikzcd}
	{\tprod{(\tprod{C}{\abcobj})}{\words[n]}} && {\tprod{(\tprod{C}{\abcobj})}{\words}} & {\tprod{C}{(\tprod{\abcobj}{\words})}} & {\tprod{C}{\words}} \\
	{\tprod{(\tprod{C}{\abcobj})}{\words}} \\
	{\tprod{C}{\words}} &&&& \outobj
	\arrow["{\tprod{\id{\tprod{C}{\abcobj}}}{\minj{n}}}", from=1-1, to=1-3]
	\arrow["{\tprod{\id{\tprod{C}{\abcobj}}}{\minj{n}}}"', from=1-1, to=2-1]
	\arrow["{\assoc{C}{\abcobj}{\words}}", from=1-3, to=1-4]
	\arrow["{\tprod{\id{C}}{\cons}}", from=1-4, to=1-5]
	\arrow["{\mcoalg{\coalg[C]}'}", from=1-5, to=3-5]
	\arrow["{\tprod{\delta_C}{\id{\words}}}"', from=2-1, to=3-1]
	\arrow["{\mcoalg{\coalg[C]}'}"', from=3-1, to=3-5]
\end{tikzcd}\]
This follows from the commuting diagram below, where we used \autoref{lem:snoc-cons-append-comp}, \autoref{lem:consn-iso}, and the definition of $\mcoalg{\coalg[C],n+1}$.
\[\hspace{-1.5cm}\begin{tikzcd}
	{\tprod{(\tprod{C}{\abcobj})}{\words}} && {\tprod{C}{(\tprod{\abcobj}{\words})}} && {\tprod{C}{\words}} \\
	& {\tprod{(\tprod{C}{\abcobj})}{\words[n]}} & {\tprod{C}{(\tprod{\abcobj}{\words[n]})}} & {\tprod{C}{\words[n+1]}} & {\sum_{n \in \N}{\tprod{C}{\words[n]}}} \\
	& {\tprod{C}{\words[n]}} \\
	{\tprod{C}{\words}} & {\sum_{n \in \N}{\tprod{C}{\words[n]}}} &&& \outobj
	\arrow["{\assoc{C}{\abcobj}{\words}}", from=1-1, to=1-3]
	\arrow["{\tprod{\delta_C}{\id{\words}}}"', from=1-1, to=4-1]
	\arrow["{\tprod{\id{C}}{\cons}}", from=1-3, to=1-5]
	\arrow["\cong", from=1-5, to=2-5]
	\arrow["{\tprod{\id{\tprod{C}{\abcobj}}}{\minj{n}}}"{description}, from=2-2, to=1-1]
	\arrow["{\assoc{C}{\abcobj}{\words[n]}}", shift left, from=2-2, to=2-3]
	\arrow["{\tprod{\delta_C}{\id{\words[n]}}}"', from=2-2, to=3-2]
	\arrow["{\tprod{\id{C}}{(\tprod{\id{\abcobj}}{\minj{n}})}}", from=2-3, to=1-3]
	\arrow["{\inv{\assoc{C}{\abcobj}{\words[n]}}}", shift left, from=2-3, to=2-2]
	\arrow["{\tprod{\id{C}}{\cons[n]}}", shift left, from=2-3, to=2-4]
	\arrow["{\tprod{\id{C}}{\minj{n+1}}}"{description}, from=2-4, to=1-5]
	\arrow["{\tprod{\id{C}}{\inv{\cons[n]}}}", shift left, from=2-4, to=2-3]
	\arrow["{\minj{n+1}}"', from=2-4, to=2-5]
	\arrow["{\mcoalg{\coalg[C],n+1}}"', from=2-4, to=4-5]
	\arrow["{\mfcoprod{n \in \N}{\mcoalg{\coalg[C],n}}}", from=2-5, to=4-5]
	\arrow["{\tprod{\id{C}}{\minj{n}}}"', from=3-2, to=4-1]
	\arrow["{\minj{n}}"', from=3-2, to=4-2]
	\arrow["{\mcoalg{\coalg[C],n}}", from=3-2, to=4-5]
	\arrow["\cong"', from=4-1, to=4-2]
	\arrow["{\mfcoprod{n \in \N}{\mcoalg{\coalg[C],n}}}"', from=4-2, to=4-5]
\end{tikzcd}\]

Finally, we need to show uniqueness of $\mcoalg{\coalg[C]}$. For this, let $h \colon C \to \langs$ be a $\coalgfunc$-coalgebra homomorphism from $\coalg[C]$ to $\mprod{\evemp}{\langderiv}$. Then, by applying similar reasoning as above (using the universal property of the product and taking adjuncts), the condition of being a $\coalgfunc$-coalgebra morphism is equivalent to the commutativity of the following two diagrams.
\begin{equation} \label{eq:fin-coalg-1}
\begin{tikzcd}
	& C & \langs \\
	& {\tprod{C}{\tunit}} & {\tprod{\langs}{\tunit}} \\
	& {\tprod{C}{\words}} & {\tprod{\langs}{\words}} \\
	C && \outobj
	\arrow["h", from=1-2, to=1-3]
	\arrow["{\inv{\runit{C}}}", from=1-2, to=2-2]
	\arrow[Rightarrow, no head, from=1-2, to=4-1]
	\arrow["{\inv{\runit{\langs}}}", from=1-3, to=2-3]
	\arrow["{\tprod{h}{\id{\tunit}}}", from=2-2, to=2-3]
	\arrow["{\tprod{\id{C}}{\minj{0}}}", from=2-2, to=3-2]
	\arrow["{\tprod{\id{\langs}}{\minj{0}}}", from=2-3, to=3-3]
	\arrow["{\tprod{h}{\id{\words}}}", from=3-2, to=3-3]
	\arrow["{\adj*{\words}{h}}"'{pos=0.2}, from=3-2, to=4-3]
	\arrow["{\ev{\words}{\outobj}}", from=3-3, to=4-3]
	\arrow["{f_C}"', from=4-1, to=4-3]
\end{tikzcd}
\end{equation}
\[\begin{tikzcd}
	{\tprod{(\tprod{C}{\abcobj})}{\words}} &&& {\tprod{(\tprod{\langs}{\abcobj})}{\words}} \\
	& {\tprod{C}{(\tprod{\abcobj}{\words})}} && {\tprod{\langs}{(\tprod{\abcobj}{\words})}} \\
	& {\tprod{C}{\words}} && {\tprod{\langs}{\words}} \\
	{\tprod{C}{\words}} &&& \outobj
	\arrow["{\tprod{(\tprod{h}{\id{\abcobj}})}{\id{\words}}}", from=1-1, to=1-4]
	\arrow["{\assoc{C}{\abcobj}{\words}}"', from=1-1, to=2-2]
	\arrow["{\tprod{\delta_C}{\id{\words}}}"', from=1-1, to=4-1]
	\arrow["{\assoc{\langs}{\abcobj}{\words}}", from=1-4, to=2-4]
	\arrow["{\tprod{h}{\id{\tprod{\abcobj}{\words}}}}", from=2-2, to=2-4]
	\arrow["{\tprod{\id{C}}{\cons}}"', from=2-2, to=3-2]
	\arrow["{\tprod{\id{\langs}}{\cons}}", from=2-4, to=3-4]
	\arrow["{\tprod{h}{\id{\words}}}", from=3-2, to=3-4]
	\arrow["{\adj*{\words}{h}}"'{pos=0.3}, from=3-2, to=4-4]
	\arrow["{\ev{\words}{\outobj}}", from=3-4, to=4-4]
	\arrow["{\adj*{\words}{h}}"', from=4-1, to=4-4]
\end{tikzcd}\]
Precomposing the left pentagon in the second diagram with $\tprod{\id{\tprod{C}{\abcobj}}}{\minj{n}}$ and applying \autoref{lem:snoc-cons-append-comp}, we get the following commuting diagram.
\begin{equation} \label{eq:fin-coalg-2}
\hspace{-1cm}\begin{tikzcd}
	{\tprod{(\tprod{C}{\abcobj})}{\words}} && {\tprod{C}{(\tprod{\abcobj}{\words})}} && {\tprod{C}{\words}} \\
	& {\tprod{(\tprod{C}{\abcobj})}{\words[n]}} & {\tprod{C}{(\tprod{\abcobj}{\words[n]})}} & {\tprod{C}{\words[n+1]}} \\
	& {\tprod{C}{\words[n]}} \\
	{\tprod{C}{\words}} &&&& \outobj
	\arrow["{\assoc{C}{\abcobj}{\words}}", from=1-1, to=1-3]
	\arrow["{\tprod{\delta_C}{\id{\words}}}"', from=1-1, to=4-1]
	\arrow["{\tprod{\id{C}}{\cons}}", from=1-3, to=1-5]
	\arrow["{\adj*{\words}{h}}", from=1-5, to=4-5]
	\arrow["{\tprod{\id{\tprod{C}{\abcobj}}}{\minj{n}}}"{description}, from=2-2, to=1-1]
	\arrow["{\assoc{C}{\abcobj}{\words[n]}}"', from=2-2, to=2-3]
	\arrow["{\tprod{\delta_C}{\id{\words[n]}}}"', from=2-2, to=3-2]
	\arrow["{\tprod{\id{C}}{(\tprod{\id{\abcobj}}{\minj{n}})}}"', from=2-3, to=1-3]
	\arrow["{\tprod{\id{C}}{\cons[n]}}"', from=2-3, to=2-4]
	\arrow["{\tprod{\id{C}}{\minj{n+1}}}"{description}, from=2-4, to=1-5]
	\arrow["{\tprod{\id{C}}{\minj{n}}}"', from=3-2, to=4-1]
	\arrow["{\adj*{\words}{h}}"', from=4-1, to=4-5]
\end{tikzcd}
\end{equation}
To show $h = \mcoalg{\coalg[C]}$, it suffices to show $\adj*{\words}{h} = \mcoalg{\coalg[C]}'$. Since $\mfcoprod{n \in \N}{\tprod{\id{C}}{\minj{n}}}$ is an isomorphism, this is equivalent to showing $\adj*{\words}{h} \mfcoprod{n \in \N}{\tprod{\id{C}}{\minj{n}}} = \mfcoprod{n \in \N}{\mcoalg{\coalg[C],n}}$. Thus, we need to show $\adj*{\words}{h} (\tprod{\id{C}}{\minj{n}}) = \mcoalg{\coalg[C],n}$ for all $n \in \N$. We do this by induction on $n$.
\begin{items}
\item Case $0$: we need to prove $\adj*{\words}{h} (\tprod{\id{C}}{\minj{0}}) = \mcoalg{\coalg[C],0} = f_C \runit{C}$. Since $\runit{C}$ is an isomorphism, this is equivalent to $\adj*{\words}{h} (\tprod{\id{C}}{\minj{0}}) \inv{\runit{C}} = f_C$. This follows from diagram~\eqref{eq:fin-coalg-1}.

\item Case $n+1$: we need to prove
\[ \adj*{\words}{h} (\tprod{\id{C}}{\minj{n+1}}) = \mcoalg{\coalg[C],n+1} = \mcoalg{\coalg[C],n} (\tprod{\delta_C}{\id{\words[n]}}) \inv{\assoc{C}{\abcobj}{\words[n]}} (\tprod{\id{C}}{\inv{\cons[n]}}). \]
This is equivalent to
\[ \adj*{\words}{h} (\tprod{\id{C}}{\minj{n+1}}) (\tprod{\id{C}}{\cons[n]}) \assoc{C}{\abcobj}{\words[n]} = \mcoalg{\coalg[C],n} (\tprod{\delta_C}{\id{\words[n]}}). \]
This follows from diagram~\eqref{eq:fin-coalg-2} and the induction hypothesis. \qed
\end{items}

\subsection{Proof of \autoref{prop:rec-lang-equiv}}

We start by defining the generalized transition function of an automaton.
\begin{definition}
Let $\aut{A} = (Q, \initmor{Q}, \transmor{Q}, \finmor{Q})$ be an automaton.
\begin{enum}
\item We define a family or morphisms $\transmor*{Q,n} \colon \tprod{Q}{\words[n]} \to Q$ by recursion on $n$.
\begin{align*}
\transmor*{Q,0} &= \tprod{Q}{\tunit} \xto{\runit{Q}} Q \\
\transmor*{Q,n+1} &= \tprod{Q}{(\tprod{\words[n]}{\abcobj})} \xto{\inv{\assoc{Q}{\words[n]}{\abcobj}}} \tprod{(\tprod{Q}{\words[n]})}{\abcobj} \xto{\tprod{\transmor*{Q,n}}{\id{\abcobj}}} \tprod{Q}{\abcobj} \xto{\transmor{Q}} Q
\end{align*}

\item We define the morphism $\transmor*{Q} \colon \tprod{Q}{\words} \to Q$ as the composite
\[ \tprod{Q}{\words} \xto{\cong} \sum_{n \in \N}{\tprod{Q}{\words[n]}} \xto{\mfcoprod{n \in \N}{\transmor*{Q,n}}} Q. \]
\end{enum}
\end{definition}

\begin{lemma} \label{lem:transmor-mcoalg-comp}
Let $\aut{A} = (Q, \initmor{Q}, \transmor{Q}, \finmor{Q})$ be an automaton. Then the diagrams
\[\begin{tikzcd}
	{\tprod{Q}{\words[n]}} && {\tprod{Q}{\words[n]}} \\
	{\tprod{Q}{\words}} & Q & {\tprod{Q}{\words}} & \outobj
	\arrow["{\tprod{\id{Q}}{\minj{n}}}"', from=1-1, to=2-1]
	\arrow["{\transmor*{Q,n}}", from=1-1, to=2-2]
	\arrow["{\tprod{\id{Q}}{\minj{n}}}"', from=1-3, to=2-3]
	\arrow["{\mcoalg{\aut{A},n}}", from=1-3, to=2-4]
	\arrow["{\transmor*{Q}}"', from=2-1, to=2-2]
	\arrow["{\mcoalg{\aut{A}}'}"', from=2-3, to=2-4]
\end{tikzcd}\]
commute (see~\autoref{sec:proof-lang-fin-coalg} for the definition of $\mcoalg{\aut{A}}'$).
\end{lemma}
\begin{proof}
The first equation follows from the following commutative diagram.
\[\begin{tikzcd}
	& {\tprod{Q}{\words[n]}} \\
	{\tprod{Q}{\words}} & {\sum_{n \in \N}{\tprod{Q}{\words[n]}}} && Q
	\arrow["{\tprod{\id{Q}}{\minj{n}}}"', from=1-2, to=2-1]
	\arrow["{\minj{n}}"', from=1-2, to=2-2]
	\arrow["{\transmor*{Q,n}}", from=1-2, to=2-4]
	\arrow["\cong", from=2-1, to=2-2]
	\arrow["{\transmor*{Q}}"', curve={height=18pt}, from=2-1, to=2-4]
	\arrow["{\mfcoprod{n \in \N}{\transmor*{Q}}}"{pos=0.3}, from=2-2, to=2-4]
\end{tikzcd}\]
Similarly, the second one follows from the following commuting diagram.
\[\begin{tikzcd}
	& {\tprod{Q}{\words[n]}} \\
	{\tprod{Q}{\words}} & {\sum_{n \in \N}{\tprod{Q}{\words[n]}}} && \outobj
	\arrow["{\tprod{\id{Q}}{\minj{n}}}"', from=1-2, to=2-1]
	\arrow["{\minj{n}}"', from=1-2, to=2-2]
	\arrow["{\mcoalg{\aut{A},n}}", from=1-2, to=2-4]
	\arrow["\cong", from=2-1, to=2-2]
	\arrow["{\mcoalg{\aut{A}}'}"', curve={height=18pt}, from=2-1, to=2-4]
	\arrow["{\mfcoprod{n \in \N}{\mcoalg{\aut{A},n}}}"{pos=0.3}, from=2-2, to=2-4]
\end{tikzcd}\]
\qed
\end{proof}

\begin{proposition} \label{prop:transmor-cons}
For any automaton $\aut{A} = (Q, \initmor{Q}, \transmor{Q}, \finmor{Q})$, the following diagram commutes.
\[\begin{tikzcd}
	{\tprod{Q}{(\tprod{\abcobj}{\words[n]})}} && {\tprod{Q}{\words[n+1]}} \\
	{\tprod{(\tprod{Q}{\abcobj})}{\words[n]}} \\
	{\tprod{Q}{\words[n]}} && Q
	\arrow["{\tprod{\id{Q}}{\cons[n]}}", from=1-1, to=1-3]
	\arrow["{\inv{\assoc{Q}{\abcobj}{\words[n]}}}"', from=1-1, to=2-1]
	\arrow["{\transmor*{Q,n+1}}", from=1-3, to=3-3]
	\arrow["{\tprod{\transmor{Q}}{\id{\words[n]}}}"', from=2-1, to=3-1]
	\arrow["{\transmor*{Q,n}}"', from=3-1, to=3-3]
\end{tikzcd}\]
\end{proposition}
\begin{proof}
We prove the claim by induction on $n$. The base case follows from the commuting diagram below.
\[\begin{tikzcd}
	{\tprod{Q}{(\tprod{\abcobj}{\tunit})}} & {\tprod{Q}{\abcobj}} & {\tprod{Q}{(\tprod{\tunit}{\abcobj})}} \\
	&& {\tprod{(\tprod{Q}{\tunit})}{\abcobj}} \\
	{\tprod{(\tprod{Q}{\abcobj})}{\tunit}} && {\tprod{Q}{\abcobj}} \\
	{\tprod{Q}{\tunit}} && Q
	\arrow["{\tprod{\id{Q}}{\runit{\abcobj}}}", from=1-1, to=1-2]
	\arrow["{\inv{\assoc{Q}{\abcobj}{\tunit}}}"', shift right, from=1-1, to=3-1]
	\arrow["{\tprod{\id{Q}}{\runit{\abcobj}}}"', from=1-1, to=3-3]
	\arrow["{\tprod{\id{Q}}{\inv{\lunit{\abcobj}}}}", from=1-2, to=1-3]
	\arrow["{\tprod{\inv{\runit{Q}}}{\id{\abcobj}}}"', from=1-2, to=2-3]
	\arrow["{\inv{\assoc{Q}{\tunit}{\abcobj}}}", from=1-3, to=2-3]
	\arrow["{\tprod{\runit{Q}}{\id{\abcobj}}}", from=2-3, to=3-3]
	\arrow["{\assoc{Q}{\abcobj}{\tunit}}"', shift right, from=3-1, to=1-1]
	\arrow["{\runit{\tprod{Q}{\abcobj}}}"', from=3-1, to=3-3]
	\arrow["{\tprod{\transmor{Q}}{\id{\tunit}}}"', from=3-1, to=4-1]
	\arrow["{\transmor{Q}}", from=3-3, to=4-3]
	\arrow["{\runit{Q}}"', from=4-1, to=4-3]
\end{tikzcd}\]
The inductive case follows from the commutativity of the diagram
\[\hspace{-1.5cm}\begin{tikzcd}
	{\tprod{Q}{(\tprod{\abcobj}{(\tprod{\words[n]}{\abcobj})})}} && {\tprod{Q}{(\tprod{(\tprod{\abcobj}{\words[n]})}{\abcobj})}} && {\tprod{Q}{(\tprod{\words[n+1]}{\abcobj})}} \\
	&& {\tprod{(\tprod{Q}{(\tprod{\abcobj}{\words[n]})})}{\abcobj}} && {\tprod{(\tprod{Q}{\words[n+1]})}{\abcobj}} \\
	{\tprod{(\tprod{Q}{\abcobj})}{(\tprod{\words[n]}{\abcobj})}} && {\tprod{(\tprod{(\tprod{Q}{\abcobj})}{\words[n]})}{\abcobj}} && {\tprod{Q}{\abcobj}} \\
	{\tprod{Q}{(\tprod{\words[n]}{\abcobj})}} && {\tprod{(\tprod{Q}{\words[n]})}{\abcobj}} & {\tprod{Q}{\abcobj}} & Q
	\arrow["{\tprod{\id{Q}}{\inv{\assoc{\abcobj}{\words[n]}{\abcobj}}}}", from=1-1, to=1-3]
	\arrow["{\inv{\assoc{Q}{\abcobj}{\tprod{\words[n]}{\abcobj}}}}"', from=1-1, to=3-1]
	\arrow["{\tprod{\id{Q}}{(\tprod{\cons[n]}{\id{\abcobj}})}}", from=1-3, to=1-5]
	\arrow["{\inv{\assoc{Q}{\tprod{\abcobj}{\words[n]}}{\abcobj}}}"', from=1-3, to=2-3]
	\arrow["{\inv{\assoc{Q}{\words[n+1]}{\abcobj}}}", from=1-5, to=2-5]
	\arrow["{\tprod{(\tprod{\id{Q}}{\cons[n]})}{\id{\abcobj}}}", from=2-3, to=2-5]
	\arrow["{\tprod{\inv{\assoc{Q}{\abcobj}{\words[n]}}}{\id{\abcobj}}}"', from=2-3, to=3-3]
	\arrow["{\tprod{\transmor*{Q,n+1}}{\id{\abcobj}}}", from=2-5, to=3-5]
	\arrow["{\inv{\assoc{\tprod{Q}{\abcobj}}{\words[n]}{\abcobj}}}", from=3-1, to=3-3]
	\arrow["{\tprod{\transmor{Q}}{\id{\tprod{\words[n]}{\abcobj}}}}"', from=3-1, to=4-1]
	\arrow["{\tprod{(\tprod{\transmor{Q}}{\id{\words[n]}})}{\id{\abcobj}}}"', from=3-3, to=4-3]
	\arrow["{\transmor{Q}}", from=3-5, to=4-5]
	\arrow["{\inv{\assoc{Q}{\words[n]}{\abcobj}}}"', from=4-1, to=4-3]
	\arrow["{\tprod{\transmor*{Q,n}}{\id{\abcobj}}}"', from=4-3, to=4-4]
	\arrow[Rightarrow, no head, from=4-4, to=3-5]
	\arrow["{\transmor{Q}}"', from=4-4, to=4-5]
\end{tikzcd}\]
where we applied the functor $\tprod{-}{\abcobj}$ to the induction hypothesis. \qed
\end{proof}

The following proposition provides a link between $\malg{\aut{A}}$, $\mcoalg{\aut{A}}$, and $\transmor*{Q}$.
\begin{proposition} \label{prop:malg-mcoalg-transmor}
Let $\aut{A} = (Q, \initmor{Q}, \transmor{Q}, \finmor{Q})$ be an automaton.
\begin{enum}
\item The morphism $\malg{\aut{A}}$ is equal to the composite
\[ \words \xto{\inv{\lunit{\words}}} \tprod{\tunit}{\words} \xto{\tprod{\initmor{Q}}{\id{\words}}} \tprod{Q}{\words} \xto{\transmor*{Q}} Q. \]
\item The morphism $\mcoalg{\aut{A}}$ is the adjunct of the composite
\[ \tprod{Q}{\words} \xto{\transmor*{Q}} Q \xto{\finmor{Q}} \outobj. \]
\end{enum}
\end{proposition}
\begin{proof}
\begin{enum}
\item The desired equality holds if and only if the two morphisms precomposed with $\minj{n}$ are equal for all $n \in \N$. This follows from the commutativity of the following diagram.
\[\begin{tikzcd}
	{\words[n]} &&& Q \\
	\words & {\tprod{\tunit}{\words[n]}} & {\tprod{Q}{\words[n]}} \\
	{\tprod{\tunit}{\words}} &&& {\tprod{Q}{\words}}
	\arrow["{\malg{\aut{A},n}}", from=1-1, to=1-4]
	\arrow["{\minj{n}}"', from=1-1, to=2-1]
	\arrow["{\inv{\lunit{\words[n]}}}"'{pos=0.6}, from=1-1, to=2-2]
	\arrow["{\inv{\lunit{\words}}}"', from=2-1, to=3-1]
	\arrow["{\tprod{\initmor{Q}}{\id{\words[n]}}}", from=2-2, to=2-3]
	\arrow["{\tprod{\id{\tunit}}{\minj{n}}}"{pos=0.3}, from=2-2, to=3-1]
	\arrow["{\transmor*{Q,n}}"'{pos=0.4}, from=2-3, to=1-4]
	\arrow["{\tprod{\id{Q}}{\minj{n}}}"'{pos=0.3}, from=2-3, to=3-4]
	\arrow["{\tprod{\initmor{Q}}{\id{\words}}}"', from=3-1, to=3-4]
	\arrow["{\transmor*{Q}}"', from=3-4, to=1-4]
\end{tikzcd}\]
where we used \autoref{lem:transmor-mcoalg-comp}. We still need to show the commutativity of the upper trapezoid. We do this by induction on $n$.
\begin{items}
\item Case $0$: we have
\[ \transmor*{Q,0} (\tprod{\initmor{Q}}{\id{\tunit}}) \inv{\lunit{\tunit}} =
    \runit{Q} (\tprod{\initmor{Q}}{\id{\tunit}}) \inv{\runit{\tunit}} =
    \initmor{Q} \runit{\tunit} \inv{\runit{\tunit}} = \initmor{Q} = \malg{\aut{A},0}. \]

\item Case $n+1$: follows from the commuting diagram below, applying the functor $\tprod{-}{\abcobj}$ to the induction hypothesis.
\[\begin{tikzcd}
	{\tprod{\words[n]}{\abcobj}} & {\tprod{Q}{\abcobj}} && Q \\
	&&& {\tprod{Q}{\abcobj}} \\
	& {\tprod{(\tprod{\tunit}{\words[n]})}{\abcobj}} && {\tprod{(\tprod{Q}{\words[n]})}{\abcobj}} \\
	{\tprod{\tunit}{(\tprod{\words[n]}{\abcobj})}} &&& {\tprod{Q}{(\tprod{\words[n]}{\abcobj})}}
	\arrow["{\tprod{\malg{\aut{A},n}}{\id{\abcobj}}}", from=1-1, to=1-2]
	\arrow["{\tprod{\inv{\lunit{\words[n]}}}{\id{\abcobj}}}", from=1-1, to=3-2]
	\arrow["{\inv{\lunit{\tprod{\words[n]}{\abcobj}}}}"', from=1-1, to=4-1]
	\arrow["{\transmor{Q}}", from=1-2, to=1-4]
	\arrow[Rightarrow, no head, from=1-2, to=2-4]
	\arrow["{\transmor{Q}}"', from=2-4, to=1-4]
	\arrow["{\tprod{(\tprod{\initmor{Q}}{\id{\words[n]}})}{\id{\abcobj}}}"', from=3-2, to=3-4]
	\arrow["{\tprod{\transmor*{Q,n}}{\id{\abcobj}}}"', from=3-4, to=2-4]
	\arrow["{\inv{\assoc{\tunit}{\words[n]}{\abcobj}}}"{pos=0.7}, from=4-1, to=3-2]
	\arrow["{\tprod{\initmor{Q}}{\id{\tprod{\words[n]}{\abcobj}}}}"', from=4-1, to=4-4]
	\arrow["{\inv{\assoc{Q}{\words[n]}{\abcobj}}}"', from=4-4, to=3-4]
\end{tikzcd}\]
\end{items}

\item The claim holds if and only if $\mcoalg{\aut{A}}' = \finmor{Q} \circ \transmor*{Q}$ (see \autoref{sec:proof-lang-fin-coalg} for the definition of $\mcoalg{\aut{A}}'$). Since $\mfcoprod{n \in \N}{\tprod{\id{Q}}{\minj{n}}}$ is an isomorphism, this is true if and only if $\mcoalg{\aut{A}}' \circ \mfcoprod{n \in \N}{\tprod{\id{Q}}{\minj{n}}} = \finmor{Q} \circ \transmor*{Q} \circ \mfcoprod{n \in \N}{\tprod{\id{Q}}{\minj{n}}}$. Using \autoref{lem:transmor-mcoalg-comp}, this holds if and only if $\mcoalg{\aut{A},n} = \finmor{Q} \circ \transmor*{Q,n}$ for all $n \in \N$. We prove the latter claim by induction on $n$.
\begin{items}
\item Case $0$: we have
\[ \mcoalg{\aut{A},0} = \finmor{Q} \circ \runit{Q} = \finmor{Q} \circ \transmor*{Q,0}. \]

\item Case $n+1$: we need to prove that
\[ \mcoalg{\aut{A},n} (\tprod{\transmor{Q}}{\id{\abcobj}}) \inv{\assoc{Q}{\abcobj}{\words[n]}} (\tprod{\id{Q}}{\inv{\cons[n]}}) = \finmor{Q} \transmor*{Q,n+1}. \]
This holds if and only if
\[ \mcoalg{\aut{A},n} (\tprod{\transmor{Q}}{\id{\abcobj}}) \inv{\assoc{Q}{\abcobj}{\words[n]}} = \finmor{Q} \transmor*{Q,n+1} (\tprod{\id{Q}}{\cons[n]}). \]
But the latter holds since
\begin{align*}
\mcoalg{\aut{A},n} (\tprod{\transmor{Q}}{\id{\abcobj}}) \inv{\assoc{Q}{\abcobj}{\words[n]}}
	&= \finmor{Q} \transmor*{Q,n} (\tprod{\transmor{Q}}{\id{\abcobj}}) \inv{\assoc{Q}{\abcobj}{\words[n]}} \\
	&= \finmor{Q} \transmor*{Q,n+1} (\tprod{\id{Q}}{\cons[n]})
\end{align*}
using the induction hypothesis and \autoref{prop:transmor-cons}. \qed
\end{items}
\end{enum}
\end{proof}

\begin{proof}[\autoref{prop:rec-lang-equiv}]
The bijection $b \colon \Hom[\cat{C}]{\words}{\outobj} \xto{\cong} \Hom[\cat{C}]{\tunit}{\inthom{\words}{\outobj}}$ is given by $L \mapsto \adj{\words}{L \circ \lunit{\words}}$. Thus, we have
\begin{align*}
b(\reclang{\aut{A}})
    &= \adj{\words}{\reclang{\aut{A}} \lunit{\words}} = \adj{\words}{\finmor{Q} \malg{\aut{A}} \lunit{\words}} = \adj{\words}{\finmor{Q} \transmor*{Q} (\tprod{\initmor{Q}}{\id{\words}}) \inv{\lunit{\words}} \lunit{\words}} \\
    &= \adj{\words}{\finmor{Q} \transmor*{Q} (\tprod{\initmor{Q}}{\id{\words}})} = \adj{\words}{\finmor{Q} \transmor*{Q}} \circ \initmor{Q} = \mcoalg{\aut{A}} \initmor{Q} = \reclang{\aut{A}}'
\end{align*}
using the naturality of $\adj{\words}{-}$ and \autoref{prop:malg-mcoalg-transmor}. \qed
\end{proof}

\section{Proofs for \autoref{sec:test-suites}}

\subsection{Proof of \autoref{thm:main}}

The following proof is a categorification of the proof of~\cite[Lemma 3.8]{DBLP:conf/tacas/KrugerJR24}. In the proof, we use the notion of an \emph{AM-bisimulation}~\cite{DBLP:journals/corr/Staton11}, which we define below.
\begin{definition}
Let $H \colon \cat{C} \to \cat{C}$ be a functor. Moreover, let $b \colon B \to HB$ and $c \colon C \to HC$ be two $H$-coalgebras. An \emph{AM-bisimulation} between $(B, b)$ and $(C, c)$ is a morphism $\mprod{r_1}{r_2} \colon R \to B \times C$ such that there exists a morphism $R \to HR$ making the following diagram commute.
\[\begin{tikzcd}
	B & R & C \\
	HB & HR & HC
	\arrow["b"', from=1-1, to=2-1]
	\arrow["{r_1}"', from=1-2, to=1-1]
	\arrow["{r_2}", from=1-2, to=1-3]
	\arrow[dashed, from=1-2, to=2-2]
	\arrow["c", from=1-3, to=2-3]
	\arrow["{Hr_1}", from=2-2, to=2-1]
	\arrow["{Hr_2}"', from=2-2, to=2-3]
\end{tikzcd}\]
\end{definition}
\begin{remark}
Usually, an AM-bisimulation is required to be a \emph{relation}, i.e. the morphism $\mprod{r_1}{r_2}$ is required to be a mono (or an element of the right class of a factorization system). We choose to omit this requirement; this makes the proof simpler.
\end{remark}

We recall that being related by an AM-bisimulation implies equivalence in the final coalgebra (if there exists one).
\begin{proposition} \label{prop:bisim-fin-coalg}
Let $H \colon \cat{C} \to \cat{C}$ be a functor. Moreover, let $b \colon B \to HB$ and $c \colon C \to HC$ be two $H$-coalgebras, and let $\mprod{r_1}{r_2} \colon R \to B \times C$ be an AM-bisimulation between $(B, b)$ and $(C, c)$. Suppose that $H$ has a final coalgebra $(Z, z)$. Then the following diagram commutes, where $!$ denotes the unique homomorphism to the final coalgebra.
\[\begin{tikzcd}
	R & C \\
	B & Z
	\arrow["{r_2}", from=1-1, to=1-2]
	\arrow["{r_1}"', from=1-1, to=2-1]
	\arrow["{!}", from=1-2, to=2-2]
	\arrow["{!}"', from=2-1, to=2-2]
\end{tikzcd}\]
\end{proposition}
\begin{proof}
See~\cite{DBLP:journals/tcs/Rutten00}. \qed
\end{proof}

For the special case when $H = \coalgfunc$, we can characterize AM-bisimulations in more elementary terms.
\begin{proposition} \label{prop:bisim-coalgfunc}
Let $\mprod{f_B}{\overline{\delta_B}} \colon B \to \outobj \times \inthom{\abcobj}{B}$ and $\mprod{f_C}{\overline{\delta_C}} \colon C \to \outobj \times \inthom{\abcobj}{C}$ be two $\coalgfunc$-coalgebras and write $\delta_B = \adj*{\abcobj}{\overline{\delta_B}}$ and $\delta_C = \adj*{\abcobj}{\overline{\delta_C}}$. Furthermore, let $\mprod{r_1}{r_2} \colon R \to B \times C$ be a morphism. Then $\mprod{r_1}{r_2}$ is an AM-bisimulation between $\mprod{f_B}{\overline{\delta_B}}$ and $\mprod{f_C}{\overline{\delta_C}}$ if and only if
\begin{enum}
\item the square
\[\begin{tikzcd}
	R & C \\
	B & \outobj
	\arrow["{r_2}", from=1-1, to=1-2]
	\arrow["{r_1}"', from=1-1, to=2-1]
	\arrow["{f_C}", from=1-2, to=2-2]
	\arrow["{f_B}"', from=2-1, to=2-2]
\end{tikzcd}\]
commutes, and

\item there exists a morphism $\delta_R \colon \tprod{R}{\abcobj} \to R$ making the following diagram commute.
\[\begin{tikzcd}
	{\tprod{B}{\abcobj}} & {\tprod{R}{\abcobj}} & {\tprod{C}{\abcobj}} \\
	B & R & C
	\arrow["{\delta_B}"', from=1-1, to=2-1]
	\arrow["{\tprod{r_1}{\id{\abcobj}}}"', from=1-2, to=1-1]
	\arrow["{\tprod{r_2}{\id{\abcobj}}}", from=1-2, to=1-3]
	\arrow["{\delta_R}"', from=1-2, to=2-2]
	\arrow["{\delta_C}", from=1-3, to=2-3]
	\arrow["{r_1}", from=2-2, to=2-1]
	\arrow["{r_2}"', from=2-2, to=2-3]
\end{tikzcd}\]
\end{enum}
\end{proposition}
\begin{proof}
A morphism $R \to \outobj \times \inthom{\abcobj}{R}$ can be written as $\mprod{f_R}{\overline{\delta_R}}$ for some morphism $f_R \colon R \to \outobj$ and $\overline{\delta_R} \colon R \to \inthom{\abcobj}{R}$. The morphism $\mprod{r_1}{r_2}$ is an AM-bisimulation if and only if the diagram
\[\begin{tikzcd}
	B && R && C \\
	{\outobj \times \inthom{\abcobj}{B}} && {O \times \inthom{\abcobj}{R}} && {\outobj \times \inthom{\abcobj}{C}}
	\arrow["{\mprod{f_B}{\overline{\delta_B}}}"', from=1-1, to=2-1]
	\arrow["{r_1}"', from=1-3, to=1-1]
	\arrow["{r_2}", from=1-3, to=1-5]
	\arrow["{\mprod{f_R}{\overline{\delta_R}}}"', from=1-3, to=2-3]
	\arrow["{\mprod{f_C}{\overline{\delta_C}}}", from=1-5, to=2-5]
	\arrow["{\id{\outobj} \times \inthom{\id{\abcobj}}{r_1}}", from=2-3, to=2-1]
	\arrow["{\id{\outobj} \times \inthom{\id{\abcobj}}{r_2}}"', from=2-3, to=2-5]
\end{tikzcd}\]
commutes. Taking projections and adjuncts, the diagram above commutes if and only if the diagrams
\[\begin{tikzcd}
	B & R & C & {\tprod{B}{\abcobj}} & {\tprod{R}{\abcobj}} & {\tprod{C}{\abcobj}} \\
	& O && B & R & C
	\arrow["{f_B}"', from=1-1, to=2-2]
	\arrow["{r_1}"', from=1-2, to=1-1]
	\arrow["{r_2}", from=1-2, to=1-3]
	\arrow["{f_R}"'{pos=0.3}, from=1-2, to=2-2]
	\arrow["{f_C}", from=1-3, to=2-2]
	\arrow["{\delta_B}"', from=1-4, to=2-4]
	\arrow["{\tprod{r_1}{\id{\abcobj}}}"', from=1-5, to=1-4]
	\arrow["{\tprod{r_2}{\id{\abcobj}}}", from=1-5, to=1-6]
	\arrow["{\delta_R}"', from=1-5, to=2-5]
	\arrow["{\delta_C}", from=1-6, to=2-6]
	\arrow["{r_1}", from=2-5, to=2-4]
	\arrow["{r_2}"', from=2-5, to=2-6]
\end{tikzcd}\]
commute. The commutativity of the left diagram is equivalent to condition (i), and the right diagram is precisely condition (ii). \qed
\end{proof}

\begin{proposition} \label{prop:words-monoid}
The triple $(\words, \append, \minj{0})$ is a monoid in $\cat{C}$, i.e. the following diagrams commute.
\[\begin{tikzcd}
	{\tprod{(\tprod{\words}{\words})}{\words}} && {\tprod{\words}{(\tprod{\words}{\words})}} && {\tprod{\words}{\words}} \\
	{\tprod{\words}{\words}} &&&& \words
	\arrow["{\assoc{\words}{\words}{\words}}", from=1-1, to=1-3]
	\arrow["{\tprod{\append}{\id{\words}}}"', from=1-1, to=2-1]
	\arrow["{\tprod{\id{\words}}{\append}}", from=1-3, to=1-5]
	\arrow["\append", from=1-5, to=2-5]
	\arrow["\append"', from=2-1, to=2-5]
\end{tikzcd}\]
\[\begin{tikzcd}
	{\tprod{\words}{\tunit}} && {\tprod{\words}{\words}} && {\tprod{\tunit}{\words}} \\
	&& \words
	\arrow["{\tprod{\id{\words}}{\minj{0}}}", from=1-1, to=1-3]
	\arrow["{\runit{\words}}"', from=1-1, to=2-3]
	\arrow["\append"', from=1-3, to=2-3]
	\arrow["{\tprod{\minj{0}}{\id{\words}}}"', from=1-5, to=1-3]
	\arrow["{\lunit{\words}}", from=1-5, to=2-3]
\end{tikzcd}\]
\end{proposition}
\begin{proof}
We first show the commutativity of the upper rectangle, then that of the lower two triangles.
\begin{enum}
\item Since $\mfcoprod{l,m,n \in \N}{\tprod{(\tprod{\minj{l}}{\minj{m}})}{\minj{n}}} \colon \sum_{l,m,n \in \N}{\tprod{(\tprod{\words[l]}{\words[m]})}{\words[n]}} \to \tprod{(\tprod{\words}{\words})}{\words}$ is an isomorphism, it suffices to show that the two sides become equal after precomposing with this isomorphism. This holds if and only if the two sides become equal after precomposing with $\tprod{(\tprod{\minj{l}}{\minj{m}})}{\minj{n}}$ for all $l,m,n \in \N$. This follows from the commutativity of the diagram below, using \autoref{lem:snoc-cons-append-comp}.
\[\hspace{-2.25cm}\begin{tikzcd}
	{\tprod{(\tprod{\words}{\words})}{\words}} && {\tprod{\words}{(\tprod{\words}{\words})}} && {\tprod{\words}{\words}} \\
	& {\tprod{(\tprod{\words[l]}{\words[m]})}{\words[n]}} & {\tprod{\words[l]}{(\tprod{\words[m]}{\words[n]})}} & {\tprod{\words[l]}{\words[m+n]}} \\
	& {\tprod{\words[l+m]}{\words[n]}} && {\words[l+m+n]} \\
	{\tprod{\words}{\words}} &&&& \words
	\arrow["{\assoc{\words}{\words}{\words} }", from=1-1, to=1-3]
	\arrow["{\tprod{\append}{\id{\words}}}"', from=1-1, to=4-1]
	\arrow["{\tprod{\id{\words}}{\append}}", from=1-3, to=1-5]
	\arrow["\append", from=1-5, to=4-5]
	\arrow["{\tprod{(\tprod{\minj{l}}{\minj{m}})}{\minj{n}}}"{description}, from=2-2, to=1-1]
	\arrow["{\assoc{\words[l]}{\words[m]}{\words[n]}}"', from=2-2, to=2-3]
	\arrow["{\tprod{\append[l,m]}{\id{\words[n]}}}"', from=2-2, to=3-2]
	\arrow["{\tprod{\minj{l}}{(\tprod{\minj{m}}{\minj{n}})}}"{description}, from=2-3, to=1-3]
	\arrow["{\tprod{\id{\words[l]}}{\append[m,n]}}"', from=2-3, to=2-4]
	\arrow["{\tprod{\minj{l}}{\minj{m+n}}}"{description}, from=2-4, to=1-5]
	\arrow["{\append[l,m+n]}", from=2-4, to=3-4]
	\arrow["{\append[l+m,n]}"', from=3-2, to=3-4]
	\arrow["{\tprod{\minj{l+m}}{\minj{n}}}"{description}, from=3-2, to=4-1]
	\arrow["{\minj{l+m+n}}"{description}, from=3-4, to=4-5]
	\arrow["\append"', from=4-1, to=4-5]
\end{tikzcd}\]

It remains to show that the rectangle
\[\begin{tikzcd}[column sep = 3em]
	{\tprod{(\tprod{\words[l]}{\words[m]})}{\words[n]}} & {\tprod{\words[l]}{(\tprod{\words[m]}{\words[n]})}} & {\tprod{\words[l]}{\words[m+n]}} \\
	{\tprod{\words[l+m]}{\words[n]}} && {\words[l+m+n]}
	\arrow["{\assoc{\words[l]}{\words[m]}{\words[n]}}", from=1-1, to=1-2]
	\arrow["{\tprod{\append[l,m]}{\id{\words[n]}}}"', from=1-1, to=2-1]
	\arrow["{\tprod{\id{\words[l]}}{\append[m,n]}}", from=1-2, to=1-3]
	\arrow["{\append[l,m+n]}", from=1-3, to=2-3]
	\arrow["{\append[l+m,n]}"', from=2-1, to=2-3]
\end{tikzcd}\]
commutes. We do this by induction on $n$.
\begin{items}
\item Case $0$: follows from the commuting diagram below.
\[\begin{tikzcd}
	& {} \\
	{\tprod{(\tprod{\words[l]}{\words[m]})}{\tunit}} & {\tprod{\words[l]}{(\tprod{\words[m]}{\tunit})}} & {\tprod{\words[l]}{\words[m]}} \\
	{\tprod{\words[l+m]}{\words[n]}} && {\words[l+m]}
	\arrow["{\assoc{\words[l]}{\words[m]}{\tunit}}", from=2-1, to=2-2]
	\arrow["{\runit{\tprod{\words[l]}{\words[m]}}}"', curve={height=12pt}, from=2-1, to=2-3]
	\arrow["{\tprod{\append[l,m]}{\id{\tunit}}}"', from=2-1, to=3-1]
	\arrow["{\tprod{\id{\words[l]}}{\runit{\words[m]}}}", from=2-2, to=2-3]
	\arrow["{\append[l,m]}", from=2-3, to=3-3]
	\arrow["{\runit{\words[l+m]}}"', from=3-1, to=3-3]
\end{tikzcd}\]

\item Case $n+1$: follows from the commuting diagram below, where we have applied the functor $\tprod{-}{\abcobj}$ to the induction hypothesis.
\[\hspace{-5.25cm}\begin{tikzcd}[column sep = 5em]
	& {} \\
	{\tprod{(\tprod{\words[l]}{\words[m]})}{(\tprod{\words[n]}{\abcobj})}} & {\tprod{\words[l]}{(\tprod{\words[m]}{(\tprod{\words[n]}{\abcobj})})}} & {\tprod{\words[l]}{(\tprod{(\tprod{\words[m]}{\words[n]})}{\abcobj})}} & {\tprod{\words[l]}{(\tprod{\words[m+n]}{\abcobj})}} \\
	& {\tprod{(\tprod{(\tprod{\words[l]}{\words[m]})}{\words[n]})}{\abcobj}} & {\tprod{(\tprod{\words[l]}{(\tprod{\words[m]}{\words[n]})})}{\abcobj}} & {\tprod{(\tprod{\words[l]}{\words[m+n]})}{\abcobj}} \\
	{\tprod{\words[l+m]}{(\tprod{\words[n]}{\abcobj})}} & {\tprod{(\tprod{\words[l+m]}{\words[n]})}{\abcobj}} && {\tprod{\words[l+m+n]}{\abcobj}}
	\arrow["{\assoc{\words[l]}{\words[m]}{\tprod{\words[n]}{\abcobj}}}", from=2-1, to=2-2]
	\arrow["{\inv{\assoc{\tprod{\words[l]}{\words[m]}}{\words[n]}{\abcobj}}}"{description}, from=2-1, to=3-2]
	\arrow["{\tprod{\append[l,m]}{\id{\tprod{\words[n]}{\abcobj}}}}"', from=2-1, to=4-1]
	\arrow["{\tprod{\id{\words[l]}}{\inv{\assoc{\words[m]}{\words[n]}{\abcobj}}}}", from=2-2, to=2-3]
	\arrow["{\tprod{\id{\words[l]}}{(\tprod{\append[m,n]}{\id{\abcobj}})}}", from=2-3, to=2-4]
	\arrow["{\inv{\assoc{\words[l]}{\tprod{\words[m]}{\words[n]}}{\abcobj}}}"', from=2-3, to=3-3]
	\arrow["{\inv{\assoc{\words[l]}{\words[m+n]}{\abcobj}}}", from=2-4, to=3-4]
	\arrow["{\tprod{\assoc{\words[l]}{\words[m]}{\words[n]}}{\id{\abcobj}}}"', from=3-2, to=3-3]
	\arrow["{\tprod{(\tprod{\append[l,m]}{\id{\words[n]}})}{\id{\abcobj}}}"', from=3-2, to=4-2]
	\arrow["{\tprod{(\tprod{\id{\words[l]}}{\append[m,n]})}{\id{\abcobj}}}"', from=3-3, to=3-4]
	\arrow["{\tprod{\append[l,m+n]}{\id{\abcobj}}}", from=3-4, to=4-4]
	\arrow["{\inv{\assoc{\words[l+m]}{\words[n]}{\abcobj}}}"', from=4-1, to=4-2]
	\arrow["{\tprod{\append[l+m,n]}{\id{\abcobj}}}"', from=4-2, to=4-4]
\end{tikzcd}\]
\end{items}

\item For the left triangle, note that $\mfcoprod{n \in \N}{\tprod{\minj{n}}{\id{\tunit}}} \colon \sum_{n \in \N}{\tprod{\words[n]}{\tunit}} \to \tprod{\words}{\tunit}$ is an isomorphism. Hence, it suffices to show
\[ \append \circ (\tprod{\id{\words}}{\minj{0}}) \circ \mfcoprod{n \in \N}{\tprod{\minj{n}}{\id{\tunit}}} = \runit{\words} \circ \mfcoprod{n \in \N}{\tprod{\minj{n}}{\id{\tunit}}}. \]
This is equivalent to
\[ \append \circ (\tprod{\id{\words}}{\minj{0}}) \circ (\tprod{\minj{n}}{\id{\tunit}}) = \runit{\words} \circ (\tprod{\minj{n}}{\id{\tunit}}) \]
for all $n \in \N$. The latter holds since
\begin{multline*}
\append \circ (\tprod{\id{\words}}{\minj{0}}) \circ (\tprod{\minj{n}}{\id{\tunit}}) \\
	= \append \circ (\tprod{\minj{n}}{\minj{0}}) = \minj{n} \circ \append[n,0] = \minj{n} \circ \runit{\words[n]} = \runit{\words} \circ (\tprod{\minj{n}}{\id{\tunit}})
\end{multline*}
using \autoref{lem:snoc-cons-append-comp} and the definition of $\append[n,0]$.

The commutativity of the right triangle is analogous, using instead the isomorphism $\mfcoprod{n \in \N}{\tprod{\id{\tunit}}{\minj{n}}} \colon \sum_{n \in \N}{\tprod{\tunit}{\words[n]}} \to \tprod{\tunit}{\words}$. We need to show
\[ \append \circ (\tprod{\minj{0}}{\id{\words}}) \circ (\tprod{\id{\tunit}}{\minj{n}}) = \lunit{\words} \circ (\tprod{\id{\tunit}}{\minj{n}}). \]
for all $n \in \N$. This holds since
\begin{multline*}
\append \circ (\tprod{\minj{0}}{\id{\words}}) \circ (\tprod{\id{\tunit}}{\minj{n}}) \\
	= \append \circ (\tprod{\minj{0}}{\minj{n}}) = \minj{n} \circ \append[0,n] = \minj{n} \circ \lunit{\words[n]} = \lunit{\words} \circ (\tprod{\id{\tunit}}{\minj{n}})
\end{multline*}
using \autoref{lem:snoc-cons-append-comp} and \autoref{lem:append-zero}. \qed
\end{enum}
\end{proof}

\begin{proposition} \label{prop:malg-append}
Let $\aut{A} = (Q, \initmor{Q}, \transmor{Q}, \finmor{Q})$ be an automaton. Then the following diagram commutes.
\[\begin{tikzcd}
	{\tprod{\words}{\words}} & \words \\
	{\tprod{Q}{\words}} & Q
	\arrow["\append", from=1-1, to=1-2]
	\arrow["{\tprod{\malg{\aut{A}}}{\id{\words}}}"', from=1-1, to=2-1]
	\arrow["{\malg{\aut{A}}}", from=1-2, to=2-2]
	\arrow["{\transmor*{Q}}"', from=2-1, to=2-2]
\end{tikzcd}\]
\end{proposition}
\begin{proof}
Since $\mfcoprod{m,n \in \N}{\tprod{\minj{m}}{\minj{n}}} \colon \sum_{m,n \in \N}{\tprod{\words[m]}{\words[n]}} \to \tprod{\words}{\words}$ is an isomorphism, it suffices to show that the two sides become equal after precomposing with this isomorphism. This holds if and only if the two sides become equal after precomposing with $\tprod{\minj{m}}{\minj{n}}$ for all $m,n \in \N$. This follows from the commutativity of the diagram below, using \autoref{lem:snoc-cons-append-comp} and \autoref{lem:transmor-mcoalg-comp}.
\[\begin{tikzcd}[column sep = 4em]
	&& {\tprod{\words}{\words}} \\
	& {\tprod{\words[m]}{\words[n]}} & {\words[m+n]} & \words \\
	{\tprod{\words}{\words}} & {\tprod{\words[m]}{\words}} & {\tprod{Q}{\words[n]}} \\
	& {\tprod{Q}{\words}} && Q
	\arrow["\append", from=1-3, to=2-4]
	\arrow["{\tprod{\minj{m}}{\minj{n}}}", from=2-2, to=1-3]
	\arrow["{\append[m,n]}"', from=2-2, to=2-3]
	\arrow["{\tprod{\minj{m}}{\minj{n}}}"', from=2-2, to=3-1]
	\arrow["{\tprod{\id{\words[m]}}{\minj{n}}}"{description}, from=2-2, to=3-2]
	\arrow["{\tprod{\malg{\aut{A},m}}{\id{\words[n]}}}"{description}, from=2-2, to=3-3]
	\arrow["{\minj{m+n}}"', from=2-3, to=2-4]
	\arrow["{\malg{\aut{A},m+n}}"{description}, from=2-3, to=4-4]
	\arrow["{\malg{\aut{A}}}", from=2-4, to=4-4]
	\arrow["{\tprod{\malg{\aut{A}}}{\id{\words}}}"', from=3-1, to=4-2]
	\arrow["{\tprod{\minj{m}}{\id{\words}}}", from=3-2, to=3-1]
	\arrow["{\tprod{\malg{\aut{A},m}}{\id{\words}}}"{description}, from=3-2, to=4-2]
	\arrow["{\tprod{\id{Q}}{\minj{n}}}"{pos=0.3}, from=3-3, to=4-2]
	\arrow["{\transmor*{Q,n}}"'{pos=0.3}, from=3-3, to=4-4]
	\arrow["{\transmor*{Q}}"', from=4-2, to=4-4]
\end{tikzcd}\]

It remains to show that the diagram
\[\begin{tikzcd}
	{\tprod{\words[m]}{\words[n]}} & {\words[m+n]} \\
	{\tprod{Q}{\words[n]}} & Q
	\arrow["{\append[m,n]}", from=1-1, to=1-2]
	\arrow["{\tprod{\malg{\aut{A},m}}{\id{\words[n]}}}"', from=1-1, to=2-1]
	\arrow["{\malg{\aut{A},m+n}}", from=1-2, to=2-2]
	\arrow["{\transmor*{Q,n}}"', from=2-1, to=2-2]
\end{tikzcd}\]
commutes for all $m,n \in \N$. We do this by induction on $n$.
\begin{items}
\item Case $0$: follows from the commutativity of the diagram below.
\[\begin{tikzcd}
	{\tprod{\words[m]}{\tunit}} & {\words[m]} \\
	{\tprod{Q}{\tunit}} & Q
	\arrow["{\runit{\words[m]}}", from=1-1, to=1-2]
	\arrow["{\tprod{\malg{\aut{A},m}}{\id{\tunit}}}"', from=1-1, to=2-1]
	\arrow["{\malg{\aut{A},m}}", from=1-2, to=2-2]
	\arrow["{\runit{Q}}"', from=2-1, to=2-2]
\end{tikzcd}\]

\item Case $n+1$: follows from the commutativity of the diagram below, where we have applied the functor $\tprod{-}{\abcobj}$ to the induction hypothesis.
\[\hspace{-1cm}\begin{tikzcd}[column sep = 3em]
	{\tprod{\words[m]}{(\tprod{\words[n]}{\abcobj})}} & {\tprod{(\tprod{\words[m]}{\words[n]})}{\abcobj}} && {\tprod{\words[m+n]}{\abcobj}} \\
	&&& {\tprod{Q}{\abcobj}} \\
	{\tprod{Q}{(\tprod{\words[n]}{\abcobj})}} & {\tprod{(\tprod{Q}{\words[n]})}{\abcobj}} & {\tprod{Q}{\abcobj}} & Q
	\arrow["{\inv{\assoc{\words[m]}{\words[n]}{\abcobj}}}", from=1-1, to=1-2]
	\arrow["{\tprod{\malg{\aut{A},m}}{\id{\tprod{\words[n]}{\abcobj}}}}"', from=1-1, to=3-1]
	\arrow["{\tprod{\append[m,n]}{\id{\abcobj}}}", from=1-2, to=1-4]
	\arrow["{\tprod{(\tprod{\malg{\aut{A},m}}{\id{\words[n]}})}{\id{\abcobj}}}"', from=1-2, to=3-2]
	\arrow["{\tprod{\malg{\aut{A},m+n}}{\id{\abcobj}}}", from=1-4, to=2-4]
	\arrow["{\transmor{Q}}", from=2-4, to=3-4]
	\arrow["{\inv{\assoc{Q}{\words[n]}{\abcobj}}}"', from=3-1, to=3-2]
	\arrow["{\tprod{\transmor*{Q,n}}{\id{\abcobj}}}"', from=3-2, to=3-3]
	\arrow[Rightarrow, no head, from=3-3, to=2-4]
	\arrow["{\transmor{Q}}"', from=3-3, to=3-4]
\end{tikzcd}\]
\qed
\end{items}
\end{proof}

We fix a specification automaton $\aut{\specname} = (\specname, \initmor{\specname}, \transmor{\specname}, \finmor{\specname})$ and an implementation automaton $\aut{\implname} = (\implname, \initmor{\implname}, \transmor{\implname}, \finmor{\implname})$ for the remainder of this subsection.

\begin{lemma} \label{lem:thm-main-bisim-1}
Let $c \colon C \to \words$ and $w \colon W \to \words$ be two morphisms containing the empty word. Let $t = c \cdot \wincl{1} \cdot w$, and suppose $\agreets{t}{\aut{\specname}}{\aut{\implname}}$. Then the diagram
\[\begin{tikzcd}
	C & \words & \implname \\
	\words \\
	\specname && \outobj
	\arrow["c", from=1-1, to=1-2]
	\arrow["c"', from=1-1, to=2-1]
	\arrow["{\malg{\aut{\implname}}}", from=1-2, to=1-3]
	\arrow["{\finmor{\implname}}", from=1-3, to=3-3]
	\arrow["{\malg{\aut{\specname}}}"', from=2-1, to=3-1]
	\arrow["{\finmor{\specname}}"', from=3-1, to=3-3]
\end{tikzcd}\]
commutes.
\end{lemma}
\begin{proof}
Since $\agreets{t}{\aut{\specname}}{\aut{\implname}}$, we have $\reclang{\aut{\specname}} \circ t = \reclang{\aut{\implname}} \circ t$, i.e. the following diagram commutes.
\[\begin{tikzcd}
	{\tprod{(\tprod{C}{\words[\le 1]})}{W}} & \words & \implname \\
	\words \\
	\specname && \outobj
	\arrow["t", from=1-1, to=1-2]
	\arrow["t"', from=1-1, to=2-1]
	\arrow["{\malg{\aut{\implname}}}", from=1-2, to=1-3]
	\arrow["{\finmor{\implname}}", from=1-3, to=3-3]
	\arrow["{\malg{\aut{\specname}}}"', from=2-1, to=3-1]
	\arrow["{\finmor{\specname}}"', from=3-1, to=3-3]
\end{tikzcd}\]
Thus, it suffices to show that $t \circ f = c$ for some morphism $f$. Since $w$ contains the empty word, there is a morphism $i_W \colon \tunit \to W$ with $w \circ i_W = \minj{0}$. Take as $f$ the composite
\[ C \xto{\inv{\runit{C}}} \tprod{C}{\tunit} \xto{\inv{\runit{\tprod{C}{\tunit}}}} \tprod{(\tprod{C}{\tunit})}{\tunit} \xto{\tprod{(\tprod{\id{C}}{\minj{0}})}{i_W}} \tprod{(\tprod{C}{\words[\le 1]})}{W}. \]
It remains to show that $t \circ f = c$. This follows from the following commuting diagram, using \autoref{prop:words-monoid}.
\[\hspace{-1.5cm}\begin{tikzcd}[column sep = 4em]
	C & {\tprod{C}{\tunit}} & {\tprod{(\tprod{C}{\tunit})}{\tunit}} && {\tprod{(\tprod{C}{\words[\le 1]})}{W}} \\
	\words & {\tprod{\words}{\tunit}} & {\tprod{(\tprod{\words}{\tunit})}{\tunit}} & {\tprod{(\tprod{\words}{\tunit})}{\words}} & {\tprod{(\tprod{\words}{\words})}{\words}} \\
	& \words & {\tprod{\words}{\tunit}} && {\tprod{\words}{\words}} \\
	&&&& \words
	\arrow["{\inv{\runit{C}}}", from=1-1, to=1-2]
	\arrow["c"', from=1-1, to=2-1]
	\arrow["{\inv{\runit{\tprod{C}{\tunit}}}}", from=1-2, to=1-3]
	\arrow["{\tprod{c}{\id{\tunit}}}"', from=1-2, to=2-2]
	\arrow["{\tprod{(\tprod{\id{C}}{\minj{0}})}{i_W}}", from=1-3, to=1-5]
	\arrow["{\tprod{(\tprod{c}{\id{\tunit}})}{\id{\tunit}}}"', from=1-3, to=2-3]
	\arrow["{\tprod{(\tprod{c}{\minj{0}})}{\minj{0}}}"{description}, from=1-3, to=2-5]
	\arrow["{\tprod{(\tprod{c}{\wincl{1}})}{w}}", from=1-5, to=2-5]
	\arrow["{\inv{\runit{\words}}}"', from=2-1, to=2-2]
	\arrow["{\inv{\runit{\tprod{\words}{\tunit}}}}"', from=2-2, to=2-3]
	\arrow["{\runit{\words}}"', from=2-2, to=3-2]
	\arrow["{\tprod{\id{\tprod{\words}{\tunit}}}{\minj{0}}}"', from=2-3, to=2-4]
	\arrow["{\tprod{\runit{\words}}{\id{\tunit}}}", from=2-3, to=3-3]
	\arrow["{\tprod{(\tprod{\id{\words}}{\minj{0}})}{\id{\words}}}"', from=2-4, to=2-5]
	\arrow["{\tprod{\runit{\words}}{\id{\words}}}"{description}, from=2-4, to=3-5]
	\arrow["{\tprod{\append}{\id{\words}}}", from=2-5, to=3-5]
	\arrow["{\inv{\runit{\words}}}"', from=3-2, to=3-3]
	\arrow["{\tprod{\id{\words}}{\minj{0}}}", from=3-3, to=3-5]
	\arrow["{\runit{\words}}"', from=3-3, to=4-5]
	\arrow["\append", from=3-5, to=4-5]
\end{tikzcd}\]
\qed
\end{proof}

\begin{lemma} \label{lem:thm-main-bisim-2}
Let $(c, \delta_C)$ be a weak state cover for $\aut{\implname}$ and let $w \colon W \to \words$ be a characterization morphism for $\aut{\specname}$. Let $t = c \cdot \wincl{1} \cdot w$ and suppose $\agreets{t}{\aut{\specname}}{\aut{\implname}}$. Then the diagram
\[\begin{tikzcd}
	{\tprod{\specname}{\abcobj}} & {\tprod{\words}{\abcobj}} & {\tprod{C}{\abcobj}} & {\tprod{\words}{\abcobj}} & {\tprod{\implname}{\abcobj}} \\
	\specname & \words & C & \words & \implname
	\arrow["{\transmor{\specname}}"', from=1-1, to=2-1]
	\arrow["{\tprod{\malg{\aut{\specname}}}{\id{\abcobj}}}"', from=1-2, to=1-1]
	\arrow["{\tprod{c}{\id{\abcobj}}}"', from=1-3, to=1-2]
	\arrow["{\tprod{c}{\id{\abcobj}}}", from=1-3, to=1-4]
	\arrow["{\delta_C}"', from=1-3, to=2-3]
	\arrow["{\tprod{\malg{\aut{\implname}}}{\id{\abcobj}}}", from=1-4, to=1-5]
	\arrow["{\transmor{\implname}}", from=1-5, to=2-5]
	\arrow["{\malg{\aut{\specname}}}", from=2-2, to=2-1]
	\arrow["c", from=2-3, to=2-2]
	\arrow["c"', from=2-3, to=2-4]
	\arrow["{\malg{\aut{\implname}}}"', from=2-4, to=2-5]
\end{tikzcd}\]
commutes.
\end{lemma}
\begin{proof}
We prove that both rectangles commute. The right hand rectangle commutes since $\transmor{\implname} \circ (\tprod{\malg{\aut{\implname}}}{\id{\abcobj}}) = \malg{\aut{\implname}} \circ \snoc$ by the fact that $\malg{\aut{A}}$ is a $\algfunc$-algebra homomorphism, and $\malg{\aut{\implname}} \circ c \circ \delta_C = \malg{\aut{\implname}} \circ \snoc \circ (\tprod{c}{\id{\abcobj}})$ by the definition of weak state cover.

For the left hand rectangle, we again use that $\transmor{\aut{\specname}} \circ (\tprod{\malg{\aut{\specname}}}{\id{\abcobj}}) = \malg{\aut{\specname}} \circ \snoc$ by the homomorphism property of $\malg{\aut{\specname}}$. Thus, we need to show that
\[\begin{tikzcd}
	{\tprod{C}{\abcobj}} & C & \words \\
	{\tprod{\words}{\abcobj}} \\
	\words && \specname
	\arrow["{\delta_C}", from=1-1, to=1-2]
	\arrow["{\tprod{c}{\id{\abcobj}}}"', from=1-1, to=2-1]
	\arrow["c", from=1-2, to=1-3]
	\arrow["{\malg{\aut{\specname}}}", from=1-3, to=3-3]
	\arrow["\snoc"', from=2-1, to=3-1]
	\arrow["{\malg{\aut{\specname}}}"', from=3-1, to=3-3]
\end{tikzcd}\]
commutes. Since $\aut{\specname}$ is minimal, it suffices to show that
\[\begin{tikzcd}
	{\tprod{C}{\abcobj}} & C & \words \\
	{\tprod{\words}{\abcobj}} && \specname \\
	\words & \specname & \langs
	\arrow["{\delta_C}", from=1-1, to=1-2]
	\arrow["{\tprod{c}{\id{\abcobj}}}"', from=1-1, to=2-1]
	\arrow["c", from=1-2, to=1-3]
	\arrow["{\malg{\aut{\specname}}}", from=1-3, to=2-3]
	\arrow["\snoc"', from=2-1, to=3-1]
	\arrow["{\mcoalg{\aut{\specname}}}", from=2-3, to=3-3]
	\arrow["{\malg{\aut{\specname}}}"', from=3-1, to=3-2]
	\arrow["{\mcoalg{\aut{\specname}}}"', from=3-2, to=3-3]
\end{tikzcd}\]
commutes. Since $w$ is a characterization morphism, it suffices to show that
\[\begin{tikzcd}
	{\tprod{C}{\abcobj}} & C && \words \\
	{\tprod{\words}{\abcobj}} &&& \specname \\
	&&& \langs \\
	\words & \specname & \langs & {\inthom{W}{\outobj}}
	\arrow["{\delta_C}", from=1-1, to=1-2]
	\arrow["{\tprod{c}{\id{\abcobj}}}"', from=1-1, to=2-1]
	\arrow["c", from=1-2, to=1-4]
	\arrow["{\malg{\aut{\specname}}}", from=1-4, to=2-4]
	\arrow["\snoc"', from=2-1, to=4-1]
	\arrow["{\mcoalg{\aut{\specname}}}", from=2-4, to=3-4]
	\arrow["{\inthom{w}{\id{\outobj}}}", from=3-4, to=4-4]
	\arrow["{\malg{\aut{\specname}}}"', from=4-1, to=4-2]
	\arrow["{\mcoalg{\aut{\specname}}}"', from=4-2, to=4-3]
	\arrow["{\inthom{w}{\id{\outobj}}}"', from=4-3, to=4-4]
\end{tikzcd}\]
commutes. This follows from the commuting diagram below, where we use that $(c, \delta_C)$ is a weak state cover.
\[\begin{tikzcd}
	{\tprod{C}{\abcobj}} & C &&& \words \\
	{\tprod{\words}{\abcobj}} & \words & \implname & \langs & \specname \\
	& {\tprod{\words}{\abcobj}} & \words && \langs \\
	\words & \specname & \langs && {\inthom{W}{\outobj}}
	\arrow["{\delta_C}", from=1-1, to=1-2]
	\arrow["{\tprod{c}{\id{\abcobj}}}"', from=1-1, to=2-1]
	\arrow["{\tprod{c}{\id{\abcobj}}}"'{pos=0.7}, from=1-1, to=3-2]
	\arrow[""{name=0, anchor=center, inner sep=0}, "c", from=1-2, to=1-5]
	\arrow["c", from=1-2, to=2-2]
	\arrow["{\malg{\aut{\specname}}}", from=1-5, to=2-5]
	\arrow["\snoc"', from=2-1, to=4-1]
	\arrow["{\malg{\aut{\implname}}}", from=2-2, to=2-3]
	\arrow[""{name=1, anchor=center, inner sep=0}, "{\mcoalg{\aut{\implname}}}", from=2-3, to=2-4]
	\arrow["{\inthom{w}{\id{\outobj}}}"', from=2-4, to=4-5]
	\arrow["{\mcoalg{\aut{\specname}}}", from=2-5, to=3-5]
	\arrow[""{name=2, anchor=center, inner sep=0}, "\snoc"', from=3-2, to=3-3]
	\arrow["{\malg{\aut{\implname}}}"', from=3-3, to=2-3]
	\arrow["{\inthom{w}{\id{\outobj}}}", from=3-5, to=4-5]
	\arrow["{\malg{\aut{\specname}}}"', from=4-1, to=4-2]
	\arrow[""{name=3, anchor=center, inner sep=0}, "{\mcoalg{\aut{\specname}}}"', from=4-2, to=4-3]
	\arrow["{\inthom{w}{\id{\outobj}}}"', from=4-3, to=4-5]
	\arrow["{(ii)}"{description}, draw=none, from=0, to=1]
	\arrow["{(i)}"{description}, draw=none, from=3, to=2]
\end{tikzcd}\]

We still need to show the commutativity of the areas labelled (i) and (ii).
\begin{enum}
\item Taking adjuncts, we need to show
\begin{multline*}
\ev{\words}{\outobj} \circ (\tprod{\id{\langs}}{w}) \circ (\tprod{(\mcoalg{\aut{\specname}} \circ \malg{\aut{\specname}} \circ \snoc \circ (\tprod{c}{\id{\abcobj}}))}{\id{W}}) = \\
\ev{\words}{\outobj} \circ (\tprod{\id{\langs}}{w}) \circ (\tprod{(\mcoalg{\aut{\implname}} \circ \malg{\aut{\implname}} \circ \snoc \circ (\tprod{c}{\id{\abcobj}}))}{\id{W}}).
\end{multline*}
Moving $\tprod{\id{\langs}}{w}$ to the end, this is equivalent to
\begin{multline*}
\ev{\words}{\outobj} \circ (\tprod{\mcoalg{\aut{\specname}}}{\id{\words}}) \circ (\tprod{\malg{\aut{\specname}}}{\id{\words}}) \circ (\tprod{\snoc}{\id{\words}}) \circ (\tprod{(\tprod{c}{\id{\abcobj}})}{w}) = \\
\ev{\words}{\outobj} \circ (\tprod{\mcoalg{\aut{\implname}}}{\id{\words}}) \circ (\tprod{\malg{\aut{\implname}}}{\id{\words}}) \circ (\tprod{\snoc}{\id{\words}}) \circ (\tprod{(\tprod{c}{\id{\abcobj}})}{w}).
\end{multline*}
Using \autoref{prop:malg-mcoalg-transmor} (ii) and that $\ev{\words}{\outobj} \circ (\tprod{h}{\id{\words}}) = \adj*{\words}{h}$ for any $h \colon Q \to \langs$, this is equivalent to
\begin{multline*}
\finmor{\specname} \circ \transmor*{\specname} \circ (\tprod{\malg{\aut{\specname}}}{\id{\words}}) \circ (\tprod{\snoc}{\id{\words}}) \circ (\tprod{(\tprod{c}{\id{\abcobj}})}{w}) = \\
\finmor{\implname} \circ \transmor*{\implname} \circ (\tprod{\malg{\aut{\implname}}}{\id{\words}}) \circ (\tprod{\snoc}{\id{\words}}) \circ (\tprod{(\tprod{c}{\id{\abcobj}})}{w}).
\end{multline*}

We show that both sides of the equation are equal to the composite 
\[ \tprod{(\tprod{C}{\abcobj})}{W} \xto{\tprod{(\tprod{\id{C}}{\minj{1} \inv{\lunit{\abcobj}}})}{\id{W}}} \tprod{(\tprod{C}{\words[\le 1]})}{W} \xto{t} \words \xto{\malg{\aut{\specname}}} \specname \xto{\finmor{\specname}} \outobj. \]
Equality of the left hand side follows from the commuting diagram below, using \autoref{prop:malg-append} and \autoref{lem:append-snoc}.
\[\begin{tikzcd}[column sep = 3.5em]
	{\tprod{(\tprod{C}{\abcobj})}{W}} && {\tprod{(\tprod{C}{\words[\le 1]})}{W}} \\
	{\tprod{(\tprod{\words}{\abcobj})}{\words}} && {\tprod{(\tprod{\words}{\words})}{\words}} \\
	&& {\tprod{\words}{\words}} & \words \\
	&& {\tprod{\specname}{\words}} & \specname
	\arrow["{\tprod{(\tprod{\id{C}}{\minj{1} \inv{\lunit{\abcobj}}})}{\id{W}}}", from=1-1, to=1-3]
	\arrow["{\tprod{(\tprod{c}{\id{\abcobj}})}{w}}"', from=1-1, to=2-1]
	\arrow["{\tprod{(\tprod{c}{\wincl{1}})}{w}}"', from=1-3, to=2-3]
	\arrow["t", curve={height=-12pt}, from=1-3, to=3-4]
	\arrow["{\tprod{(\tprod{\id{\words}}{\minj{1} \inv{\lunit{\abcobj}}})}{\id{\words}}}", from=2-1, to=2-3]
	\arrow["{\tprod{\snoc}{\id{\words}}}"', from=2-1, to=3-3]
	\arrow["{\tprod{\append}{\id{\words}}}"', from=2-3, to=3-3]
	\arrow["\append", from=3-3, to=3-4]
	\arrow["{\tprod{\malg{\aut{\specname}}}{\id{\words}}}"', from=3-3, to=4-3]
	\arrow["{\malg{\aut{\specname}}}", from=3-4, to=4-4]
	\arrow["{\transmor*{\specname}}"', from=4-3, to=4-4]
\end{tikzcd}\]
Equality of the right hand side is analogous, using also that $\finmor{\specname} \circ \malg{\aut{\specname}} \circ t = \finmor{\implname} \circ \malg{\aut{\implname}} \circ t$ (since $\agreets{t}{\aut{\specname}}{\aut{\implname}}$).

\item Taking adjuncts, we need to show
\begin{multline*}
\ev{\words}{\outobj} \circ (\tprod{\id{\langs}}{w}) \circ (\tprod{(\mcoalg{\aut{\specname}} \circ \malg{\aut{\specname}} \circ c)}{\id{W}})  = \\
\ev{\words}{\outobj} \circ (\tprod{\id{\langs}}{w}) \circ (\tprod{(\mcoalg{\aut{\implname}} \circ \malg{\aut{\implname}} \circ c)}{\id{W}}).
\end{multline*}
Moving $\tprod{\id{\langs}}{w}$ to the end, this is equivalent to
\begin{multline*}
\ev{\words}{\outobj} \circ (\tprod{\mcoalg{\aut{\specname}}}{\id{\words}}) \circ (\tprod{\malg{\aut{\specname}}}{\id{\words}}) \circ (\tprod{c}{w}) = \\
\ev{\words}{\outobj} \circ (\tprod{\mcoalg{\aut{\implname}}}{\id{\words}}) \circ (\tprod{\malg{\aut{\implname}}}{\id{\words}}) \circ (\tprod{c}{w}).
\end{multline*}
Using \autoref{prop:malg-mcoalg-transmor} (ii) and that $\ev{\words}{\outobj} \circ (\tprod{h}{\id{\words}}) = \adj*{\words}{h}$ for any $h \colon Q \to \langs$, this is equivalent to
\[ \finmor{\specname} \circ \transmor*{\specname} \circ (\tprod{\malg{\aut{\specname}}}{\id{\words}}) \circ (\tprod{c}{w}) = \\
\finmor{\implname} \circ \transmor*{\implname} \circ (\tprod{\malg{\aut{\implname}}}{\id{\words}}) \circ (\tprod{c}{w}). \]

We show that both sides of the equation are equal to the composite
\[ \tprod{C}{W} \xto{\tprod{\inv{\runit{C}}}{\id{W}}} \tprod{(\tprod{C}{\tunit})}{W} \xto{\tprod{(\tprod{\id{C}}{\minj{0}})}{\id{W}}} \tprod{(\tprod{C}{\words[\le 1]})}{W} \xto{t} \words \xto{\malg{\aut{\specname}}} \specname \xto{\finmor{\specname}} \outobj. \]
Equality of the left hand side follows from the commuting diagram below, using \autoref{prop:malg-append} and \autoref{prop:words-monoid}.
\[\hspace{-0.15cm}\begin{tikzcd}
	{\tprod{C}{W}} & {\tprod{(\tprod{C}{\tunit})}{W}} && {\tprod{(\tprod{C}{\words[\le 1]})}{W}} \\
	{\tprod{\words}{W}} & {\tprod{(\tprod{\words}{\tunit})}{W}} && {\tprod{(\tprod{\words}{\words})}{\words}} \\
	{\tprod{\words}{\words}} & {\tprod{(\tprod{\words}{\tunit})}{\words}} && {\tprod{\words}{\words}} & \words \\
	&&& {\tprod{\specname}{\words}} & \specname \\
	&&& {}
	\arrow["{\tprod{\inv{\runit{C}}}{\id{W}}}", from=1-1, to=1-2]
	\arrow["{\tprod{c}{\id{W}}}", from=1-1, to=2-1]
	\arrow["{\tprod{c}{w}}"', curve={height=30pt}, from=1-1, to=3-1]
	\arrow["{\tprod{(\tprod{\id{C}}{\minj{0}})}{\id{W}}}", from=1-2, to=1-4]
	\arrow["{\tprod{(\tprod{c}{\id{\tunit}})}{\id{W}}}"', from=1-2, to=2-2]
	\arrow["{\tprod{(\tprod{c}{\wincl{1}})}{w}}"', from=1-4, to=2-4]
	\arrow["t", curve={height=-12pt}, from=1-4, to=3-5]
	\arrow["{\tprod{\inv{\runit{\words}}}{\id{W}}}"', from=2-1, to=2-2]
	\arrow["{\tprod{\id{\words}}{w}}", from=2-1, to=3-1]
	\arrow["{\tprod{\id{\tprod{\words}{\tunit}}}{w}}"', from=2-2, to=3-2]
	\arrow["{\tprod{\append}{\id{\words}}}"', from=2-4, to=3-4]
	\arrow["{\tprod{\inv{\runit{\words}}}{\id{\words}}}"', from=3-1, to=3-2]
	\arrow["{\tprod{(\tprod{\id{\words}}{\minj{0}})}{\id{\words}}}"{description}, from=3-2, to=2-4]
	\arrow["{\tprod{\runit{\words}}{\id{\words}}}"', from=3-2, to=3-4]
	\arrow["\append", from=3-4, to=3-5]
	\arrow["{\tprod{\malg{\aut{\specname}}}{\id{\words}}}"', from=3-4, to=4-4]
	\arrow["{\malg{\aut{\specname}}}", from=3-5, to=4-5]
	\arrow["{\transmor*{\specname}}"', from=4-4, to=4-5]
\end{tikzcd}\]
Equality of the right hand side is analogous, using also that $\finmor{\specname} \circ \malg{\aut{\specname}} \circ t = \finmor{\implname} \circ \malg{\aut{\implname}} \circ t$ (since $\agreets{t}{\aut{\specname}}{\aut{\implname}}$). \qed
\end{enum}
\end{proof}

\begin{proof}[\autoref{thm:main}]
Let $\mprod{r_1}{r_2} \colon C \to \specname \times \implname$ be defined as the morphism $\mprod{\malg{\aut{\specname}} \circ c}{\malg{\aut{\implname}} \circ c}$. By \autoref{lem:thm-main-bisim-1}, \autoref{lem:thm-main-bisim-2}, and \autoref{prop:bisim-coalgfunc}, $\mprod{r_1}{r_2}$ is an AM-bisimulation between $(\specname, \mprod{\finmor{\specname}}{\adj{\abcobj}{\transmor{\specname}}})$ and $(\implname, \mprod{\finmor{\implname}}{\adj{\abcobj}{\transmor{\implname}}}$. Since $c$ contains the empty word, there is a morphism $i_C \colon \tunit \to C$ such that $c \circ i_C = \minj{0}$. Then we get the following commutative diagram
\[\begin{tikzcd}
	\tunit \\
	& C & \words & \implname \\
	& \words \\
	& \specname && \langs
	\arrow["{i_C}"{description}, from=1-1, to=2-2]
	\arrow["{\minj{0}}", from=1-1, to=2-3]
	\arrow["{\initmor{\implname}}", curve={height=-18pt}, from=1-1, to=2-4]
	\arrow["{\minj{0}}"', from=1-1, to=3-2]
	\arrow["{\initmor{\specname}}"', curve={height=18pt}, from=1-1, to=4-2]
	\arrow["c", from=2-2, to=2-3]
	\arrow["c"', from=2-2, to=3-2]
	\arrow["{\malg{\aut{\implname}}}", from=2-3, to=2-4]
	\arrow["{\mcoalg{\aut{\implname}}}", from=2-4, to=4-4]
	\arrow["{\malg{\aut{\specname}}}"', from=3-2, to=4-2]
	\arrow["{\mcoalg{\aut{\specname}}}"', from=4-2, to=4-4]
\end{tikzcd}\]
where the square commutes due to \autoref{prop:bisim-fin-coalg} and \autoref{prop:lang-fin-coalg}. Thus, $\reclang{\aut{\specname}}' = \mcoalg{\aut{\specname}} \circ \initmor{\specname} = \mcoalg{\aut{\implname}} \circ \initmor{\implname} = \reclang{\aut{\implname}}'$. By \autoref{prop:rec-lang-equiv}, this is equivalent to $\reclang{\aut{\specname}} = \reclang{\aut{\implname}}$. Hence, $\autequiv{\aut{\specname}}{\aut{\implname}}$, as desired. \qed
\end{proof}

\subsection{Proof of \autoref{cor:gen-W-comp}}

\begin{proposition} \label{prop:concat-assoc}
For all $f \colon X \to \words$, $g \colon Y \to \words$, and $h \colon Z \to \words$, we have
\[ (f \cdot g) \cdot h = (f \cdot (g \cdot h)) \assoc{X}{Y}{Z}. \]
\end{proposition}
\begin{proof}
We calculate using the definition of $\cdot$ and \autoref{prop:words-monoid}.
\begin{align*}
(f \cdot g) \cdot h &= \append (\tprod{f \cdot g}{h}) = \append (\tprod{\append (\tprod{f}{g})}{h})
     = \append (\tprod{\append}{\id{\words}}) (\tprod{(\tprod{f}{g})}{h}) \\
    &= \append (\tprod{\id{\words}}{\append}) \assoc{\words}{\words}{\words} (\tprod{(\tprod{f}{g})}{h}) \\
    &= \append (\tprod{\id{\words}}{\append}) (\tprod{f}{(\tprod{g}{h})}) \assoc{X}{Y}{Z} \\
    &= \append (\tprod{f}{\append (\tprod{g}{h})}) \assoc{X}{Y}{Z} = \append (\tprod{f}{g \cdot h}) \assoc{X}{Y}{Z} = (f \cdot (g \cdot h)) \assoc{X}{Y}{Z}
\end{align*}
\qed
\end{proof}

For the next proposition, we need the morphism $\append*[k,l] \colon \tprod{\words[\le k]}{\words[\le l]} \to \words[\le k+l]$ defined as the composite
\[\hspace{-0.5cm} \tprod{\words[\le k]}{\words[\le l]} \xto{\cong} \sum_{m \le k, n \le l}{\tprod{\words[m]}{\words[n]}} \xto{\sum_{m \le k, n \le l}{\append[m,n]}} \sum_{m \le k, n \le l}{\words[m+n]} \xto{\mfcoprod{m \le k, n \le l}{\minj{m+n}}} \words[\le k+l]. \]
\begin{proposition} \label{prop:combine-wincl}
For all $k, l \in \N$, we have $\wincl{k} \cdot \wincl{l} = \wincl{k+l} \append*[k,l]$.
\end{proposition}
\begin{proof}
Since $\mfcoprod{m \le k, n \le l}{\tprod{\minj{m}}{\minj{n}}} \colon \sum_{m \le k, n \le l}{\tprod{\words[m]}{\words[n]}} \to \tprod{\words[\le k]}{\words[\le l]}$ is an isomorphism, $\wincl{k} \cdot \wincl{l} = \wincl{k+l} \append*[k,l]$ if and only if
\[ (\wincl{k} \cdot \wincl{l}) \circ \mfcoprod{m \le k, n \le l}{\tprod{\minj{m}}{\minj{n}}} = \wincl{k+l} \circ \mfcoprod{m \le k, n \le l}{\minj{m+n}} \circ \sum_{m \le k, n \le l}{\append[m,n]}. \]
This holds if and only if the two sides become equal after precomposing with $\minj{m,n}$ for all $m, n \in \N$. The latter is true since
\begin{align*}
(\wincl{k} \cdot \wincl{l}) \mfcoprod{m \le k, n \le l}{\tprod{\minj{m}}{&\minj{n}}} \circ \minj{m,n}
	= (\wincl{k} \cdot \wincl{l}) (\tprod{\minj{m}}{\minj{n}}) \\
	&= \append (\tprod{\wincl{k}}{\wincl{l}}) (\tprod{\minj{m}}{\minj{n}}) = \append (\tprod{\minj{m}}{\minj{n}}) = \minj{m+n} \append[m,n]
\end{align*}
using \autoref{lem:snoc-cons-append-comp}, and
\begin{align*}
 \wincl{k+l} \mfcoprod{m \le k, n \le l}{\minj{m+n}} &\circ \sum_{m \le k, n \le l}{\append[m,n]} \circ \minj{m,n}
	= \wincl{k+l} \mfcoprod{m \le k, n \le l}{\minj{m+n}} \circ \minj{m,n} \append[m,n] \\
	&= \wincl{k+l} \minj{m+n} \append[m,n] = \minj{m+n} \append[m,n].
\end{align*}
\qed
\end{proof}

\begin{lemma} \label{lem:w-rearrange}
For all $p \colon P \to \words$, $w \colon W \to \words$, and $k \in \N$, we have
\[ \mW{k}{p}{w} (\tprod{(\tprod{\id{P}}{\append*[k,1]}) \assoc{P}{\words[\le k]}{\words[\le 1]}}{\id{W}}) = p \cdot \wincl{k} \cdot \wincl{1} \cdot w. \]
\end{lemma}
\begin{proof}
We calculate using \autoref{prop:concat-assoc} and \autoref{prop:combine-wincl}.
\begin{align*}
p \cdot \wincl{k} \cdot \wincl{1}
    &= (p \cdot (\wincl{k} \cdot \wincl{1})) \assoc{P}{\words[\le k]}{\words[\le 1]}
     = (p \cdot \wincl{k+1} \append*[k,1]) \assoc{P}{\words[\le k]}{\words[\le 1]} \\
    &= \append (\tprod{p}{\wincl{k+1} \append*[k,1]}) \assoc{P}{\words[\le k]}{\words[\le 1]}
     = \append (\tprod{p}{\wincl{k+1}}) (\tprod{\id{P}}{\append*[k,1]}) \assoc{P}{\words[\le k]}{\words[\le 1]} \\
    &= (p \cdot \wincl{k+1}) (\tprod{\id{P}}{\append*[k,1]}) \assoc{P}{\words[\le k]}{\words[\le 1]}
\end{align*}
Thus, we get
\begin{align*}
\mW{k}{p}{w} (\tprod{(\tprod{\id{P}}{\append*[k,1]}) &\assoc{P}{\words[\le k]}{\words[\le 1]}}{\id{W}}) \\
    &= (p \cdot \wincl{k+1} \cdot w) (\tprod{(\tprod{\id{P}}{\append*[k,1]}) \assoc{P}{\words[\le k]}{\words[\le 1]}}{\id{W}}) \\
    &= \append (\tprod{p \cdot \wincl{k+1}}{w}) (\tprod{(\tprod{\id{P}}{\append*[k,1]}) \assoc{P}{\words[\le k]}{\words[\le 1]}}{\id{W}}) \\
    &= \append (\tprod{(p \cdot \wincl{k+1}) (\tprod{\id{P}}{\append*[k,1]}) \assoc{P}{\words[\le k]}{\words[\le 1]}}{w}) \\
    &= \append (\tprod{p \cdot \wincl{k} \cdot \wincl{1}}{w}) = p \cdot \wincl{k} \cdot \wincl{1} \cdot w.
\end{align*}
\qed
\end{proof}

\begin{proof}[\autoref{cor:gen-W-comp}]
Let $\aut{\implname} \in \fdomgcomp[k]{p}$, and suppose $\agreets{\mW{k}{p}{w}}{\aut{\specname}}{\aut{\implname}}$. Then we have $\reclang{\aut{\specname}} \circ \mW{k}{p}{w} = \reclang{\aut{\implname}} \circ \mW{k}{p}{w}$ by definition. Precomposing this equation with the morphism $\tprod{(\tprod{\id{P}}{\append*[k,1]})\assoc{P}{\words[\le k]}{\words[\le 1]}}{\id{W}}$, we obtain $\reclang{\specname} (p \cdot \wincl{k} \cdot \wincl{1} \cdot w) = \reclang{\implname} (p \cdot \wincl{k} \cdot \wincl{1} \cdot w)$ by \autoref{lem:w-rearrange}. That is, writing $t = p \cdot \wincl{k} \cdot \wincl{1} \cdot w$, we have $\agreets{t}{\specname}{\implname}$. By the definition of $\fdomgcomp[k]{p}$, there exists a $\delta \colon \tprod{(\tprod{P}{\words[k]})}{\abcobj} \to \tprod{P}{\words[k]}$ such that $(p \cdot \wincl{k}, \delta)$ is a weak state cover for $\aut{\implname}$. Thus, by \autoref{thm:main}, we get $\autequiv{\aut{\specname}}{\aut{\implname}}$. \qed
\end{proof}

\section{Details and proofs for \autoref{sec:applications}} \label{sec:applications-details}

\subsection{Equivalent characterizations of WAs} \label{sec:wa-equiv-char}

To show the correspondence of WAs and automata in $\Vect{\fld}$, it is convenient to introduce linear weighted automata~\cite{DBLP:journals/iandc/BonchiBBRS12,DBLP:conf/concur/Boreale09}. They can be seen as determinized versions of weighted automata, where the state space consists of linear combinations of states. In this section, $\abcobj$ denotes a finite set.
\begin{definition}
A \emph{linear weighted automaton} (or LWA for short) is a tuple $(V, v_0, (\delta_a)_{a \in \abcobj}, f)$, where $V$ is a finite-dimensional vector space called the \emph{state space}, $v_0 \in V$ is an \emph{initial state}, $(\delta_a \colon V \to V)_{a \in \abcobj}$ is a family of linear maps called the \emph{transition maps}, and $f \colon V \to \fld$ is a linear map called the \emph{output map}.
\end{definition}
For $\aut{A} = (V, v_0, (\delta_a)_{a \in \abcobj}, f)$, we write $v \in \aut{A}$ for $v \in V$.

For an LWA $\aut{A} = (V, v_0, (\delta_a)_{a \in \abcobj}, f)$ and word $w \in \words$, define $\delta_w \colon V \to V$ by recursion on $w$: $\delta_\epsilon = \id{V}$ and $\delta_{wa} = \delta_a \circ \delta_w$. The \emph{recognized language} $\reclang{\aut{A}}[v]$ of a state $v \in V$ is defined as $\reclang{\aut{A}}[v](w) = f(\delta_w(v))$. The \emph{recognized language} of $\aut{A}$ is $\reclang{\aut{A}} = \reclang{\aut{A}}[v_0]$.

We claim that WAs and LWAs are equivalent. Given a WA $(Q, s_0, \delta, f)$, we can construct an associated LWA as follows. Take the free vector space $V = \funcset{Q}{\fld}$ generated by $Q$ as the state space, define the initial state as $v_0 = s_0$, define the transition maps $\delta_a$ via matrix multiplication as $\delta_a(v) = \wamatrix{\delta}{a} v$, and define the output map $\overline{f}$ as $\overline{f}(s) = \transp{f} s$. Then the recognized language of $(V, v_0, (\delta_a)_{a \in \abcobj}, \overline{f})$ coincides with the recognized language of $(Q, s_0, \delta, f)$.

Conversely, given an LWA $(V_0, v_0, (\delta_a)_{a \in \abcobj}, f)$, we choose a basis $Q \subseteq V$ of $V$. Then the set $Q$ is the state space of the corresponding WA, and the initial weight function $s_0$ consists of the coordinates of $v_0$ in the basis $Q$. The transition function $\delta$ is given by $\delta(p, a, q) = \wamatrix{\delta}{a}(q, p)$, where $\wamatrix{\delta}{a}$ is the matrix corresponding to the linear map $\delta_a$ in the basis $Q$. Finally, the output weight function $\dot{f}$ is the restriction of $f$ to $Q$. Again, the recognized language of $(V_0, v_0, (\delta_a)_{a \in \abcobj}, f)$ is the same as that of $(Q, s_0, \delta, \dot{f})$.

Moreover, LWAs correspond to automata in $\Vect{\fld}$. To see this, note that a bilinear map $\transmor{V}' \colon V \times \abcobj' \to V$ is completely determined by its values on the basis vectors. Thus, such a $\transmor{V}'$ corresponds to a family of linear maps $\delta_a = \transmor{V}'(-, a) \colon V \to V$ indexed by $a \in \abcobj$. Furthermore, the initial state morphism $\initmor{V}' \colon \fld \to V$ corresponds to an element $v_0 = \initmor{V}'(1)$. In the following, we identify WAs, LWAs, and automata in $\Vect{\fld}$ via the correspondences given above.

\subsection{Proof of \autoref{prop:wa-defs-wla}}

Throughout the proof, we denote by $(\funcset{Q}{\fld}, \initmor{Q}, \transmor{Q}, \finmor{Q})$ the automaton in $\Vect{\fld}$ corresponding to the WA $\aut{A}$. Recall that $\initmor{Q} \colon \fld \to \funcset{Q}{\fld}$ is defined as $\initmor{Q}(k) = ks_0$, that $\transmor{Q} \colon \tprod{\funcset{Q}{\fld}}{\abcobj'} \to \funcset{Q}{\fld}$ corresponds to the bilinear extension of the map $(s, a) \mapsto \wamatrix{\delta}{a} s$, and that $\finmor{Q} \colon \funcset{Q}{\fld} \to \fld$ is defined as $\finmor{Q}(s) = \transp{f} s$.

\begin{enum}
\item We only need to check that $\append(\tprod{u}{v}) = u v$ for all $u, v \in \words$. Recall that $\append$ is the composite
\[ \tprod{\words}{\words} \xto{\cong} \bigoplus_{m,n \in \N}{\tprod{\words[m]}{\words[n]}} \xto{\bigoplus_{m,n \in \N}{\append[m,n]}} \bigoplus_{m,n \in \N}{\words[m+n]} \xto{\mfcoprod{m,n \in \N}{\minj{m+n}}} \words. \]
Writing $m = \card{u}$ and $n = \card{v}$, we have $\append(\tprod{u}{v}) = \append[m,n](\tprod{u}{v})$. Thus, we need to prove that $\append[m,n](\tprod{u}{v}) = u v$ for all $m, n \in \N$ and $u \in \words[m], v \in \words[n]$. We do this by induction on $n$.
\begin{items}
\item Case $0$: then $v$ is of the form $\epsilon$. Recall that $\append[m,0] = \runit{(\abcobj')^m}$. Thus,
\[ \append[m,0](\tprod{u}{\epsilon}) = \runit{(\abcobj')^m}(\tprod{u}{1}) = u \cdot 1 = u = u \epsilon. \]

\item Case $n+1$: then $v$ is of the form $va$ for $v \in \words[n], a \in \abcobj$. Recall that $\append[m,n+1] = (\tprod{\append[m,n]}{\id{\abcobj'}}) \circ \inv{\assoc{(\abcobj')^m}{(\abcobj')^n}{\abcobj'}}$. Thus,
\begin{align*}
\append[m,n+1](\tprod{u}{va}) &= (\tprod{\append[m,n]}{\id{\abcobj'}})(\inv{\assoc{(\abcobj')^m}{(\abcobj')^n}{\abcobj'}}(\tprod{u}{(\tprod{v}{a})})) \\
	&= (\tprod{\append[m,n]}{\id{\abcobj'}})(\tprod{(\tprod{u}{v})}{a})
 	 = \tprod{\append[m,n](\tprod{u}{v})}{a} \\
	&= \tprod{uv}{a} = (uv)a = u(va).
\end{align*}
\end{items}

\item We only need to prove $\malg{\aut{A}}(w) = \wamatrix{\delta}{w} s_0$. Since $\malg{\aut{A}} = \mfcoprod{n \in \N}{\malg{\aut{A},n}}$, we have $\malg{\aut{A}}(w) = \malg{\aut{A},n}(w)$ for $n = \card{w}$. Thus, we need to prove that $\malg{\aut{A},n}(w) = \wamatrix{\delta}{w} s_0$ for all $n \in \N$ and $w \in \words[n]$. We do this by induction on $n$.
\begin{items}
\item Case $0$: then $w = \epsilon$. Recall that $\malg{\aut{A},0} = \initmor{Q}$. Thus, we have
\[ \malg{\aut{A},0}(\epsilon) = \initmor{Q}(1) = 1 s_0 = s_0 = \wamatrix{\delta}{\epsilon} s_0. \]

\item Case $n+1$: then $w$ is of the form $wa$ for some $w \in \words[n], a \in \abcobj$. Recall that $\malg{\aut{A},n+1} = \transmor{Q} (\tprod{\malg{\aut{A},n}}{\id{\abcobj'}})$. Thus, we have
\begin{align*}
\malg{\aut{A},n+1}(wa)
	&= \transmor{Q}((\tprod{\malg{\aut{A},n}}{\id{\abcobj'}})(\tprod{w}{a})) = \transmor{Q} (\tprod{\malg{\aut{A},n}(w)}{a}) \\
	&= \transmor{Q}(\tprod{\wamatrix{\delta}{w} s_0}{a}) = \wamatrix{\delta}{a} (\wamatrix{\delta}{w} s_0) = \wamatrix{\delta}{wa} s_0.
\end{align*}
\end{items}

\item The recognized language of $\aut{A}$ in the categorical sense is the composite $\finmor{Q} \malg{\aut{A}}$. Hence, it suffices to show that $\finmor{Q}(\malg{\aut{A}}(w)) = \reclang{\aut{A}}(w)$. This holds due to following calculation using the previous part.
\[ \finmor{Q}(\malg{\aut{A}}(w)) = \finmor{Q}(\wamatrix{\delta}{w} s_0) = \transp{f} \wamatrix{\delta}{w} s_0 = \reclang{\aut{A}}(w) \]

\item Recall that $\mcoalg{\aut{A}}$ is the adjunct of the composite
\[ \tprod{\funcset{Q}{\fld}}{(\abcobj')^*} \xto{\cong} \bigoplus_{n \in \N}{\tprod{\funcset{Q}{\fld}}{(\abcobj')^n}} \xto{\mfcoprod{n \in \N}{\mcoalg{\aut{A},n}}} \fld. \]
Thus, $\mcoalg{\aut{A}}(s)(w) = \mcoalg{\aut{A},n}(\tprod{s}{w})$, where $n = \card{w}$. Hence, we need to prove that $\mcoalg{\aut{A},n}(\tprod{s}{w}) = \reclang{\aut{A}}[s](w)$ for all $n \in \N, s \in \funcset{Q}{\fld}$, and $w \in \words[n]$. We do this by induction on $n$.
\begin{items}
\item Case $0$: then $w = \epsilon$. Recall that $\mcoalg{\aut{A},0} = \finmor{Q} \runit{\funcset{Q}{\fld}}$. Hence,
\begin{align*}
\mcoalg{\aut{A},0}(\tprod{s}{\epsilon})
	&= \finmor{Q}(\runit{\funcset{Q}{\fld}}(\tprod{s}{1})) = \finmor{Q}(1 s) \\
	&= \finmor{Q}(s) = \transp{f} s = \transp{f} \wamatrix{\delta}{\epsilon} s = \reclang{\aut{A}}[s](\epsilon).
\end{align*}

\item Case $n+1$: then $w$ is of the form $aw$ for some $a \in \abcobj, w \in \words[n]$. Recall that $\mcoalg{\aut{A},n+1} = \mcoalg{\aut{A},n} (\tprod{\transmor{Q}}{\id{(\abcobj')^n}}) \inv{\assoc{\funcset{Q}{\fld}}{\abcobj'}{(\abcobj')^n}} (\tprod{\id{\funcset{Q}{\fld}}}{\inv{\cons[n]}})$. Thus, we have
\begin{align*}
\mcoalg{\aut{A},n+1}&(\tprod{s}{aw}) \\
	&= \mcoalg{\aut{A},n}((\tprod{\transmor{Q}}{\id{(\abcobj')^n}})(\inv{\assoc{\funcset{Q}{\fld}}{\abcobj'}{(\abcobj')^n}}((\tprod{\id{\funcset{Q}{\fld}}}{\inv{\cons[n]}})(\tprod{s}{aw})))) \\
	&= \mcoalg{\aut{A},n}((\tprod{\transmor{Q}}{\id{(\abcobj')^n}})(\inv{\assoc{\funcset{Q}{\fld}}{\abcobj'}{(\abcobj')^n}}(\tprod{s}{(\tprod{a}{w})}))) \\
	&= \mcoalg{\aut{A},n}((\tprod{\transmor{Q}}{\id{(\abcobj')^n}})(\tprod{(\tprod{s}{a})}{w}))
	= \mcoalg{\aut{A},n}(\tprod{\transmor{Q}(\tprod{s}{a})}{w}) \\
	&= \mcoalg{\aut{A},n}(\tprod{\wamatrix{\delta}{a} s}{w}) = \reclang{\aut{A}}[\wamatrix{\delta}{a} s](w) = \transp{f} \wamatrix{\delta}{w} (\wamatrix{\delta}{a} s) \\
	&= \transp{f} \wamatrix{\delta}{aw} s = \reclang{\aut{A}}[s](aw). 
\end{align*}
\end{items}

\item Recall that the corresponding automaton in $\Vect{\fld}$ is minimal if and only if $\mcoalg{\aut{A}}$ is a monomorphism. Since in $\Vect{\fld}$, monomorphisms coincide with injective linear maps, this holds if and only if $\mcoalg{\aut{A}}(s) = 0$ implies $s = 0$. But since $\mcoalg{\aut{A}}(s)$ is the linear extension of $\reclang{\aut{A}}[s]$ by the previous part, $\mcoalg{\aut{A}}(s) = 0$ if and only if $\reclang{\aut{A}}[s] = 0$. \qed
\end{enum}

\subsection{Proof of \autoref{prop:wa-agree-ts}}

By definition, $\agreets{t}{\aut{\specname}}{\aut{\implname}}$ if and only if $\reclang{\aut{\specname}} \circ t = \reclang{\aut{\implname}} \circ t$. Since $t$ is a subspace inclusion, this is equivalent to $\reclang{\aut{\specname}}|_{\vspan{T}} = \reclang{\aut{\implname}}|_{\vspan{T}}$. Since $\vspan{T}$ is generated by $T$, we need only check equality on the generators. Thus, the latter is equivalent to $\reclang{\aut{\specname}}|_T = \reclang{\aut{\implname}}|_T$. \qed

\subsection{Proof of \autoref{prop:wa-sc-char}}

\begin{lemma} \label{lem:wa-empty-word}
Let $A \subseteq \words$ and let $a \colon \vspan{A} \to (\abcobj')^*$ be the corresponding subspace inclusion. Then $a$ contains the empty word if and only if $\epsilon \in A$.
\end{lemma}
\begin{proof}
For any linear map $i_A \colon \tunit \to \vspan{A}$, we have $ai_A = \minj{0}$ if and only if $i_A(k) = k\epsilon$ for any $k \in \fld$, if and only if $i_A(1) = \epsilon$. Thus, $a$ contains the empty word if and only if $\epsilon \in \vspan{A}$, which holds if and only if $\epsilon \in A$. \qed
\end{proof}

\begin{proof}[\autoref{prop:wa-sc-char}]
\begin{enum}
\item Let $p \colon \vspan{P} \to (\abcobj')^*$ be the subspace inclusion. By \autoref{lem:wa-empty-word}, $p$ contains the empty word if and only if $\epsilon \in P$. Furthermore, we have
\begin{align*}
	\malg{\aut{\specname}} \circ p&\text{ is a split epi}
	 \iff \malg{\aut{\specname}} \circ p\text{ is an epi}
	 \iff \malg{\aut{\specname}} \circ p\text{ is surjective} \\
	&\iff \forall s \in \funcset{Q}{\fld}. \exists \omega \in \vspan{P}.\ \malg{\aut{\specname}}(\omega) = s \\
	&\iff \forall s \in \funcset{Q}{\fld}. \exists k_i \in \fld, w_i \in P\ (i = 1, \ldots ,n).\ \malg{\aut{\specname}}\left(\sum_{i=1}^n{k_i w_i}\right) = s \\
	&\iff \forall s \in \funcset{Q}{\fld}. \exists k_i \in \fld, w_i \in P\ (i = 1, \ldots ,n).\ \sum_{i=1}^n{k_i \malg{\aut{\specname}}(w_i)} = s \\
	&\iff \forall s \in \funcset{Q}{\fld}. \exists k_i \in \fld, w_i \in P\ (i = 1, \ldots ,n).\ \sum_{i=1}^n{k_i (\wamatrix{\delta}{w_i} s_0)} = s \\
	&\iff \funcset{Q}{\fld}\text{ is generated by $\setof{\wamatrix{\delta}{w} s_0}{w \in P}$}
\end{align*}
using that every epi in the category $\Vect{\fld}$ is split and \autoref{prop:wa-defs-wla} (ii).

\item Let $w \colon \vspan{W} \to (\abcobj')^*$ be the subspace inclusion. By \autoref{lem:wa-empty-word}, $w$ contains the empty word if and only if $\epsilon \in W$. By \autoref{rem:char-mor-def}, it remains to check that the kernel pair $(e_1^w, e_2^w)$ of $\inthom{w}{\id{\fld}} \circ \mcoalg{\aut{\specname}}$ is included in the kernel pair $(e_1, e_2)$ of $\mcoalg{\aut{\specname}}$. In this case, the domain of the kernel pair of $f \colon X \to Y$ is the subspace $\setof{(x_1,x_2) \in X \times X}{f(x_1) = f(x_2)}$ and the kernel pair itself is the pair of projection morphisms. Thus, $(e_1^w, e_2^w)$ factors through $(e_1, e_2)$ if and only if
\[ \setof{(s_1, s_2)}{\inthom{w}{\id{\fld}}(\mcoalg{\aut{\specname}}(s_1)) = \inthom{w}{\id{\fld}}(\mcoalg{\aut{\specname}}(s_2))} \subseteq \setof{(s_1, s_2)}{\mcoalg{\aut{\specname}}(s_1) = \mcoalg{\aut{\specname}}(s_2)}. \]
We have, for all $s_1, s_2 \in \funcset{Q}{\fld}$, that
\begin{align*}
\inthom{w}{\id{\fld}}&(\mcoalg{\aut{\specname}}(s_1)) = \inthom{w}{\id{\fld}}(\mcoalg{\aut{\specname}}(s_2)) \\
&\iff \mcoalg{\aut{\specname}}(s_1) \circ w = \mcoalg{\aut{\specname}}(s_2) \circ w \\
&\iff \mcoalg{\aut{\specname}}(s_1)|_{\vspan{W}} = \mcoalg{\aut{\specname}}(s_2)|_{\vspan{W}} \\
&\iff \forall \omega \in \vspan{W}.\ \mcoalg{\aut{\specname}}(s_1)(\omega) = \mcoalg{\aut{\specname}}(s_2)(\omega) \\
&\iff \forall w \in W.\ \mcoalg{\aut{\specname}}(s_1)(w) = \mcoalg{\aut{\specname}}(s_2)(w) \\
&\iff \forall w \in W.\ \reclang{\aut{\specname}}[s_1](w) = \reclang{\aut{\specname}}[s_2](w) \\
&\iff \reclang{\aut{\specname}}[s_1]|_W = \reclang{\aut{\specname}}.[s_2]|_W
\end{align*}
We also have
\begin{align*}
\mcoalg{\aut{\specname}}(s_1) = \mcoalg{\aut{\specname}}(s_2)
&\iff \forall \omega \in (\abcobj')^*.\ \mcoalg{\aut{\specname}}(s_1)(\omega) = \mcoalg{\aut{\specname}}(s_2)(\omega) \\
&\iff \forall w \in \words.\ \mcoalg{\aut{\specname}}(s_1)(w) = \mcoalg{\aut{\specname}}(s_2)(w) \\
&\iff \forall w \in \words.\ \reclang{\aut{\specname}}[s_1](w) = \reclang{\aut{\specname}}[s_2](w) \\
&\iff \reclang{\aut{\specname}}[s_1] = \reclang{\aut{\specname}}[s_2]
\end{align*}
using that $\vspan{W}$ is generated by $W$, that $(\abcobj')^*$ is generated by $\words$, and \autoref{prop:wa-defs-wla} (iv). Thus, $w$ is a characterization morphism if and only if $\epsilon \in W$ and
\[ \forall s_1, s_2 \in \funcset{Q}{\fld}.\ \reclang{\aut{\specname}}[s_1]|_W = \reclang{\aut{\specname}}[s_2]|_W \implies \reclang{\aut{\specname}}[s_1] = \reclang{\aut{\specname}}[s_2]. \]
Since $\reclang{\aut{\specname}} \colon \funcset{Q}{\fld} \to \funcset{\words}{\fld}$ is linear, the latter condition is equivalent to
\[ \forall s \in \funcset{Q}{\fld}.\ \reclang{\aut{\specname}}[s]|_W = 0 \implies \reclang{\aut{\specname}}[s] = 0. \]
\qed
\end{enum}
\end{proof}

\subsection{Proof of \autoref{lem:wa-ts-image}}

To simplify notation, we introduce concatenation of subspaces. Given two subspaces $X, Y \subseteq (\abcobj')^*$, denote by $X \cdot Y$ the subspace $\im{a \cdot b} \subseteq (\abcobj')^*$, where $a \colon X \to (\abcobj')^*$ and $b \colon Y \to (\abcobj')^*$ are the subspace inclusions and $\im{-}$ denotes the image of a map.

Concatenation of subspaces and concatenation of subsets is connected by the following proposition.
\begin{proposition} \label{prop:span-concat}
Let $A, B \subseteq \words$. Then $\vspan{A} \cdot \vspan{B} = \vspan{A \cdot B}$.
\end{proposition}
\begin{proof}
Let $a \colon \vspan{A} \to (\abcobj')^*$ and $b \colon \vspan{B} \to (\abcobj')^*$ denote the subspace inclusions. Then $\tprod{a}{b}$ is the inclusion of the subspace $\tprod{\vspan{A}}{\vspan{B}}$ into $\tprod{(\abcobj')^*}{(\abcobj')^*}$. We use the following two facts about spans and tensor products, which can be readily verified from the definitions.
\begin{enum}
\item If $f \colon X \to Y$ is a linear map and $A \subseteq X$, then $\im{f}[\vspan{A}] = \vspan{\im{f}[A]}$ (where $\im{f}[A] = \setof{f(x)}{x \in A}).$
\item If $A \subseteq X$ and $B \subseteq Y$ are two subsets of vector spaces $X$ and $Y$, then $\tprod{\vspan{A}}{\vspan{B}} = \vspan{\setof{\tprod{a}{b}}{a \in A, b \in B}}$ as subsets of $\tprod{X}{Y}$.
\end{enum}
With these ingredients at hand, the proposition follows from a straightforward calculation, using \autoref{prop:wa-defs-wla} (i).
\begin{align*}
\vspan{A} \cdot \vspan{B}
	&= \im{a \cdot b} = \im{\append \circ (\tprod{a}{b})} = \im{\append|_{\tprod{\vspan{A}}{\vspan{B}}}} \\
	&= \im{\append}[\tprod{\vspan{A}}{\vspan{B}}] = \im{\append}[\vspan{\setof{\tprod{a}{b}}{a \in A, b \in B}}] \\
	&= \vspan{\im{\append}[\setof{\tprod{a}{b}}{a \in A, b \in B}]} \\
	&= \vspan{\setof{\append(\tprod{a}{b})}{a \in A, b \in B}} \\
	&= \vspan{\setof{ab}{a \in A, b \in B}} = \vspan{A \cdot B}
\end{align*}
\qed
\end{proof}

\begin{lemma} \label{lem:image-concat}
Let $a \colon A \to (\abcobj')^*$, $b \colon B \to (\abcobj')^*$ be two morphisms, and let $d \colon \im{a} \to (\abcobj')^*$ be the inclusion. Then $\im{a \cdot b} = \im{d \cdot b}$.
\end{lemma}
\begin{proof}
We prove the inclusions in both directions.
\begin{enum}
\item Let $\omega \in \im{a \cdot b}$. Then there is a $\tau \in \tprod{A}{B}$ such that $\omega = (a \cdot b)(\tau)$, where $\tau$ can be written as $\tau = \sum_{i=1}^n{k_i (\tprod{x_i}{y_i})}$ for $k_i \in \fld$, $x_i \in A$, and $y_i \in B$. Thus, $\omega = (a \cdot b)(\tau) = \sum_{i=1}^n{k_i (a \cdot b)(\tprod{x_i}{y_i})}$. Hence, to show $\omega \in \im{d \cdot b}$, it suffices to show $(a \cdot b)(\tprod{x_i}{y_i}) \in \im{d \cdot b}$ for all $i = 1, \ldots, n$. We have
\begin{align*}
(a \cdot b)(\tprod{x_i}{y_i})
	&= \append((\tprod{a}{b})(\tprod{x_i}{y_i})) = \append(\tprod{a(x_i)}{b(y_i)}) \\
	&= \append(\tprod{d(a(x_i))}{b(y_i)}) = \append((\tprod{d}{b})(\tprod{a(x_i)}{y_i})) \\
	&= (d \cdot b)(\tprod{a(x_i)}{y_i}).
\end{align*}
Thus, $(a \cdot b)(\tprod{x_i}{y_i}) \in \im{d \cdot b}$.

\item Let $\omega \in \im{d \cdot b}$. Then there is a $\tau \in \tprod{\im{a}}{B}$ such that $\omega = (d \cdot b)(\tau)$, where $\tau$ can be written as $\tau = \sum_{i=1}^n{k_i (\tprod{t_i}{y_i})}$ for $k_i \in \fld$, $t_i \in \im{a}$, and $y_i \in B$. Thus, $\omega = (d \cdot b)(\tau) = \sum_{i=1}^n{k_i (d \cdot b)(\tprod{t_i}{y_i})}$. Hence, to show $\omega \in \im{a \cdot b}$, it suffices to show $(d \cdot b)(\tprod{t_i}{y_i}) \in \im{a \cdot b}$ for all $i = 1, \ldots, n$. Since $t_i \in \im{a}$, $t_i = a(x_i)$ for some $x_i \in A$. Then we have
\begin{align*}
(d \cdot b)(\tprod{t_i}{y_i})
	&= \append((\tprod{d}{b})(\tprod{a(x_i)}{y_i})) = \append(\tprod{d(a(x_i))}{b(y_i)}) \\
	&= \append(\tprod{a(x_i)}{b(y_i)}) = \append((\tprod{a}{b})(\tprod{x_i}{y_i})) \\
	&= (a \cdot b)(\tprod{x_i}{y_i}).
\end{align*}
Thus, $(d \cdot b)(\tprod{t_i}{y_i}) \in \im{a \cdot b}$. \qed
\end{enum}
\end{proof}

\begin{proof}[\autoref{lem:wa-ts-image}]
Note that for any $k \in \N$, $(\abcobj')^{\le k} = \vspan{\words[\le k]}$. Thus, $\wincl{k}$ is the inclusion of $\vspan{\words[\le k]}$ into $(\abcobj')^*$. Let $d \colon \vspan{P} \cdot \vspan{\words[\le k+1]} \to (\abcobj')^*$ be the inclusion of $\vspan{P} \cdot \vspan{\words[\le k+1]} = \im{p \cdot \wincl{k+1}}$. Then we have
\begin{align*}
\vspan{\W{k}{P}{W}} &= \vspan{P \cdot \words[\le k+1] \cdot W} = (\vspan{P} \cdot \vspan{\words[\le k+1]}) \cdot \vspan{W} \\
&= \im{d \cdot w} = \im{p \cdot \wincl{k+1} \cdot w} = \im{\mW{k}{p}{w}}
\end{align*}
using \autoref{prop:span-concat} and \autoref{lem:image-concat}. \qed
\end{proof}

\subsection{Proof of \autoref{thm:w-method-lwa}}

\begin{lemma} \label{lem:wa-fdom-incl}
Let $P \subseteq \words$, $k \in \N$, and let $p \colon \vspan{P} \to (\abcobj')^*$ be the corresponding subspace inclusion. Then we have $\fdomwa{k}{P} \subseteq \fdomgcomp[k]{p}$.
\end{lemma}
\begin{proof}
Let $\aut{\implname} \in \fdomwa{k}{P}$. Then $P \cdot \words[\le k]$ is a state cover for $\aut{\implname}$. Let us denote the image factorization of $p \cdot \wincl{k}$ by $\tprod{\vspan{P}}{(\abcobj')^{\le k}} \xto{f} \im{p \cdot \wincl{k}} \xto{g} (\abcobj')^*$. We have
\[ \vspan{P \cdot \words[\le k]} = \vspan{P} \cdot \vspan{\words[\le k]} = \vspan{P} \cdot (\abcobj')^{\le k} = \im{p \cdot \wincl{k}} \]
by \autoref{prop:span-concat}. Thus, $g$ is the inclusion of $\vspan{P \cdot \words[\le k]}$ into $(\abcobj')^*$. Since $\vspan{P \cdot \words[\le k]}$ is a state cover for $\aut{\implname}$ (i.e. $g$ is a state cover for $\aut{\implname}$), we have that $\malg{\aut{\implname}} \circ g$ is a split epi. Since $f$ is epi (and hence split epi), also $\malg{\aut{\implname}} \circ g \circ f = \malg{\aut{\implname}} \circ (p \cdot \wincl{k})$ is a split epi. Thus, $p \cdot \wincl{k}$ is a state cover for $\aut{\implname}$. By \autoref{prop:sc-is-weak-sc}, $p \cdot \wincl{k}$ is a weak state cover for $\aut{\implname}$. Hence, $\aut{\implname} \in \fdomgcomp[k]{p}$. \qed
\end{proof}

\begin{proof}[\autoref{thm:w-method-lwa}]
Let $\aut{\implname} \in \fdomwa{k}{P}$ such that $\reclang{\aut{\specname}}|_{\W{k}{P}{W}} = \reclang{\aut{\implname}}|_{\W{k}{P}{W}}$. Then we have $\agreets{t}{\aut{\specname}}{\aut{\implname}}$, where $t \colon \vspan{\W{k}{P}{W}} \to (\abcobj')^*$ is the inclusion, by \autoref{prop:wa-agree-ts}. Let $p \colon \vspan{P} \to (\abcobj')^*$ and $w \colon \vspan{W} \to (\abcobj')^*$ denote the inclusions. By \autoref{cor:gen-W-comp}, $\mW{k}{p}{w}$ is complete for $\aut{\specname}$ with respect to $\fdomgcomp[k]{p}$. By \autoref{cor:fact-comp} and \autoref{lem:wa-ts-image}, $t \colon \vspan{\W{k}{P}{W}} = \im{\mW{k}{p}{w}} \to (\abcobj')^*$ is also complete for $\aut{\specname}$ with respect to $\fdomgcomp[k]{p}$. Since $\aut{\implname} \in \fdomwa{k}{P}$, we have $\aut{\implname} \in \fdomgcomp[k]{p}$ by \autoref{lem:wa-fdom-incl}. Thus, since $\agreets{t}{\aut{\specname}}{\aut{\implname}}$, we get $\autequiv{\aut{\specname}}{\aut{\implname}}$. Hence, $\reclang{\aut{\specname}} = \reclang{\aut{\implname}}$, as desired. \qed
\end{proof}

\subsection{An implementation WA} \label{sec:ex-wa-impl}

We now return to the example in \autoref{sec:applications-wa}. Consider the faulty implementation $\aut{\implname}$ depicted in \autoref{fig:lwa-impl}.
\begin{figure}[ht]
\centering
\resizebox{100pt}{!}{
\begin{tikzpicture}[auto,accepting/.style=accepting by arrow]
	\node[state,initial below,initial text={$1$}] (0)              {$q_0$};
	\node[state,accepting,accepting text={$1$}]   (1) [right=of 0] {$q_1$};
	\node[state]                                  (2) [above=of 0] {$q_2$};
	\node[state,accepting,accepting text={$1$}]   (3) [below=of 1] {$q_3$};
	\node[state,accepting,accepting text={$1$}]   (4) [above=of 1] {$q_4$};
	\path[->]
				(0) edge              node {$a/1$}     (2)
						edge [loop left]  node {$b/1$}     ()
						edge              node {$b/1$}     (1)
				(1) edge              node {$a/2$}     (3)
						edge              node {$b/2$}     (4)
				(2) edge [loop above] node {$a/1,b/1$} ()
						edge              node {$b/1$}     (4)
				(3) edge [loop below] node {$a/2,b/2$} ()
						edge              node {$a/2$}     (0)
				(4) edge [loop above] node {$a/2,b/2$} ();
\end{tikzpicture}
}
\caption{A faulty implementation WA $\aut{\implname}$}
\label{fig:lwa-impl}
\end{figure}
We have $\wamatrix{\delta}{a}(k_0q_0 + k_1q_1 + k_2q_2 + k_3q_3 + k_4q_4) = 2k_3q_0 + (k_0 + k_2)q_2 + (2k_1 + 2k_3)q_3 + 2k_4q_4$ and $\wamatrix{\delta}{b}(k_0q_0 + k_1q_1 + k_2q_2 + k_3q_3 + k_4q_4) = k_0q_0 + k_0q_1 + k_2q_2 + 2k_3q_3 + (2k_1 + k_2 + 2k_4)q_4$. Hence, we have
\begin{align*}
\wamatrix{\delta}{\epsilon} q_0 &= q_0, \\
\wamatrix{\delta}{b} q_0 - \wamatrix{\delta}{\epsilon} q_0 &= (q_0 + q_1) - q_0 = q_1, \\
\wamatrix{\delta}{a} q_0 &= q_2, \\
\frac{1}{2} \wamatrix{\delta}{ba} q_0 - \frac{1}{2} \wamatrix{\delta}{a} q_0 &= \frac{1}{2}(q_2 + 2q_3) - \frac{1}{2}q_2 = q_3,\text{ and} \\
\frac{1}{2} \wamatrix{\delta}{bb} q_0 - \frac{1}{2} \wamatrix{\delta}{b} q_0 &= \frac{1}{2}(q_0 + q_1 + 2q_4) - \frac{1}{2}(q_0 + q_1) = q_4.
\end{align*}
Since $\{q_0, q_1, q_2, q_3, q_4\}$ is a basis for the state space of $\aut{\implname}$, by \autoref{prop:wa-sc-char} (i), $P \cdot \words[\le 1] = \{ \epsilon, a, b, ba, bb \}$ is a state cover for $\aut{\implname}$. Hence, $\aut{\implname} \in \fdomwa{1}{P}$. Indeed, the test case $baab$ (and only this) rejects $\aut{\implname}$ since $\reclang{\aut{\specname}}(baab) = 9 \ne 13 = \reclang{\aut{\implname}}(baab)$.